\documentclass[11pt]{article}
\usepackage[a4paper,hmargin=1.0in,vmargin=1.0in]{geometry}
\usepackage{amsmath,amsthm,amssymb,setspace,sectsty,bbm,graphicx,mathtools}
\usepackage[round]{natbib}
\usepackage[usenames,dvipsnames]{xcolor}
\usepackage[linktocpage=true,pagebackref=true,colorlinks,allcolors=blue,bookmarks=true,bookmarksopen,bookmarksnumbered]{hyperref}

\usepackage{booktabs}
\usepackage{pdfpages}
\newtheoremstyle{DStheorem}
{\topsep}
{\topsep}
{\itshape}
{0pt}
{\scshape}
{.}
{ }
{\thmname{#1}\thmnumber{ #2}\thmnote{ (#3)}}
\theoremstyle{DStheorem}
\newtheorem{theorem}{Theorem}[section]
\newtheorem{lemma}[theorem]{Lemma}
\newtheorem{proposition}[theorem]{Proposition}
\newtheorem{definition}[theorem]{Definition}
\newtheorem{claim}[theorem]{Claim}

\newtheorem{remark}[theorem]{Observation}
\newtheorem{fact}[theorem]{Fact}
\newtheorem{example}[theorem]{Example}
\newtheorem{assumption}[theorem]{Assumption}

\let\oldproofname=\proofname
\renewcommand{\proofname}{\rm\sc{\oldproofname}}

\usepackage{bm}
\usepackage{multirow}
\usepackage{array}
\newcolumntype{H}{>{\setbox0=\hbox\bgroup}c<{\egroup}@{}}
\usepackage{mathtools}
\usepackage{xcolor}
\usepackage{bbm}
\usepackage{algorithm}
\usepackage[noend]{algpseudocode}
\usepackage{wrapfig}
\usepackage{multirow}
\usepackage{booktabs}
\usepackage{makecell}
\usepackage{subcaption}
\usepackage{bibunits}

\makeatletter
\def\BState{\State\hskip-\ALG@thistlm}
\makeatother
\def\uno{{\mathbbm{1}}}
\def\RR{{\mathbb R}}
\def\R{{\mathcal R}}

\def\P{{\mathcal P}}
\def\EE{{\mathbb E}}

\def\PP{{\mathbb P}}

\def\N{{\mathcal N}}

\def\C{{\mathcal C}}

\def\U{{\mathcal U}}

\def\V{{\mathcal V}}
\def\E{{\mathcal E}}

\def\M{{\mathcal M}}
\def\A{{\mathcal A}}
\def\I{{\mathcal I}}

\def\S{{\mathcal S}}
\def\P{{\mathcal P}}

\def\argmax{\text{argmax\,}}

\DeclareMathOperator{\OPT}{OPT}

\newcommand{\Ma}{\mathbf{M}}

\newcommand{\revised}[1]{{\color{black} #1}}

\allowdisplaybreaks

\usepackage{hyperref}

\sectionfont{\large} \subsectionfont{\normalsize}
\allowdisplaybreaks
\makeindex
\newcommand{\changelocaltocdepth}[1]{%
  \addtocontents{toc}{\protect\setcounter{tocdepth}{#1}}%
  \setcounter{tocdepth}{#1}%
}
\begin{document}
\begin{titlepage}

\title{Multi-agent Assortment Optimization \\ in Sequential Matching Markets}

\author{%
\and Alfredo Torrico\thanks{ Daniel J. Epstein Department of Industrial and Systems Engineering, University of Southern California, Los Angeles, CA 90089, United States. Email: {\tt atorrico@usc.edu}.}
\and Margarida Carvalho \thanks{CIRRELT and DIRO, Université de Montr\'eal. Email: {\tt carvalho@iro.umontreal.ca}.}%
\and Andrea Lodi \thanks{Jacobs Technion-Cornell Institute, Cornell Tech. Email: {\tt andrea.lodi@cornell.edu}. }
}

\date{}
\maketitle

\setcounter{page}{200}
\thispagestyle{empty}

\begin{abstract}

Motivated by the challenges faced by choice-based matching platforms, in this work, we study a generalization of the foundational model introduced by Ashlagi et al. (2022). Specifically, this model consists of two sides (e.g., customers and suppliers) who have preferences over the opposite side in the form of discrete choice models. 
 A matching is realized through a two-step process where the platform's decision is to initially present to each agent on one side an assortment of users of the opposite side. The ultimate goal of the platform is to maximize the total expected revenue collected from the matching. Due to the computational complexity of this problem, we focus on the design of approximation algorithms with constant-factor guarantees for the constrained version of the problem under general choice preferences. First, to the best of our knowledge, we are the first to draw connections with the literature of submodular optimization and show that one-sided choice-oblivious approaches perform arbitrarily bad. Second, we show that, for the general setting, there exists a two-sided choice-aware randomized algorithm that achieves a $(1-1/e)$-approximation factor, which is the best possible. Third, we design an algorithm that is asymptotically optimal when one side follows the well-known multinomial-logit (MNL) choice model under the small probability regime. Fourth, we provide a simple and effective algorithm with a $1/4$-approximation guarantee for the unconstrained setting with both sides following MNL models. Finally, we assess the performance of our proposed methods with a detailed computational study and discuss several extensions.
\end{abstract}


\end{titlepage}

\newpage
\pagestyle{plain}
\setcounter{page}{1}

\section{Introduction}\label{sec:introduction}
Two-sided marketplaces have increasingly grown in popularity in the last decade and, nowadays, can be found in multiple areas such as labor (Upwork, TaskRabbit), accommodation (Airbnb, Vrbo), dating (Tinder, Bumble) and carpooling (Blablacar). Because of this, two-sided markets have been extensively studied over the recent years, e.g., \cite{rochet_tirole03,horton10,fradkin15,banerjee_etal15,rios2022improving,aouad2020online,besbes21,manshadi22,rodilitz22}. Choice-based matching platforms, such as the above, are generally characterized by: (i) having two sides (e.g., customers and suppliers) who are looking to match or make a transaction, (ii) being decentralized, i.e., the platform serves as a mediator between both sides by controlling what users see and how they interact, but not their choices. An immediate challenge arises in this two-sided decentralized setting: If the sets of alternatives (e.g., agents on the opposite side) that will be shown to the users are not appropriately designed then, due to correlated preferences, popular alternatives may get more request than they can actually handle; a problem known as \emph{choice congestion}. Therefore, the platform must correctly balance between showing enough alternatives to keep users engage and managing popular options to minimize choice congestion.

Assortment optimization \cite{talluri04,kok_etal08} appears as a natural algorithmic tool to control  choice congestion. However, this literature has mostly focused on single-customer settings where  the preferences of only one side are accounted for. The adaptation of this literature to two-sided scenarios is not immediate. \cite{ashlagi_etal19} recently introduced the foundational model for the literature of offline two-sided assortment optimization where matches between customers and  suppliers occur when they sequentially select each other during a two-step process. In this setting, they show that the problem of maximizing the expected number of matches is strongly NP-hard, even when both sides have a simple preference structure, which opens the door for the development of efficient algorithms with provable guarantees.  In particular, \cite{ashlagi_etal19} provide constant factor approximation guarantees for a simple unconstrained setting and, to the best of our knowledge, their techniques cannot be generalized to more complex scenarios. 
Our main goal in this work is to provide strong approximation guarantees for this foundational model under general choice preferences and constraints.

\subsection{Problem Formulation}\label{sec:formulation}
We study a generalization of the offline two-sided assortment optimization model originally introduced by \cite{ashlagi_etal19}. Formally, consider a two-sided market represented by a set $\C$ with $n$ customers and a set $\S$ with $m$ suppliers. The compatibility between customers and suppliers can be captured by a bipartite graph in which each edge represents that both agents find each other acceptable. For simplicity, we assume that every supplier is compatible with every customer; however, our results easily extend to the case in which suppliers are compatible to a subset of customers.
Our goal is to construct an assortment of suppliers $S_i\subseteq \S$ for each customer $i\in \C$ in order to maximize the expected revenue of the matching between customers and suppliers that is obtained as a result of a two-step process. 

\paragraph{Choice Preferences.} Following the terminology in the assortment optimization literature, we assume that the preferences of customers and suppliers are governed by \emph{discrete choice models}. Formally, the choices of each customer $i\in \C$ and each supplier $j\in \S$ are determined by functions $\phi_i:\S\cup\{0\}\times 2^\S\to[0,1]$ and $\phi_j:\C\cup\{0\}\times 2^\C\to[0,1]$, respectively, where $0$ denotes the \emph{outside option} of leaving the market unmatched. This means, for instance, that given a subset of options $S\subseteq \S$, a customer $i\in \C$ chooses $j\in S\cup\{0\}$ with probability $\phi_i(j,S)$ and for $j\notin S\cup\{0\}$ we have $\phi_i(j,S) = 0$.
In other words, for a given subset $S\subseteq\S$, the terms $\phi_i(\cdot,S)$ define a probability distribution over $S\cup\{0\}$, namely,
\(f_i(S) + \phi_{i}(0,S) = 1,\)
where $f_i(S) =\sum_{j\in S}\phi_{i}(j,S)$ is called the \emph{demand function} of customer $i$.
Similarly, given a subset of customers $C\subseteq \C$, we denote by $f_j(C)$ the demand function of supplier $j$.
\paragraph{Matching Process.} The process occurs in two steps: First, an assortment of suppliers $S_i\subseteq \S$ is offered to each customer $i\in \C$ from which they choose--simultaneously, independently and irrevocably--with probability $\phi_{i}(j,S_i)$ a  supplier $j\in S_i$  or leave the system with probability $\phi_{i}(0,S_i)$. 
For every supplier $j\in \C$, let $C_j$ be the random subset of customers (possibly empty) that chose $j$ in the first step. \revised{This set $C_j$ is generally referred to as the backlog of supplier $j$.} In the second step of the process, each supplier $j\in \S$ selects--simultaneously, independently and irrevocably--a single customer $i$ from $C_j$ with probability $\phi_{j}(i,C_j)$ or decides to remain unmatched with probability $\phi_{j}(0,C_j)$. 
Thus, a match occurs when both a customer and a supplier sequentially select each other. 

\paragraph{Optimization Problem.}  As we mentioned above, we would like to obtain an \emph{assortment profile},  $\mathbf{S}=(S_1,\ldots, S_n)$ that maximizes the expected revenue of the final matching. Let $r_j\in \RR_+$ be the revenue obtained when supplier $j\in \C$ is matched. Let us denote by $\I_i\subseteq 2^\S$ the family of feasible assortments for customer $i\in \C$ and $\I = \{\mathbf{S}\in 2^{\C\times \S}: \mathbf{S} = (S_1,\ldots, S_n), \ S_i\in \I_i \ \text{for all} \ i\in \C\}$ the collection of all feasible assortment profiles.   
We denote by $\Ma_{\mathbf{S}}$ the random variable that indicates the revenue of the final matching and whose distribution depends on the assortment profile $\mathbf{S}\in\I$. Therefore, we aim to solve the following revenue maximization problem 
\begin{equation}\label{def:opt_problem}
\max\{\EE[\Ma_{\mathbf{S}}]: \ {\mathbf{S}}\in \I\}.
\end{equation}
In the remainder of the manuscript, we denote by $\OPT$ the optimal value of Problem~\eqref{def:opt_problem} and $\mathbf{S}^\star$ an optimal assortment profile. 
Problem~\eqref{def:opt_problem} corresponds to a generalization of the setting introduced in \cite{ashlagi_etal19}. Specifically, the authors assume that customers are homogeneous with the same \emph{multinomial logit} (MNL) choice model---a well studied model in the assortment optimization literature~\cite{kok_etal08}---where the choice probabilities are proportional to the preference weights of the options in the assortment plus the weight of the outside option; for more details, see Definition~\ref{def:MNL} in Section~\ref{sec:contributions}. On the supplier side, \cite{ashlagi_etal19} assume that users choose \emph{uniformly} over their potential customers. Even in the simple setting with revenues $r_j=1$ for all $j\in\S$, \cite{ashlagi_etal19} show that Problem~\eqref{def:opt_problem} is strongly NP-hard. For more details, we refer the interested reader to Section \ref{sec:related_work}. 
Due to this hardness result, we focus on the design of polynomial-time \emph{approximation algorithms}. We say that a polynomial time algorithm achieves an approximation factor $\alpha\in(0,1]$ for Problem \eqref{def:opt_problem} if it outputs a feasible solution $\mathbf{S}\in \I$ such that $\EE[\Ma_{\mathbf{S}}] \geq \alpha\cdot \OPT$, where the expectation is also taken over the possible randomization of the algorithm. \revised{With a slight abuse of notation, we refer to algorithms with the best possible approximation guarantees as optimal algorithms and, similarly, to algorithms with suboptimal guarantees.}

\vspace{1em}
\noindent{\bf Notation.} In the remainder of the manuscript, we write vectors, matrices and collection of sets in bold and their components in italic, for example, $\mathbf{z}\in[0,1]^{\C\times\S}$ and $z_{i,j}$. For any set function $f$, we use the shorthand $f(S+e)$ to denote $f(S\cup\{e\})$. For two vectors $\mathbf{z}$ and $\mathbf{z}'$ we denote by $\mathbf{z}\vee\mathbf{z}'$, the vector whose components are $\max\{z_{i,j},z'_{i,j}\}$. Finally, we denote the dot product as $\langle\cdot,\cdot\rangle$.

\subsubsection*{Our Main Assumptions}\label{subsubsec:main_assumptions}
The model described above is sufficiently general in that if we do not impose any assumptions on the agents' choice structure, then one can easily construct an instance where  Problem~\eqref{def:opt_problem} cannot be approximated within any reasonable factor.\footnote{Consider one customer and $m$ suppliers. The customer's preferences follow a \emph{mixed multinomial-logit model} and suppliers' choices are deterministic, i.e., they choose the customer with probability one if present in the assortment. \cite{desir2022capacitated} show that in this case (suppliers are instead items), the unconstrained (single customer) assortment optimization problem cannot be approximated within a factor better than $O(1/K^{1-\delta})$ for any $\delta>0$, unless NP$\;\subseteq\;$BPP, where $K$ is the number of mixtures.} In the following, we briefly discuss our main assumptions that, we emphasize, are not restrictive and still allow for a wide range of choice preferences. On the customers' side, we do not impose any specific structure on their choice models, however, we make the standard assumption that we have access to an efficient exact optimization oracle for the single-customer revenue optimization problem.  
\begin{assumption}[Oracle]\label{assumption:oracle}
For every customer $i\in \C$, there exists a polynomial-time algorithm that solves the single-customer assortment optimization problem. Formally, given values $\theta_{i,j}\geq0$ for all $j\in \S$, the algorithm returns a set $S\in\argmax\Big\{\sum_{j\in S}\theta_{i,j}\cdot\phi_i(j,S): \ S\in \mathcal{I}_i\Big\}$ in polynomial time.
\end{assumption}
Assumption~\ref{assumption:oracle} is not arbitrary as it has been previously considered in the literature of one-sided and two-sided online assortment optimization, see e.g.,~\cite{golrezaei2014real,goyal2020asymptotically,ma2020algorithms,aouad2020online}.
More importantly, it is not restrictive since several single-customer assortment optimization problems can be solved efficiently, for example, for logit-based models~\cite{talluri04,rusmevich_etal10,davis14,gallego2014constrained} and the Markov-chain choice model~\cite{blanchet16}.
Moreover, our results easily extend to the case in which we have access to an approximate oracle instead of an exact oracle.  

On the suppliers' side, we impose the following mild assumption for their choice models.
\begin{assumption}[Monotone Submodular Suppliers' Demand Functions]\label{assumption:suppliers_monotone_submod}
The demand function of any supplier $j\in\S$ is a monotone submodular set function. Formally, for each $j\in\S$, the function $f_j:2^\C\to[0,1]$ satisfies the following two properties: (i) for every $C\subseteq C'\subseteq \C$, we have $f_j(C)\leq f_j(C')$; (ii) for every $i\in\C$ and $C\subseteq C'\subseteq \C\setminus\{i\}$, we have $f_j(C+i)-f_j(C)\geq f_j(C'+i) - f_j(C')$.
\end{assumption}
We emphasize that this assumption is satisfied by a large class of choice models. First, any choice model that satisfies the \emph{substitutability axiom},\footnote{This axiom states that for any given $C\subseteq C'\subseteq\C$ and $i\in C\cup\{0\}$, then $\phi_j(i,C)\geq \phi_j(i,C')$, i.e., the probability of choosing a specific option does not increase when the assortment is enlarged. Then $f_j(C)=1-\phi_j(0,C)\leq 1-\phi_j(0,C') = f_j(C')$.} standard in the assortment optimization literature~\cite{berbeglia_joret20}, has a monotone demand function. Furthermore, the demand function of the Markov chain model~\cite{desir_etal15} and the wide class of random utility based models \cite{thurstone27,luce_suppes65,mcfadden73,baltas_doyle01}, which includes logit-based models, have a submodular demand function; see e.g., \cite{berbeglia_joret20} for a proof.  We also remark that, even though we assume that suppliers' actions are determined by choice models, our main technical results are valid when we consider a general \emph{aggregate} non-negative set function $f_j:2^\C\to\RR_+$ for each supplier $j\in \S$ that is monotone, submodular and does not necessarily result from a choice model.

\subsection{Our Contributions and Results}\label{sec:contributions}

In this work, we present constant factor guarantees for the offline two-sided assortment optimization Problem \eqref{def:opt_problem} under a general heterogenous setting of the  model introduced in~\cite{ashlagi_etal19}. To the best of our knowledge, we are the first to connect the growing literature of two-sided assortment optimization and submodular optimization. A summary of our results and a comparison with the previous literature is in Table~\ref{table:summary}.
\begin{table}[htbp]
  \caption{Summary of results: Comparing approximation factors and competitive ratios in the literature of one-sided and two-sided assortment optimization, and the submodular welfare problem.}\label{table:summary}
  \centerline{\scalebox{0.65}{\begin{tabular}{cccccccc}
    \toprule
   && & & &   \multicolumn{3}{c}{$\epsilon$-small Probability Regime}\\
   \cmidrule(lr){6-8}
  	&& Hardness & \makecell{Main \\ Guarantee}   & \makecell{MNL-MNL \\ (unconstrained)}  & General-MNL & General-NL & General-Inventory\\
 \midrule
  \makecell{Offline Submodular Welfare \\ \cite{vondrak2008optimal,khot2008inapproximability}} & & $1-1/e$ & $1-1/e$  & NA & NA   &  NA  & NA \\
  \midrule
  \makecell{Offline Two-sided \\ \cite{ashlagi_etal19}} && - & - & ${\approx 10^{-5}}^*$ & -   &  -  & -\\
  \midrule
  \makecell{Offline Two-sided\\ {\bf (this work)}} & &\makecell{$1-1/e$ \\ (Prop.~\ref{prop:optimal_factor})  }& \makecell{$1-1/e$\\ (Thm.~\ref{thm:main_general})} & \makecell{$1/4^{**}$ \\(Thm.~\ref{thm:mnl_mnl}) } & \makecell{$1-\epsilon$ \\ (Thm.~\ref{thm:fw_nearoptimality_epsregime})} & \makecell{$1-O(\frac{1}{n})$ \\(Prop.~\ref{prop:approx_nl})} & \makecell{1\\ Remark~\ref{remark:inventory}} \\
  \midrule
 \makecell{Online Two-sided\\ ~\cite{aouad2020online}} & \makecell{adv. \\ i.i.d.}& \makecell{$1/2$\\  $1-1/e$}  &  \makecell{ $1/2$\\ $1-1/e$} & - &  \makecell{ - \\ $0.67-O(\epsilon)$}  & \makecell{ - \\ $\alpha\in (1-1/e , 0.7)^{***}$} & - \\
 \midrule
 \makecell{Online One-sided\\ ~\cite{golrezaei2014real}} & \makecell{adv. \\ i.i.d.}& \makecell{$1-1/e$\\  - }  &  \makecell{ $1/2$\\ -} & NA &  NA  & NA & \makecell{$1-1/e$ \\ 3/4} \\
 \bottomrule 
  \end{tabular}}
  \vspace{0.2cm}
  }
  \centerline{\begin{minipage}[t]{15cm}
\footnotesize{\emph{Note.} Types of instances are written as Model-Model specifying the model on each side, e.g., MNL-MNL means that both sides follow MNL choice models. ``NA'' means that the problem cannot capture that type of instances. $^*$Valid only for a particular case of MNL-MNL. $^{**}$The $(1-1/e)$-factor for the general case applies to the MNL-MNL, however, we present a simple and effective algorithm that achieves $1/4$.} $^{***}$The competitive ratio is strictly better than $1-1/e$ but smaller than 0.7. \revised{For description of each online setting, we refer the interested reader to Section~\ref{sec:related_work}.}
\end{minipage}} 
\end{table}

First, in Section~\ref{sec:warmup}, we provide a warm-up discussion on the connections between Problem~\eqref{def:opt_problem} and the literature of submodular optimization. 
Specifically, we compare our setting with the offline \emph{submodular welfare problem}~\cite{lehmann_etal06,vondrak2008optimal} and highlight the main differences. We first note that the submodular welfare problem is a special case of our framework, which leads to our first result:
\begin{proposition}\label{prop:optimal_factor}
Unless $P=NP$, for any $\delta>0$, there is no $(1-1/e+\delta)$-approximation algorithm (even randomized) for Problem~\eqref{def:opt_problem}. This result holds even under rank-based preferences that satisfy the $\epsilon$-small probability regime (see Definition~\ref{def:epsilon_small} below).
\end{proposition}
More importantly, we show that one-sided approaches that are oblivious to the preferences of one of the sides (and natural candidates for the submodular welfare problem) perform arbitrarily bad.
\revised{Finally, we introduce our main algorithmic approach and highlight the main aspects of our algorithms. In summary, our meta-algorithm, formalized in Algorithm~\ref{alg:meta_algorithm}, solves a given relaxation of Problem~\eqref{def:opt_problem} to obtain an assortment distribution for each customer and, subsequently, sample one from it. To some extent, all the algorithms analyzed in this work will follow this approach.}

\revised{In Section~\ref{sec:optimal_algorithms}, we focus on optimal or near-optimal algorithms.} Specifically, in Section~\ref{sec:general_case}, we study the most general constrained setting where both sides may have general, and possibly different, choice models satisfying Assumption~\ref{assumption:oracle} and~\ref{assumption:suppliers_monotone_submod}. Our main result is:
\begin{theorem}\label{thm:main_general}
There exists a randomized polynomial-time algorithm that, under Assumptions~\ref{assumption:oracle} and~\ref{assumption:suppliers_monotone_submod}, achieves (in expectation) an optimal $(1-1/e)$-approximation factor for Problem~\eqref{def:opt_problem}.
\end{theorem}
This result is a significant improvement over the initial (unconstrained) model studied in~\cite{ashlagi_etal19} for which a $\approx 10^{-5}$-factor  was shown. Our method consists of a \emph{choice-aware variant of the continuous greedy algorithm} that accounts for the preferences of both sides. Specifically, our algorithm relies on a continuous submodular relaxation of Problem~\eqref{def:opt_problem} and consists of greedily constructing two fractional points: a choice probability vector and an assortment distribution. These points are iteratively updated by following an ascent direction obtained from $n$ single-customer revenue maximization problems, which can be solved efficiently thanks to~Assumption~\ref{assumption:oracle}. At the end of this process, an assortment for each customer is sampled from the corresponding distribution.

In Section~\ref{sec:general_mnl}, we study a particular class of instances called $\epsilon$-small probability regime which usually captures large markets where suppliers are significantly ``picky'', i.e., when presented only one alternative, the probability of selecting this individual is small. Formally, we define it as follows:
\begin{definition}[{\large$\epsilon$}-Small Probability Regime]\label{def:epsilon_small}
Given $\epsilon\in(0,1)$, we define the $\epsilon$-small probability regime as the family of instances for which the suppliers' demand functions are such that $f_j(\{i\})\leq \epsilon\cdot f_j(\C)$ for all $i\in\C$ and $j\in\S$.
\end{definition}
This class of instances has been previously studied in the literature of online matching as the small bid regime, see e.g.,~\cite{mehta2007adwords,mehta2012online}.
\revised{This regime has also been studied in online assortment optimization~\cite{golrezaei2014real,goyal2020asymptotically,aouad2020online}. In the online model, customers arrive sequentially over time (possibly in adversarial manner), and the decision-maker must display irrevocably an assortment of options (capacitated resources in~\cite{golrezaei2014real} or suppliers in~\cite{aouad2020online}) to each arriving customer. The goal then is to maximize the expected total reward (one-sided model) or size of the matching (two-sided model). In this setting, the $\epsilon$-small probability regime translates either in the smallest item capacity going to infinity (one-sided model) or picky suppliers (two-sided model).}  
In our model, we aim to show that guarantees better than $1-1/e$ can be obtained in the small probability regime. We remark that, given Proposition~\ref{prop:optimal_factor}, $1-1/e$ is the best possible even under rank-based preferences that satisfy Definition~\ref{def:epsilon_small}. Therefore, we focus on a simpler class of choice models, arguably the most used in practice, called multinomial-logit.
\begin{definition}[MNL]\label{def:MNL}
Fix $j\in\S$. Assume that $j$ has a preference weight $w_{j,i}\geq0$ for every customer $i\in\C$ and $w_{j,0}=1$\footnote{Without loss of generality, we normalized the value of the outside option.} is the value for the outside option.\footnote{The value $w_{j,i}$ can be interpreted as how attractive customer $i$ is to supplier $j$ and $w_{j,0}$ is how attractive the outside option is, e.g., a competing platform.} Then, a supplier $j\in\S$ follows a MNL choice model if, for any assortment $C\subseteq\C$ we have
\(
\phi_{j}(i,C) = {w_{j,i}}/{(1+w_j(C))}, 
\)
for $i\in C\cup\{0\}$, where $w_j(C):=\sum_{\ell\in C}w_{j,\ell}$. Similarly, we define the MNL choice model for a customer $i\in \C$ with known values $v_{i,j}\geq0$ for every supplier $j\in\S$ and $v_{i,0}=1$ their value for the outside option. 
\end{definition}
\vspace{0.5em}
For this class of choice models, we show the following result:
\begin{theorem}\label{thm:fw_nearoptimality_epsregime}
Assume that each customer follows a general choice model that satisfies Assumption~\ref{assumption:oracle} and each supplier follows a MNL choice model satisfying the $\epsilon$-small probability regime with $\epsilon\in(0,1)$. Then, there exists a polynomial-time randomized algorithm that guarantees a $(1-\epsilon)$-approximation factor for Problem~\eqref{def:opt_problem}.
\end{theorem}
In other words, we can guarantee near-optimality when suppliers follow MNL choice models satisfying Definition~\ref{def:epsilon_small}. This marks a clear difference with the i.i.d. online variant of Problem~\ref{def:opt_problem} for which, under the $\epsilon$-small probability regime, \cite{aouad2020online} showed a competitive ratio of $0.67-O(\epsilon)$ slightly improving the general $1-1/e$ guarantee; see Table~\ref{table:summary} for a comparison. Our method consists of a choice-aware variant of the Frank-Wolfe algorithm that relies on a concave relaxation. As in our continuous greedy algorithm, this method iteratively updates fractional solutions and, at the end of the process, samples an assortment for each customer from the resulting distribution. Thanks to a key result in  Lemma~\ref{lemma:supplier_mnl_bound1}, satisfied by the demand functions of the MNL models, we are able to prove our guarantee. We finish Section~\ref{sec:general_mnl} by showing that our \emph{choice-aware Frank-Wolfe} algorithm guarantees a sub-optimal $1/4$-approximation factor under any regime.

\revised{In Section~\ref{sec:efficient_algorithms}, our focus is on efficient algorithms. We start  by presenting a discrete variant of the choice-aware continuous greedy algorithm. Unlike the continuous version, this algorithm construct an assortment family by iteratively simulating the choices of the $n$ customers. Formally, we show that the \emph{choice-aware discrete greedy} guarantees a (suboptimal) $1/2$-approximation factor for any type of instances, as long as Assumptions~\ref{assumption:oracle} and~\ref{assumption:suppliers_monotone_submod} are satisfied.}
In Section~\ref{sec:unconstrained_mnl_mnl}, our goal is to design  an effective method for an immediate generalization of the model in \cite{ashlagi_etal19}. Specifically, we assume that suppliers and customers follow (possibly) different MNL choice models and there are no constraints on the assortments.\footnote{Recall that \cite{ashlagi_etal19} considers the unconstrained setting with customers following the same MNL model and suppliers following a uniform model (a special case of MNL).} This setting is relevant because it represents the ``simplest'' scenario in (offline) two-sided assortment optimization relative to (offline) one-sided assortment optimization~\cite{talluri04}.
Our main result is the following:
\begin{theorem}\label{thm:mnl_mnl}
Assume that both suppliers and customers have MNL choice models and that $\I_i=2^\S$ for every customer $i\in\C$. Then, there exists an efficient polynomial-time algorithm that samples an assortment for each customer from a nested family composed by at most $m+1$ assortments that guarantees a $1/4$-approximation factor for Problem~\eqref{def:opt_problem}.
\end{theorem}
First, we remark that the $1-1/e$-factor in Theorem~\ref{thm:main_general} still applies to this setting, however, the $1/4$-approximation algorithm is significantly more efficient. This method \revised{(which we referred to as the \emph{MNL-MNL method} and formalized in Algorithm~\ref{alg:mnl_mnl})}, first, solves a (novel) polynomially-sized concave relaxation whose fractional solution is used to construct, for each customer, an assortment distribution over $m+1$ nested assortments. Then, it samples a solution from this distribution without loss in the objective value. \revised{We finish this section by presenting a detailed computational study in Section~\ref{sec:computational_study} where we empirically evaluate the performance (objective value and running time) of our methods. Surprisingly, the MNL-MNL method significantly outperforms the algorithms with optimal guarantees, both in time and objective value. On the other hand, the choice-aware discrete greedy algorithm empirically demonstrate to be a versatile and efficient alternative to its time-consuming continuous variant.

Finally, in Section~\ref{sec:other_results}, we provide several extensions and other discussions. One natural question is whether near-optimality, under the $\epsilon$-small probability regime, can be obtained for other choice models rather than MNL. In particular, we show that this is possible for the class of \emph{nested-logit} (NL) choice models, a generalization of MNL. We also draw connections between the $\epsilon$-small probability regime and the curvature of a set function. Then, we discuss in which scenarios monotonicity and submodularity are lost, and provide some insights on what approaches are suitable in these cases. Finally, we provide an additional discussion in Appendix~\ref{sec:operational_insights} where we study relevant aspects of the matching market, especially related to how the order of interactions between customers and suppliers in the matching process, along with selectivity and the size of each side, impacts the optimal expected number of matches.

\subsection{Related Work}\label{sec:related_work}
Assortment optimization has been extensively studied in the revenue management literature. The classic offline model focuses on one-sided markets in which there is a single customer who chooses an item from a given menu. The customer's behavior is determined by an underlying choice model. This problem was introduced by \cite{talluri04} who consider the MNL choice model and show that the problem can be solved in polynomial time by checking the revenue-ordered assortments. However, the problem becomes NP-hard for other choice models \cite{bront_etal09,rusmevich14,aouad18}.  \cite{berbeglia_joret20} show that the revenue-ordered assortments strategy provides provable guarantees for a general regular discrete choice model. Other examples of discrete choice models that have been studied in the assortment optimization problem are the Mixed Multinomial Logit \cite{rusmevich14}, the Nested Logit \cite{davis14}, the Rank-based \cite{aouad18} and the Markov chain-based \cite{blanchet16} models. For more details we refer to the survey on assortment optimization~\cite{kok_etal08}. Assortment optimization under capacity constraints has also been studied \cite{rusmevich_etal10,sumida2021revenue,desir_etal15}. In particular, under the MNL model the problem  is polynomial-time solvable \cite{rusmevich_etal10}, whereas under the Markov chain model, the problem becomes APX-hard \cite{desir_etal15}. The online variant of one-sided assortment optimization has also been studied under several settings and arrival models, see e.g.~\cite{golrezaei2014real,ma2020algorithms,goyal2020asymptotically,feng2022near,gong2022online}. In particular, \cite{golrezaei2014real} consider the setting where each ``supplier'' (item) has an inventory captured by a function $f_j(C) = \min\{|C|,c_j\}$ and show that under the large-inventory regime (analogous to the $\epsilon$-small probability regime) one can get a $0.75$ competitive ratio for i.i.d. arrivals; see Table~\ref{table:summary} for a comparison. 
There is a growing literature on assortment planning in two-sided markets. The foundational model of two-sided assortment optimization was introduced by \cite{ashlagi_etal19} who focus on the specific unconstrained setting with homogeneous customers following the MNL choice model and suppliers with uniform choice models.
As anticipated, under these conditions, \cite{ashlagi_etal19} show that Problem \eqref{def:opt_problem} is strongly NP-hard and provide constant-factor approximation algorithms. More precisely, the authors separate the analysis in two cases: $v_j\geq 1$ and $v_j\leq 1$, where $v_j$ is the value that a customer has over supplier $j$. For the former, they provide a $(1-e^{-1/24})/8$-approximation factor by appropriately solving a concave relaxation. For the second case, they consider a \emph{bucketing} technique to design an auxiliary linear programming problem whose fractional solution can be rounded to obtain an approximately $(e-1)/(61,952e)$-approximation factor. By appropriately combining these analyses, they show a $(e-1)/(123,904e)$-approximation factor for the general case with no restriction on the $v_j$'s. Recently, \cite{aouad2020online} studied an online two-sided assortment optimization model \revised{where customers arrive in an online fashion and suppliers' availability is also dynamic. In this model, the decision-maker must sequentially display an assortment of suppliers to each arriving customer with the goal of maximizing the expected total size of the matching.} The authors design a greedy algorithm that guarantees a $1/2$-competitive ratio with respect to the optimal clairvoyant benchmark (the dynamic program) and show that no online algorithm can improve this factor. Under known i.i.d. arrivals, they show that there exists an online algorithm that guarantees a $1-1/e$ competitive ratio and relies on a linear program relaxation of the optimal value. Furthermore, they show improved competitive ratios when the instances of the problem fall in the $\epsilon$-small probability regime and when suppliers follow the multinomial-logit model or the nested-logit model.  \cite{rios2022improving} focus on the use of behavioral information and empirical approaches to improve the assortment decisions of dating platforms with the goal of optimizing matching rates; we remark that no provable guarantees are provided.
In a recent work, \cite{shi2022endogeneous} uses assortment planning to design an optimal match recommendation policy that maximizes the total market surplus. After an earlier version of the present manuscript, there have been several works on two-sided assortment optimization~\cite{ahmed2022parameterized,rios2023platform,housni2024two}. However, we remark that they focus on either parametric guarantees, different classes of policies or different types of markets, and none of their results apply to our setting.

Another related literature stream is the stochastic matching problem in the query-commit model, also known as stochastic probing with commitment. In this problem, the goal is to sequentially query  edges of a general graph where, after each query, the edge realizes with certain known probability. The goal is to obtain a matching of maximum cardinality from the realized edges. Starting with~\cite{Chen09}, who introduced the problem (with patience constraints) and provided the first performance guarantee, most of this literature has focused on settings where only one edge is queried at a time, see e.g.~\cite{adamczyk2011improved, costello2012stochastic,bansal2012lp,gamlath2019beating,hikima2021integrated,brubach2021improved,Jeloudar21}. This stream of literature differs from our work in that the probabilities are only edge-dependent and not assortment-dependent, i.e., they do not capture substitution effects. 

The last literature stream related to our work is on platform design that focuses on other tools rather than algorithmic approaches. For example, authors study how platforms can reduce choice congestion via  matching recommendations, signalling the available capacity of the providers \cite{fradkin15,horton17,li_netessine2020}, limiting the visibility \cite{halaburda_etal16,arnosti_etal18} and the choice~\cite{immorlica2021designing}, controlling which agents can initiate contact \cite{kanoria2021facilitating,ashlagi_etal2020clearing}, among others. Finally, \cite{shi2022strategy} studies the problem of improving market's efficiency through the lens of communication complexity.

\subsection{Warm-up Discussion: Connections with Submodular Optimization and Hardness}\label{sec:warmup}
In this section, we provide a brief discussion on the connections between our problem and the literature of submodular optimization and how na\"ive methods fail to provide optimal approximation guarantees.

First, let us rewrite the objective function in~\eqref{def:opt_problem}. For any supplier $j\in \S$ and assortment profile $\mathbf{S}\in \I$, let $Y^{\mathbf{S}}_j$ be the indicator random variable that takes value 1 when supplier $j$ ends up matched at the end of the two-step process and 0 otherwise. Also, for each supplier $j\in \S$ and assortment profile $\mathbf{S}\in \I$, let $C^{\mathbf{S}}_j$ be the random set of customers who selected supplier $j$ in the first step of the  process. Observe that, conditioned on $C^{\mathbf{S}}_j\subseteq \C$, the indicator variable $Y^\mathbf{S}_j$ equals 1 with probability $f_j(C^{\mathbf{S}}_j)\in[0,1]$ and $Y^\mathbf{S}_j=0$ otherwise, where $f_j(.)$ is the demand function of supplier $j$, i.e., the probability that $j$ chooses some customer from $C^{\mathbf{S}}_j$. Thus, for every $j\in\S$ and $\mathbf{S}\in\I$, we have
\[
\EE\left[ Y_j^{\mathbf{S}}\right]= \EE\left[\EE\left[ Y_j^{\mathbf{S}}| C^{\mathbf{S}}_j\right]\right] = \EE\left[ f_j(C^{\mathbf{S}}_j)\right],
\]
where the expectation is over the randomness of $C^{\mathbf{S}}_j$, i.e., the probabilistic choices of the customers. Therefore, the objective function in~\eqref{def:opt_problem} can be re-written as
\begin{equation}\label{eq:objective_alternative}
\EE[\Ma_{\mathbf{S}}] = \EE\left[\sum_{j\in\S} r_j\cdot Y_j^{\mathbf{S}}\right] = \EE\left[\sum_{j\in\S} r_j\cdot f_j(C^{\mathbf{S}}_j)\right].
\end{equation}
Fix $\mathbf{S}\in\I$. There are two immediate observations from~\eqref{eq:objective_alternative}. First, since every customer makes a single choice (either some supplier in their assortment or the outside option), then the sets $C^{\mathbf{S}}_j$ with $j\in\S$ and $C^{\mathbf{S}}_0$ (customers that chose the outside option) define a partition over the space $\C$. Second, as a function of $C^{\mathbf{S}}_j$ with $j\in\S$, the expression $\sum_{j\in\S}r_j\cdot f_j(.)$ is a monotone submodular function. Therefore, one could be initially tempted to obtain strong approximation guarantees for Problem~\eqref{def:opt_problem} in the following way: For any assortment profile $\mathbf{S}\in\I$ define the set function $f(\mathbf{S}):=\EE\big[\sum_{j\in\S} r_j\cdot f_j(C^{\mathbf{S}}_j) \big| \mathbf{S}\big]$, i.e., $f$ returns the expected revenue conditioned on $\mathbf{S}$. Then, we could use the methods from monotone submodular optimization directly over $f$. However, as we show in Example~\ref{example:not_monotone_nor_submodular} in Appendix~\ref{app:warmup}, this function is neither monotone nor submodular. This is the first aspect that highlights the differences between Problem~\eqref{def:opt_problem} and the submodular welfare problem~\cite{lehmann_etal06,vondrak2008optimal} defined as follows:

\paragraph{The Submodular Welfare Problem.} Consider a set of items $\V$ and a set of agents $\A$. Each agent $a\in\A$ has a monotone submodular utility set function $u_a: 2^\V\to\RR_+$. The goal is to allocate each item to a single agent such that the total utility is maximized. Formally, the objective is to assign a subset of items $V_a\subseteq\V$ to each agent $a\in\A$ such that the collection $\{V_a\}_{a\in\A}$ defines a partition over $\V$ and maximizes the total utility $\sum_{a\in\A}u_a(V_a)$.

Note that the submodular welfare problem is a special case of Problem~\eqref{def:opt_problem}: customers and suppliers correspond to items and agents, i.e., $\C = \V$ and $\S = \A$, respectively. For each customer $i\in \C$, we consider the uniform choice model $\phi_i(j,S) = 1/|S|$ and $\phi_i(0,S)=0$ for $S\subseteq\S$ such that $S\neq\emptyset$.\footnote{If $S=\emptyset$, then we set $\phi_i(0,S)=1$.} To ease the exposition, suppose that the agents' utility functions are normalized, then for each supplier $j\in\S$, we consider $r_j = 1$ and $\phi_j(i,C) = u_j(C)/|C|$ and $\phi_j(0,C)=0$. Finally, for each customer $i\in\C$ consider $\I_i = \{S\subseteq\S: \ |S|\leq 1\}$. From this reduction, we can easily note that \eqref{eq:objective_alternative} coincides with the objective function of the submodular welfare problem. Since the submodular welfare problem cannot be approximated within a $(1-1/e+\delta)$-factor for any $\delta>0$, unless P $=$ NP~\cite{khot2008inapproximability,kapralov2013online}, then we can conclude the hardness of approximation of Problem~\eqref{def:opt_problem}. For completeness, we provide a formal proof of  Proposition~\ref{prop:optimal_factor} in Appendix~\ref{app:hardness}, which applies even  under the $\epsilon$-small probability regime. 

As we mentioned above, function~\eqref{eq:objective_alternative} is not necessarily monotone submodular as a function of $\mathbf{S}$, even if one thinks of the multivariate version, see e.g., ~\cite{soma2018maximizing,santiago2019multivariate}. Given this, one cannot directly apply to our setting the approximation algorithms designed for the submodular welfare problem. In fact, these methods (one-sided in nature as they do not account for the preferences of both sides) perform arbitrarily bad as we discuss in the following section. 

\subsubsection{Poor Performance of Choice-oblivious Algorithms}\label{subsubsec:choice_oblivious_greedy}
One of the simplest approaches to approximately solve Problem~\eqref{def:opt_problem} is by greedily constructing the assortment of each customer without accounting for the suppliers preferences. We formalize this method in Algorithm~\ref{alg:single_customer}.
\begin{algorithm}
\caption{Choice-oblivious Greedy Algorithm}\label{alg:single_customer}
\begin{algorithmic}[1]
\Require  Customers' choice models $\phi_i$, revenues $r_j\geq0$.
\Ensure Assortment family $\mathbf{S}$.
\State For each customer $i\in \C$, consider
\(
S_i \in \argmax\left\{\sum_{j\in S} r_j\cdot \phi_{i}(j,S): \ S\in \I_i\right\}
\)
\end{algorithmic}
\end{algorithm}
Note that Algorithm~\ref{alg:single_customer} solves the classic single-customer assortment optimization problem for each customer $i\in\C$, which can be done efficiently due to Assumption~\ref{assumption:oracle}. We note that in the literature of online one-sided assortment optimization, the adaptation of Algorithm~\ref{alg:single_customer} is known as the \emph{myopic policy} and guarantees $1/2$-competitive ratio. 
Unfortunately, in our setting, the performance of this method can be arbitrarily bad as we show next.
\begin{proposition}\label{prop:guarantee_singlecustomer}
The worst-case approximation guarantee of Algorithm \ref{alg:single_customer} is $O(1/n)$, where $n$ is the size of the market.
\end{proposition}
The counterexample presented in this proposition shows that, when we do not account for the suppliers' preferences, then customers' requests can get ``highly concentrated'' on popular suppliers, which results in a suboptimal solution since at most one of those customers will be served by the supplier.

\begin{algorithm}
\caption{Choice-oblivious Submodular Welfare}\label{alg:oblivious_submodularwelfare}
\begin{algorithmic}[1]
\Require  Suppliers' demand functions $f_j$, revenues $r_j\geq0$.
\Ensure Assortment family $\mathbf{S}$.
\State (Approximately) Solve
\(
\mathbf{C}^\star\in \argmax\left\{\sum_{j\in\S}r_j\cdot f_j(C_j):\; C_j\subseteq \C, \ C_j\cap C_{j'}\neq\emptyset\text{ for }j\neq j'\in\S\right\}
\)
\State For each $i\in\C$, set $S_i = \{j\in\C: \ i\in C_j^\star\}$.
\end{algorithmic}
\end{algorithm}

One could think of a similar simple strategy where, instead of focusing on customers, we greedily construct assortments but only accounting for the suppliers' preferences. Formally, obtain $C^\star_j$ for all $j\in\S$ by (approximately) solving the submodular welfare problem defined only by the demand functions of the suppliers with no consideration of customers' preferences. Then, set $S_i = \{j: \; i\in C^\star_j\}$.\footnote{Note that since $C^\star_j$'s define a partition, then every assortment $S_i$ is of size 1. For simplicity, we can assume that singletons are feasible in each $\I_i$.} We formalize this method in Algorithm~\ref{alg:oblivious_submodularwelfare}.
Unfortunately, one can show that this approach also performs poorly.
\begin{proposition}\label{prop:guarantee_oblivious_sw}
The worst-case approximation guarantee of Algorithm \ref{alg:oblivious_submodularwelfare} is arbitrarily close to 0.
\end{proposition}
We defer the proofs of these results to Appendix~\ref{app:poor_performance}.
The results in this section highlight the differences between the submodular welfare problem and our setting that captures a broader class of problems. The main issue with both previous approaches is that none of them consider the preference structure of both sides of the market.

\subsection{Our Main Algorithmic Approach}\label{sec:meta_approach}
Given the preliminary results in the previous section, we first note  that an algorithm with strong guarantees must account for the choices of both sides of the market. To achieve this, our approach aims to connect the customers' choices with that of suppliers via their demand functions. Let us provide some intuition: Suppose a customer $i$ chooses a supplier $j$ in the first step, then the marginal probability that $j$ chooses $i$ back, i.e., $f_j(C+i)-f_j(C)$ where $C$ is set of customers that have chosen $j$ ``so far'' (i.e., $j$'s backlog), decreases as the backlog enlarges due to the diminishing returns property of the demand function $f_j$. Therefore, by having access to a rough estimation of the marginal values (which would indicate how important $i$ is for $j$ with the current backlog), then the algorithm would be able to ``weight'' $j$ accordingly when selecting an assortment to display to customer $i$. The second aspect that we incorporate in most of our methods is randomization.
Roughly speaking, our algorithms construct an assortment distribution for each customer and then from each distribution an assortment is sampled. The key part here is that these distributions are obtained by solving a tailored relaxation of Problem~\eqref{def:opt_problem}. Finally, the first and second points are connected as follows: While iteratively constructing each assortment distribution, we obtain an approximation for the expected marginal values $f_j(C+i)-f_j(C)$ for all $j\in\S,i\in\C$, which then affects subsequent iterations of the algorithm.

Now let us formalize our meta-algorithm. For each customer $i\in\C$, let $\tau_{i,\cdot}\in[0,1]^{\I_i}$ be a distribution over the family of feasible assortment of suppliers $\I_i$. Specifically, $\tau_{i,S}\geq0$ for all $S\subseteq\S$ and $\sum_{S\in\I_i}\tau_{i,S} = 1$. Therefore, the probability that $i$ sees supplier $j$ and chooses $j$ corresponds to $\sum_{\substack{S\in\I_i:\; j\in S}}\tau_{i,S}\cdot\phi_i(j,S)$. We now connect this marginal probability with each supplier $j$ objective function. Denote the marginal probability as $z_{i,j} = \sum_{\substack{S\in\I_i:\; j\in S}}\tau_{i,S}\cdot\phi_i(j,S)$ and consider an objective function $G_j:[0,1]^{\C\times\S}\to\RR_+$ defined over the space of marginal probabilities, i.e., for each $\mathbf{z}\in [0,1]^{\C\times\S}$, $G_j(\mathbf{z})$ outputs an upper bound to the expected demand of supplier $j\in\S$. In other words, we are aiming to construct a function $G_j$ for each $j\in\S$ such that  $\sum_{j\in\S}G_j(\mathbf{z})$ upper bounds the objective function in \eqref{eq:objective_alternative}. Therefore, our meta-relaxation is
\begin{align}\label{eq:meta_relaxation}
\max &\quad \sum_{j\in \C}G_j(\mathbf{z}_j) \tag{R}\\
s.t. &\quad (\mathbf{z},\tau)\in\P, \notag
\end{align}
where $\mathbf{z}_j$ is the $j$-th column of $\mathbf{z}$ and the feasible region $\P$ corresponds to
\begin{align}\label{eq:feasible_region_meta_relaxation}
\P = \Bigg\{(\mathbf{z},\tau)\in[0,1]^{\C\times\S}\times[0,1]^{\C\times \I_i}:& \ \sum_{\substack{S\in\I_i:\; j\in S}}\tau_{i,S}\cdot\phi_i(j,S)=z_{i,j}  \quad\text{for all }i\in\C, \ j\in\S\notag\\
&\sum_{S\in\I_i}\tau_{i,S} = 1, \hspace{6.8em}\text{for all }i\in\C,\notag\\
&\tau_{i,S}\geq 0, \hspace{9em} \text{for all }i\in\C, \ S\in\I_i
\Bigg\}.
\end{align}
Given the above, our meta-algorithm solves the relaxation~\eqref{eq:meta_relaxation} by using an appropriate iterative method that accounts for the choices of both sides as we discussed at the beginning of this section. We formalize this meta-algorithm in Algorithm~\ref{alg:meta_algorithm}.
\begin{algorithm}[htpb]
\caption{Meta-Algorithm}\label{alg:meta_algorithm}
\begin{algorithmic}[1]
\Require  Choice models $\phi$, feasibility assortment family $\mathcal{I}$, and Relaxation \eqref{eq:meta_relaxation}.
\Ensure An assortment $S_i$ for each customer $i\in\C$.
\State Compute a feasible solution $(\mathbf{z},\tau)$ of the meta-relaxation \eqref{eq:meta_relaxation} by using an  iterative ascent method.
\State For each $i\in\C$, independently sample an assortment from distribution $\tau_{i,\cdot}$.
\end{algorithmic}
\end{algorithm}
The main challenge now is to design an iterative method that deals with the first step in Algorithm~\ref{alg:meta_algorithm} since the set of distributions can be exponentially large. To resolve this, we note that we can compute ascent directions for each function $G_j$ by optimizing a linear function over $\P$, which can be done efficiently thanks to Assumption~\ref{assumption:oracle}. Formally, for a non-negative weight vector $\bm\theta\in\RR_+^{\C\times\S}$, we are interested in solving the linear program $\max\{\sum_{i\in\C}\sum_{j\in\S}\theta_{i,j}z_{i,j}: \ (\mathbf{z},\tau)\in\P\}$. Note that in this linear program, the objective and feasible region are separable. In addition, $\mathbf{z}$ and $\tau$ are connected by the first equality in $\P$. Therefore, this linear program is equivalent to solving $n$ independent  single-customer revenue maximization problems of the form $\max\{\sum_{j\in\S}\theta_{i,j}\cdot\phi_i(j,S): \ S\in\I_i\}$, one for each customer $i\in\C$. Each of these subproblems can be solved efficiently thanks to Assumption~\ref{assumption:oracle}. We remark that each quantity $\theta_{i,j}$ is not arbitrary and will be designed in such a way that a supplier $j$ with a larger expected backlog has less weight for each customer. Finally, our iterative methods will use the assortment from each of these linear programs to construct a feasible fractional ascent direction that (i) improves the global objective of the relaxation, and (ii) keeps a feasible distribution over assortments.
}

\section{Optimal Algorithms}\label{sec:optimal_algorithms}

In this section, we discuss our main algorithms and their (near-)optimal guarantees, namely Theorems~\ref{thm:main_general} and~\ref{thm:fw_nearoptimality_epsregime}. For this, we follow the meta-framework described in Section~\ref{sec:meta_approach}.
\subsection{Optimal Approximation Guarantee for General Instances}\label{sec:general_case}
Our goal now is to prove Theorem~\ref{thm:main_general}. Specifically, we show that for general instances that satisfy Assumptions~\ref{assumption:oracle} and~\ref{assumption:suppliers_monotone_submod}, there is an algorithm based on a continuous submodular relaxation of Problem~\eqref{def:opt_problem} that guarantees a $1-1/e$ approximation factor. In the following, we draw connections between the objective function of Problem~\eqref{def:opt_problem} and the  multilinear extension of set functions, and present some properties of this extension when the set function is monotone and submodular. Then, in Section~\ref{sec:analysis_optimal_guarantee_general}, we present the analysis of our main algorithm.

Recall that customers' choices are independent in the first step of the matching process, thus we can re-write the objective function of Problem~\eqref{def:opt_problem} as follows.
\begin{claim}\label{claim:multilinear_expression}
Expression~\eqref{eq:objective_alternative} can be equivalently written as
\begin{equation} \label{eq:short_identity}
\EE[\Ma_{\mathbf{S}}]=\sum_{j\in \S}r_j\cdot \sum_{C\subseteq \C}f_j(C)\prod_{i\in C}\phi_{i}(j,S_i)\prod_{i\in\C\setminus C}(1-\phi_{i}(j,S_i))
\end{equation}
\end{claim}
For completeness, we include the details of this claim in Appendix~\ref{sec:missing_details_optimal}. Expression \eqref{eq:short_identity} is crucial for the analysis of our main algorithm as it is used to build a continuous submodular relaxation of Problem~\eqref{def:opt_problem}. To derive this relaxation, 
let us briefly recap the definition of the multilinear extension of a set function.

\paragraph{Multilinear Extension.}
Consider a ground set of elements $\U$ and a generic non-negative set function $f:2^{\U}\to\RR_+$. 
The multilinear extension $F: [0,1]^{\U} \rightarrow \RR_+$ of function $f(.)$ is defined for any fractional vector $\mathbf{z}\in [0,1]^{\U}$ as the expected value of $f(A)$ where $A$ is the random set generated by drawing independently each element $e\in \U$ with probability $z_e$. Formally,
\begin{equation}\label{eq:ML_def} 
F(\mathbf{z})  = \EE_{A\sim \mathbf{z}}[f(A)]=\sum_{A \subseteq \U} f(A) \prod_{e \in A} z_e \prod_{e \in\U\setminus A} (1-z_{e}).
\end{equation}
Observe that $F$ is in fact an extension of $f$, since for any subset $A\subseteq \U$, we have $f(A)=F(\uno_A)$, where $\uno_A\in\{0,1\}^{\U}$ is the indicator vector of $A$, i.e., $\uno_A(e) = 1$ if $e\in A$ and zero otherwise.
\begin{fact}{\emph{\cite{calinescu2011maximizing}}.}\label{fact:ML} 
Let $f$ be a monotone submodular function and $F$ its multilinear extension. By monotonicity of $f$, we have $\frac{\partial F}{\partial z_e} \geq 0$ for any $e\in \U$. This implies that for any $\mathbf{z}\leq \mathbf{z}'$ coordinate-wise, $F(\mathbf{z})\leq F(\mathbf{z}')$. On the other hand, by submodularity of $f$, then for any pair of vectors $\mathbf{z}, \mathbf{z}'\in[0,1]^\U$, we have that $F(\mathbf{z}\vee\mathbf{z}') -F(\mathbf{z})\leq\sum_{e\in\U}\EE_{A\sim\mathbf{z}}[f(A\cup\{e\})-f(A)]\cdot z'_e\leq \langle\nabla F(\mathbf{z}),\mathbf{z}'\rangle,$ where the expectation is over the product distribution defined by $\mathbf{z}$.
\end{fact}
Now, we return to our problem. Observe that we can write the objective function of Problem~\eqref{def:opt_problem} as a linear combination of several multilinear extensions. For each $j\in \S$, denote by $F_j:[0,1]^\C\to[0,1]$ the multilinear extension of the demand function $f_j:2^\C\to [0,1]$. Thanks to Claim~\ref{claim:multilinear_expression}, we conclude that the objective function of Problem~\eqref{def:opt_problem} corresponds to
\begin{equation}\label{eq:multilinear_version_obj}
\EE[\Ma_{\mathbf{S}}] = F(\Phi(\mathbf{S})) := \sum_{j\in \S}r_j\cdot F_j(\Phi_j(\mathbf{S})),
\end{equation}
where, for each supplier $j\in\S$, $\Phi_j(\mathbf{S})\in[0,1]^\C$ is the vector of choice probabilities whose $i$-th component corresponds to $\phi_{i}(j,S_i)$.\footnote{Recall that for $j\in\S$ we have $\phi_{i}(j,S_i)=0$ whenever $j\notin S_i$. Since the demand functions are monotone and submodular, their multilinear extensions satisfy the properties described in Fact~\ref{fact:ML}}

\subsubsection{Analysis}\label{sec:analysis_optimal_guarantee_general}

Our algorithm follows the general approach described in Section~\ref{sec:meta_approach}. Specifically, first, we design a continuous submodular function over the space of assortment distributions. Formally, the objective function of this relaxation corresponds to $\sum_{j\in \C}r_j\cdot F_j(\mathbf{z}_j)$
where, for each $j \in \S$, $\mathbf{z}_j\in[0,1]^\C$ is the vector whose $i$-th component is $z_{i,j}$ and $F_j$'s are defined as in \eqref{eq:multilinear_version_obj}. On the other hand, we consider the feasible region over the space of assortment distributions $\P$ given in \eqref{eq:feasible_region_meta_relaxation}.
Thus, our main continuous submodular relaxation is the following:
\begin{align}\label{eq:relaxation_general_case}
\max &\quad \sum_{j\in \C}r_j\cdot F_j(\mathbf{z}_j)\\
s.t. &\quad (\mathbf{z},\tau)\in\P. \notag
\end{align}

\begin{lemma}\label{lemma:relaxation_general_case}
Problem~\eqref{eq:relaxation_general_case} is a relaxation of Problem~\eqref{def:opt_problem}.
\end{lemma}
We defer the proof of this lemma to Appendix~\ref{sec:proofs_general_case}.
As we discussed earlier in Section~\ref{sec:meta_approach}, there are two main structural properties that we will use in the design of our  method. First, recall that thanks to Assumption~\ref{assumption:oracle}, we can efficiently maximize a linear function over $\P$ by solving $n$ (single customer) revenue maximization problems. Second, due to Assumption~\ref{assumption:suppliers_monotone_submod}, the demand functions of the suppliers are monotone and submodular, so we can exploit the properties of the multilinear extension shown in Fact~\ref{fact:ML}. 
Both structural properties play a key role in our \emph{choice-aware continuous greedy method} which is inspired by the algorithmic ideas used for the problem of maximizing a monotone submodular function over a single matroid constraint~\cite{vondrak2008optimal,calinescu2011maximizing}. Roughly speaking, our continuous greedy algorithm updates two objects, the choice probability vector ${\Phi}$ and the assortment distribution $\bm{\tau}$, by aggregating the fractional feasible solution built from solving $n$ single customer revenue optimization problems.
The weights $\theta_{i,j}$ in these subproblems are determined by the expected marginal probability that supplier $j$ chooses customer $i$ back. At the end of this process, our algorithm returns an assortment distribution $\bm{\tau}$ from which we can sample an assortment for each customer without loss in the objective function. We formalize our method in Algorithm~\ref{alg:continous_greedy_choice}.

To ease the exposition of the analysis, we focus on the continuous-time version and we leave the formal details of the discretized version to Appendix~\ref{sec:proofs_general_case}. For the continuous-time variant, we can think of two points $\mathbf{z}(t)\in[0,1]^{\C\times\S}$ and $\bm{\tau}(t)\in[0,1]^{\C\times 2^\S}$ making a continuous trajectory with respect to time $t\in[0,1]$ as follows: Both points start at $\mathbf{z}(0) = \mathbf{0}$ and $\bm{\tau}(0) = \mathbf{0}$, and their trajectory is determined by the following differential equations
\begin{align*}
\frac{\mathrm{d}\mathbf{z}}{\mathrm{d}t} = \Phi(\mathbf{S}(t)), \qquad \text{and} \qquad \frac{\mathrm{d}\bm{\tau}}{\mathrm{d}t} = \uno_{\mathbf{S}(t)},
\end{align*}
where $\Phi(\mathbf{S}(t))\in[0,1]^{\C\times\S}$ is the vector whose $(i,j)$-th component is $\phi_{i}(j,S_i(t))$, $\uno_{\mathbf{S}(t)}\in\{0,1\}^{\C\times2^\S}$ is the indicator vector that is equal 1 in component $(i,S_i(t))$ for $i\in \C$ and zero otherwise, and $S_i(t)$ at time $t$ is defined as an optimal solution of 
\begin{equation}\label{eq:linear_maximization_subproblem}
S_i(t) \in \argmax\left\{\sum_{j\in S}\theta_{i,j}(t)\cdot\phi_i(j,S): \ S\in\I_i\right\},
\end{equation}
where the weights in each subproblem are defined as $\theta_{i,j}(t) := r_j\cdot\EE_{C\sim\mathbf{z}_j(t)}[f_j(C+i)-f_j(C)]$. Due to Assumption~\ref{assumption:oracle}, Problem~\eqref{eq:linear_maximization_subproblem} can be solved efficiently for each $i\in\C$. 
The term $\EE_{C\sim\mathbf{z}_j(t)}[f_j(C+i)-f_j(C)]$ corresponds to the expected marginal probability that supplier $j$ chooses customer $i$ back, where the expectation is over the randomness of the backlog $C$ sampled from the independent customer choices made according to current choice probability vector $\mathbf{z}_j(t)$ whose $i$-th coordinate is defined as
\begin{equation}\label{eq:relation_z_tau}
z_{i,j}(t) = \sum_{S\in\I_i:\; j\in S}\tau_{i,S}(t)\cdot \phi_i(j,S).
\end{equation}
In other words, we have that
\begin{equation}\label{eq:expected_marginal}
\EE_{C\sim\mathbf{z}_j(t)}\big[f_j(C+i)-f_j(C)\big] = \sum_{C\subseteq\C}\big[f_j(C+i)-f_j(C)\big]\cdot\prod_{i\in C}z_{i,j}(t)\prod_{i\in\C\setminus C}(1-z_{i,j}(t)).
\end{equation}
\begin{remark}
Since Expression~\eqref{eq:expected_marginal} is defined as the sum over an exponential number of subsets, then we need to estimate this value for each $i\in\C, \ j\in\S$. This can be done by sampling a polynomial number of subsets as observed in Lemma 3.2 in~\cite{calinescu2011maximizing}. To ease the explanation of our algorithm, we assume that we have access to the exact values.
\end{remark}

\begin{algorithm}[htpb]
\caption{Discretized Choice-aware Continuous Greedy}\label{alg:continous_greedy_choice}
\begin{algorithmic}[1]
\Require  Choice models $\phi$ and feasibility assortment family $\mathcal{I}$, revenues $r_j$, and $\delta = 1/n^2$.
\Ensure Assortment family $\mathbf{S}$.
\State $\mathbf{z}^0 = \mathbf{0}$, $\tau^0_{i,S} = 0$,  for all $i\in \C$, $S\in\I_i$.
\For{$t=0,\delta,2\delta,\ldots,1-\delta$} 
\State For each $i\in\C$ obtain $S_i^t\in \argmax\left\{\sum_{j\in S}r_j\cdot\EE_{C\sim\mathbf{z}_j^t}[f_j(C+i)-f_j(C)]\cdot\phi_i(j,S): \ S\in\mathcal{I}_i\right\}$.
\State Set $\mathbf{z}^{t+\delta} \leftarrow \mathbf{z}^t + \delta\cdot \Phi(\mathbf{S}^t)$.
\State Set $\tau^{t+\delta}_{i,S_i^t} \leftarrow \tau^{t}_{i,S_i^t} + \delta$, for all $i\in \C$ and $\tau_{i,S}^{t+\delta}\leftarrow \tau_{i,S}^{t}$ for all $i\in\C$ and $S\neq S_i^{t}$.
\EndFor
\State For each $i\in\C$, independently sample an assortment from distribution $\tau_i^1$.
\end{algorithmic}
\end{algorithm}
\revised{As we can see from Equations~\eqref{eq:relation_z_tau} and~\eqref{eq:expected_marginal}, the expected marginal values $\EE_{C\sim\mathbf{z}_j(t)}\big[f_j(C+i)-f_j(C)\big]$ and $\tau(t)$ are closely related to each other, which we roughly discussed in Section~\ref{sec:meta_approach}. In other words, the current assortment distributions at time $t$ determine the values of $\mathbf{z}(t)$ via Equation~\eqref{eq:relation_z_tau} which then affect the expected marginal values in~\eqref{eq:expected_marginal}. These marginal values are used in step $t+\delta$ to obtain the next assortment family $\mathbf{S}(t+\delta)$ via~\eqref{eq:linear_maximization_subproblem}.} We now provide an sketch analysis of the continuous-time version of Algorithm~\ref{alg:continous_greedy_choice} and we defer the formal analysis of the discretized version to Appendix~\ref{sec:proofs_general_case}.
\proof[Proof sketch of Theorem~\ref{thm:main_general}.]
For any $\mathbf{z}\in[0,1]^{\C\times\S}$, we define
\(
F(\mathbf{z}) := \sum_{j\in\S}r_j\cdot F_j(\mathbf{z}_j)
\)
and let $\mathbf{S}^\star$ be an optimal solution of Problem~\eqref{def:opt_problem}. The variation of the objective function with respect to time satisfies the following
\begin{align*}
\frac{\mathrm{d}F}{\mathrm{d}t}(\mathbf{z}(t)) &= \sum_{j\in \S}r_j\cdot\frac{\mathrm{d}F_j}{\mathrm{d}t}(\mathbf{z}_j(t)) =\sum_{j\in \S}r_j\cdot\langle \nabla F_j(\mathbf{z}_j(t)), \frac{\mathrm{d}\mathbf{z}_j}{\mathrm{d}t}\rangle \\
&=\sum_{i\in\C} \sum_{j\in S_i(t)} r_j\cdot \nabla_i F_j(\mathbf{z}_j(t))\cdot\phi_i(j,S_i(t)) \\
&\geq\sum_{i\in\C} \sum_{j\in S_i(t)}  r_j\cdot\EE_{C\sim\mathbf{z}_j(t)}[f_j(C+i)-f_j(C)]\cdot \phi_i(j,S_i(t)) \\
&\geq \sum_{i\in\C} \sum_{j\in S_i^\star} r_j\cdot\EE_{C\sim\mathbf{z}_j(t)}[f_j(C+i)-f_j(C)]\cdot \phi_i(j,S_i^\star)\\
&\geq \sum_{j\in \S} r_j\cdot F_j(\mathbf{z}_j(t) \vee \Phi_j(\mathbf{S}^\star)) - \sum_{j\in \S} r_j\cdot F_j(\mathbf{z}_j(t))\\
&\geq \sum_{j\in \S} r_j\cdot F_j(\Phi_j(\mathbf{S}^\star)) - \sum_{j\in V} r_j\cdot F_j(\mathbf{z}_j(t))\\
& = \OPT - F(\mathbf{z}(t)).
\end{align*}
where the second equality is due to the chain rule, the next equality is due to the definition of $\frac{\mathrm{d}\mathbf{z}_j}{\mathrm{d}t}$, the first inequality is due to Fact~\ref{fact:ML}, the next inequality follows from the greedy optimality of $\mathbf{S}(t)$ and the feasibility of $\mathbf{S}^\star$ in the subproblem, the third inequality follows from Fact~\ref{fact:ML} and the following inequality is due to monotonicity.
The solution of this differential inequality is such that for all $t\in[0,1]$
\[
F(\mathbf{z}(t)) \geq (1-e^{-t})\cdot\OPT.
\]
In particular for $t=1$, we have
\(
F(\mathbf{z}(1)) \geq (1-e^{-1})\cdot\OPT.
\)
Now we prove that sampling according to $\tau(1)$ provides a solution to the problem without loss in the approximation factor. Recall that 
\[
\frac{\mathrm{d}\bm{\tau}}{\mathrm{d}t} = \uno_{\mathbf{S}(t)} \quad \text{and} \quad \frac{\mathrm{d}\mathbf{z}}{\mathrm{d}t} = \Phi(\mathbf{S}(t)),
\]
so if we integrate over $t\in[0,1]$ we obtain
\[
\tau(1) = \int_0^1 \uno_{\mathbf{S}(t)}\;\mathrm{d}t, \quad \text{and} \quad \mathbf{z}(1) = \int_0^1 \Phi(\mathbf{S}(t))\;\mathrm{d}t
\]
which means that $\bm{\tau}(1)$ is a convex combination of assortment profiles $\mathbf{S}(t)$. In fact, for each $i\in \C$ we have 
\[
\sum_{S\in\I_i}\tau_{i,S}(1) = \int_{0}^1\sum_{S\in\I_i}\uno_{S_i(t)}\;\mathrm{d}t = \int_{0}^11\;\mathrm{d}t =1
\]
Therefore, by sampling an assortment $S$ for each $i\in \C$ according to $\bm{\tau}_i(1)$, in expectation for each $i\in\C, \ j\in\S$ we get 
\[
\EE_{S\sim\bm{\tau}_i(1)}[\phi_i(j,S)]=\sum_{S\in\I_i:\; j\in S}\tau_{i,S}(1)\cdot\phi_{i}(j,S) = \sum_{S\in\I:\; j\in S}\int_0^1 \uno_{S_i(t)}\;\mathrm{d}t \cdot\phi_{i}(j,S) =  \int_0^1 \phi_{i}(j,S_i(t))\;\mathrm{d}t = z_{i,j}(1).
\]
Finally, since the rounding was done independently for each $i\in \C$, we obtain
\begin{align*}
\EE_{\mathbf{S}\sim\bm{\tau}(1)}\Big[F(\Phi(\mathbf{S}))\Big] &= \sum_{j\in\S}r_j\cdot \EE_{\mathbf{S}\sim\bm{\tau}(1)}\Big[F_j(\Phi_j(\mathbf{S}))\Big]\\
&=\sum_{j\in\S}r_j\cdot \sum_{C\subseteq\C}f_j(C)\prod_{i\in C}\EE_{S\sim\tau_i(1)}[\phi_{i}(j,S)]\prod_{i\in\C\setminus C}(1-\EE_{S\sim\tau_i(1)}[\phi_{i}(j,S)])\\
&=\sum_{j\in\S}r_j\cdot \sum_{C\subseteq\C}f_j(C)\prod_{i\in C}z_{i,j}(1)\prod_{i\in\C\setminus C}(1-z_{i,j}(1))\\
& = \sum_{j\in\S}r_j\cdot F_j(\mathbf{z}_j(1)) = F(\mathbf{z}(1))\geq (1-1/e)\cdot\OPT,
\end{align*}
which concludes the proof.
\endproof

\begin{remark}
To the best of our knowledge, our choice-aware continuous greedy method and its analysis are novel in the context of the two-sided assortment optimization. It differs from the standard continuous greedy for submodular maximization in three key aspects. First, our method needs to keep track of two points, the choice probabilities $\mathbf{z}(t)$ and the assortment distributions $\bm{\tau}(t)$, which are related to each other via \eqref{eq:relation_z_tau}. In contrast, the classic continuous greedy just keep tracks of a fractional solution that represents the fraction of the elements in the solution set. Second, the maximization subproblem~\eqref{eq:linear_maximization_subproblem} is two-sided in nature since it accounts for customers' and suppliers' choice preferences as opposed to the vanilla continuous greedy which  accounts for the contribution of one side (in our problem this would be the suppliers' demand). Third, we do not need a rounding procedure such as pipage rounding or dependent swap rounding~\cite{calinescu2011maximizing,chekuri2010dependent}, because we are not keeping track of fractional elements but instead we construct a distribution over assortments. This distribution, which is of size $O(n^2)$ in Algorithm~\ref{alg:continous_greedy_choice}, allows us to sample assortments without loss in the expected solution value, as shown above.
\end{remark}

\begin{remark}\label{remark:approx_oracle}
Note that the proof above can be easily adapted to the case in which we have an approximate oracle in Assumption~\ref{assumption:oracle} instead of an exact oracle. To be more precise, if for every $i\in\C$, the algorithm in Assumption~\ref{assumption:oracle} is a polynomial-time $\beta_i$-approximation algorithm, then Algorithm~\ref{alg:continous_greedy_choice} guarantees a $(1-\exp(-\min_{i\in\C}\{\beta_i\}))$-factor for Problem~\eqref{def:opt_problem}. For example, this captures the case when $\I_i$ corresponds to a single cardinality constraint and $\phi_i$ is the Markov Chain or nested-logit choice model, for which there is a $1/2$-approximation algorithm for the single-customer assortment optimization problem~\cite{desir_etal15,gallego2014constrained}. 
\end{remark}

\begin{remark}
We finally remark that the main disadvantage of Algorithm~\ref{alg:continous_greedy_choice} is its time performance due to the sampling of~\eqref{eq:expected_marginal}, which could be worse than  Algorithm~\ref{alg:choiceaware_greedy}. Both approaches present a tradeoff between their time performance and approximation guarantees. In Section~\ref{sec:computational_study}, we provide more details on the empirical performance of Algorithm~\ref{alg:choiceaware_greedy} and its running time.
\end{remark}

\subsection{Near-Optimality for General-MNL Instances under the Small Probability Regime}\label{sec:general_mnl}
In this section, we study Problem~\eqref{def:opt_problem} for the specific case when suppliers' preferences are governed by MNL choice models,  and customers follow general choice models satisfying Assumption~\ref{assumption:oracle}. Given the structure of the MNL demand function, we adapt the Frank-Wolfe algorithm to our two-sided setting and we prove that it is asymptotically optimal in the $\epsilon$-small probability regime (Theorem~\ref{thm:fw_nearoptimality_epsregime}).

As in the previous section, we follow the template given in our Meta-algorithm~\ref{alg:meta_algorithm}, so we first construct a relaxation of Problem~\ref{def:opt_problem} when suppliers follow MNL choice models.
By recalling Definition~\ref{def:MNL}, note that the demand function of supplier $j\in \S$ under the MNL choice model is
\[
f_j(C) = \sum_{i\in C}\phi_j(i,C)=\frac{w_j(C)}{1+ w_j(C)},
\]
for any $C\subseteq\C$ where the outside option is normalized $w_{j,0}=1$ and $w_j(C) := \sum_{i\in C}w_{j,i}$. \revised{Now, let us consider the following continuous function $G_j(\mathbf{z}_j) = r_j\cdot \sum_{i\in\C}z_{i,j}/(1+  \sum_{i\in\C}z_{i,j})$ for $\mathbf{z}_j\in[0,1]^{\C}$. As discussed in Section~\ref{sec:meta_approach}, our relaxation would correspond to $\max\Big\{\sum_{j\in\S}G_j(\mathbf{z}_j):\ (\mathbf{z},\tau)\in\P\Big\}$. However, since $G_j$ has a special structure, we can aggregate $\mathbf{z}_j$ by considering $z_j = \sum_{i\in\C}z_{i,j}$ and defining $G_j(z_j) = r_j\cdot z_j/(1+z_j)$. On the other hand, since we are aggregating $\mathbf{z}_j$, we can take a linear combination of the first constraints of $\P$ in \eqref{eq:feasible_region_meta_relaxation}.} Putting all these pieces together, our  main relaxation for this section is the following:
\begin{align}\label{eq:relaxation_mnl_suppliers}
\max &\quad \sum_{j\in \S}r_j\cdot \frac{z_j}{1+z_j}\\
s.t. &\quad \sum_{i\in \C}w_{j,i}\cdot\sum_{S\in\mathcal{I}_i: S\ni j}\tau_{i,S}\cdot\phi_i(j,S)=z_j, \qquad \text{for all}  \ j\in \S \notag\\
&\quad \sum_{S\in\I_i} \tau_{i,S} =1, \hspace{11em} \text{for all} \ i\in \C, \notag \\
&\quad z_j, \tau_{i,S}\geq 0 \hspace{12.2em} \text{for all} \ i\in \C, \ S\in\mathcal{I}_i, \ j\in \S. \notag
\end{align}
\begin{lemma}\label{lemma:relaxation_mnl_suppliers}
Problem~\eqref{eq:relaxation_mnl_suppliers} is a relaxation of Problem~\eqref{def:opt_problem} when suppliers have MNL choice models.
\end{lemma}
We defer the proof of this lemma to Appendix~\ref{sec:missing_details_optimal}.
Note that the objective function in \eqref{eq:relaxation_mnl_suppliers} is a separable concave function, thus Problem~\eqref{eq:relaxation_mnl_suppliers}  can be solved efficiently as long as maximizing a linear objective over the feasible region can be done efficiently, which is our case.
Following the framework of our Meta-algorithm~\ref{alg:meta_algorithm} and the ideas in Algorithm~\ref{alg:continous_greedy_choice}, our choice-aware Frank-Wolfe algorithm keeps track of two points: the choice probability vector $\mathbf{z}^t$ and the assortment distributions $\bm{\tau}^t$, however, they are updated in a slightly different way. In every iteration, our method updates these points by doing a convex combination between the incumbent points and the current ascent directions.\footnote{This convex combination depends on the step size $\gamma_t$ which we specify in the proof.} In each step $t$, these directions can be efficiently obtained (Assumption~\ref{assumption:oracle}) by solving, for each customer $i\in\C$, the subproblem $
 \max\left\{\sum_{j\in S}\theta_{i,j}^{t}\cdot\phi_i(j,S): \ S\in\mathcal{I}\right\}$ where the weights are now defined as
 \[
 \theta_{i,j}^t:= r_j\cdot \frac{w_{j,i}}{(1+z^{t-1}_j)^2}
 \]
 We remark that the expression $1/(1+z^{t-1}_j)^2$ corresponds to the derivative of the function $h(z)=z/(1+z)$ evaluated in the previous point $z_j^{t-1}$. Note that these weights do not require sampling as we needed in Algorithm~\ref{alg:continous_greedy_choice}, which makes our Frank-Wolfe method significantly faster, as demonstrated empirically in Section~\ref{sec:computational_study}. We formally present our procedure in Algorithm~\ref{alg:frankwolfe_choice}.
 
\begin{algorithm}[htpb]
\caption{Choice-aware MNL Frank-Wolfe}\label{alg:frankwolfe_choice}
\begin{algorithmic}[1]
\Require  Customers' choice models $\phi_i$, weights $\mathbf{w}$,  $T$ sufficiently large.
\Ensure Assortment family $\mathbf{S}$ and $\mathbf{z}^T$.
\State $\mathbf{z}^0 = 0$, $\tau^0_{i,S} = 0$ for all $i\in\C$, $\emptyset\neq S\in\I_i$ and $\tau_{i,\emptyset}^0 =1$ for all $i\in \C$.
\For{$t\in [T]$} 
\State For each $i\in\C$ obtain $S_i^{t}\in \argmax\left\{\sum_{j\in S}r_j\cdot\frac{w_{j,i}}{(1+z_j^{t-1})^2}\cdot\phi_i(j,S): \ S\in\mathcal{I}_i\right\}$.
\State Set $z_j^{t} \leftarrow \left(1-\gamma_t\right)\cdot z_j^{t-1} + \gamma_t\cdot \sum_{i\in \C}w_{j,i}\cdot\phi_i(j,S^{t}_i)$ for all $j\in\S$.
\State Set $\tau^t_{i,S_i^{t}} \leftarrow (1-\gamma_t)\cdot\tau^{t-1}_{i,S_i^{t}} + \gamma_t$  for all $i\in \C$ and $\tau^{t}_{i,S} = (1-\gamma_t)\cdot\tau^{t-1}_{i,S}$ for all $i\in \C$, $S\neq S_i^{t}$.
\EndFor
\State For each $i\in \C$, independently sample an assortment from distribution $\bm{\tau}^T_{i}$.
\end{algorithmic}
\end{algorithm}

In the following, we present two key technical results for the proof of Theorem~\ref{thm:fw_nearoptimality_epsregime}.  Since the proofs of these results rely on standard techniques, we defer them to Appendix~\ref{app:proofs_general_mnl}.
First, in Lemma~\ref{lemma:feasibility_fw}, we show that the points $\mathbf{z}^t$ and $\bm{\tau}^t$ constructed by our algorithm are feasible in \eqref{eq:relaxation_mnl_suppliers} at any step $t\in[T]$.\footnote{Note that we consider $\tau_{i,\emptyset}^0 =1$ purely to satisfy feasibility at $t=0$. We can also consider $\tau_{i,\emptyset}^0 =0$, as it is enough for us to have a distribution such that $\sum_{S\in\I_i,S\neq\emptyset}\tau^t_{i,S}\leq 1$ at every $t\in[T]$. Since the demand functions are monotone, we can always re-weight the distribution by removing the empty set. Namely, it is always better to show a non-empty assortment to every customer.}
\begin{lemma}\label{lemma:feasibility_fw}
For any given step $t=0,\ldots,T$, the points $\mathbf{z}^t$ and $\tau^t$ constructed by Algorithm~\ref{alg:frankwolfe_choice} are feasible in Problem~\eqref{eq:relaxation_mnl_suppliers}.
\end{lemma}
More importantly, as we show in the following proposition, the objective value of the fractional solution output by Algorithm~\ref{alg:frankwolfe_choice} asymptotically converges (for sufficiently large $T$) to the optimal value of Problem~\eqref{eq:relaxation_mnl_suppliers}, up to a vanishing additive error.
\begin{proposition}\label{prop:convergence_fw_mnl}
Algorithm~\ref{alg:frankwolfe_choice} outputs a feasible solution $(\mathbf{z}^T,\tau^T)$ such that
\[
\sum_{j\in \S}r_j\cdot\frac{z^T_j}{1+z^T_j} \geq \OPT_{\eqref{eq:relaxation_mnl_suppliers}} - \frac{4\sqrt{mRW}}{T+2},
\]
where $\OPT_{\eqref{eq:relaxation_mnl_suppliers}}$ is the optimal value of Problem~\eqref{eq:relaxation_mnl_suppliers}, $m$ is the number of suppliers, $R=\max_{j\in\S}\{r_j\}$ and $W=\sum_{j\in\S}(\sum_{i\in\C}w_{j,i})^2$. Therefore, the solution of Algorithm~\ref{alg:frankwolfe_choice} converges to the optimal value as $T$ increases.
\end{proposition}

\subsubsection{Analysis}
\label{sec:near_optimality_mnl_suppliers}
In this section, we assume that the MNL choice models of the suppliers satisfy the $\epsilon$-small probability regime with $\varepsilon\in(0,1)$. Specifically, given Definition~\ref{def:epsilon_small}, we have that the demand function of each supplier $j\in\S$ is such that for every customer $i\in\C$
\begin{equation}\label{eq:mnl_epsilonsmallregime}
\frac{w_{j,i}}{1+w_{j,i}}\leq\epsilon\cdot f_j(\C) \qquad \Rightarrow \qquad w_{j,i}\leq \frac{\epsilon}{1-\epsilon},
\end{equation}
where we used that $f_j(\C)\leq 1$. The key lemma to prove Theorem~\ref{thm:fw_nearoptimality_epsregime} is given in the next result which, roughly speaking, shows that the concave objective function in Problem~\eqref{eq:relaxation_mnl_suppliers} is $(1-\epsilon)$ close to the original objective function.
\begin{lemma}\label{lemma:supplier_mnl_bound1}
Assume that suppliers have MNL choice models satisfying~\eqref{eq:mnl_epsilonsmallregime}. Then, for any vector $\mathbf{d}\in[0,1]^{\C\times \S}$, we have
\begin{equation}\label{eq:lowerbound_mnlmnl}
\sum_{j\in \S}r_j\cdot F_j(\mathbf{d}_j)\geq (1-\epsilon)\cdot\sum_{j\in \S}r_j\cdot\frac{\sum_{i\in \C}w_{j,i}\cdot d_{i,j}}{1+\sum_{i\in \C}w_{j,i}\cdot d_{i,j}}.
\end{equation}
\end{lemma}
\proof[Proof of Lemma~\ref{lemma:supplier_mnl_bound1}.]
Fix $j\in \S$ and consider $i\in \C$ such that $d_{i,j}>0$. Then, conditioned on the event that $i$ chose $j$ (which happens with probability $d_{ij}$), the expected probability (with respect to the product distribution) that $j$ selects $i$ back can be lower bounded as follows
\begin{align*}
\EE_{C\sim\mathbf{d}_j^{-i}}\left[\phi_{j}(i,C)\big| \text{$i$ chooses $j$}\right] &= \EE_{C\sim\mathbf{d}_j^{-i}}\left[\frac{w_{j,i}}{1+w_{j,i} + \sum_{k\in C, k\neq i}w_{j,k}} \Big| \text{$i$ chooses $j$}\right]\\
& =  \EE_{C\sim\mathbf{d}_j^{-i}}\left[\frac{w_{j,i}}{1+w_{j,i} + \sum_{k\in C, k\neq i}w_{j,k}} \right]\\
&\geq \frac{w_{ji}}{1+w_{j,i} + \EE_{C\sim\mathbf{d}_j^{-i}}\left[\sum_{k\in C, k\neq i}w_{j,k}\right]}\\
&\geq \frac{w_{j,i}}{1+\frac{\epsilon}{1-\epsilon} + \sum_{k\in \C, k\neq i}w_{j,k}d_{k,j}}\\
&= (1-\epsilon)\cdot\frac{w_{j,i}}{1+ (1-\epsilon)\cdot\sum_{k\in \C, k\neq i}w_{j,k}d_{k,j}}\\
&\geq(1-\epsilon)\cdot \frac{w_{j,i}}{1+\sum_{k\in \C}w_{j,k}d_{k,j}}.
\end{align*}
where $\mathbf{d}_j^{-i}$ is the vector $\mathbf{d}_j$ without the $i$-th component. The first equality follows from the assumption that $i$ chose $j$, the second equality follows from the independent choices of the customers, the first inequality is due to Jensen's inequality, the second inequality is due to \eqref{eq:mnl_epsilonsmallregime} and the last inequality is because the expression decreases when we include more terms to the sum in the denominator. Therefore, the demand function of supplier $j$ can be lower bounded as follows
\begin{align*}
\EE_{C\sim\mathbf{d}_j}[f_j(C)] &= \EE_{C\sim\mathbf{d}_j}\left[\sum_{i\in C} \phi_{j}(i,C) \right] = \EE_{C\sim\mathbf{d}_j}\left[\sum_{i\in \C} \phi_{j}(i,C) \cdot \uno_{\{i\in C\}} \right]\\
& = \sum_{i\in \C} \EE_{C\sim\mathbf{d}^{-i}_{j}}\left[\phi_{j}(i,C)\cdot  \uno_{\{i\in C\}}  \big| \text{$i$ chooses $j$}\right]\PP(\text{$i$ chooses $j$})\\
& =\sum_{i\in \C} \EE_{C\sim\mathbf{d}^{-i}_{j}}\left[\phi_{j}(i,C)  \big| \text{$i$ chooses $j$}\right]\cdot d_{i,j} \\
&\geq(1-\epsilon)\cdot \sum_{i\in \C} \frac{w_{j,i}}{1+\sum_{k\in \C}w_{j,k}d_{k,j}}\cdot d_{ij}.
\end{align*}
Finally, by multiplying by $r_j$ and summing over $j\in \S$, we obtain the desired inequality. 
\endproof

With the above lemma, we are ready to prove our main result.

\proof[Proof of Theorem~\ref{thm:fw_nearoptimality_epsregime}.]
Consider the vector $\bm{\tau}^T$ output by Algorithm~\ref{alg:frankwolfe_choice}. Note that $\bm{\tau}^T$ has a polynomial number of non-zero entries as long as the parameters $W$ and $R$ are polynomially bounded (in the size of the input). Let $\mathbf{S}$ be the random assortment family sample from this assortment distribution. Then, we obtain 
\begin{align*}
\EE_{\mathbf{S}\sim\bm{\tau}^T}[\Ma_{\mathbf{S}} ] &=\sum_{j\in\S}\sum_{C\subseteq\C}f_j(C)\prod_{i\in C}\left(\sum_{S\in\I_i:S\ni j}\tau^T_{i,S}\cdot\phi_i(j,S)\right)\prod_{i\in\C\setminus C}\left(1-\sum_{S\in\I_i:S\ni j}\tau^T_{i,S}\cdot\phi_i(j,S)\right)\\
&\geq (1-\epsilon)\cdot\sum_{j\in \S}r_j\cdot\frac{\sum_{i\in\C}w_{j,i}\cdot\sum_{S\in\I_i:S\ni j}\tau^T_{i,S}\cdot\phi_i(j,S)}{1+\sum_{i\in\C}w_{j,i}\cdot\sum_{S\in\I_i:S\ni j}\tau^T_{i,S}\cdot\phi_i(j,S)}\\
& = (1-\epsilon)\cdot\sum_{j\in \S}r_j\cdot\frac{z_j^T}{1+z_j^T} \geq (1-\epsilon)\cdot\OPT_{\eqref{eq:relaxation_mnl_suppliers}}- \; O(1/T) \geq (1-\epsilon)\cdot\OPT- \; O(1/T)
\end{align*}
where the first equality is due the independent choices made by customers (same as the last part of the proof of Theorem~\ref{thm:main_general}), the next inequality follows from Lemma~\ref{lemma:supplier_mnl_bound1} by considering $d_{i,j} = \sum_{S\in\I_i:S\ni j}\tau^T_{i,S}\cdot\phi_i(j,S)$, the next equality from the feasibility of $\mathbf{z}^T$ and $\bm{\tau}^T$ shown in Lemma~\ref{lemma:feasibility_fw}, the second inequality follows from Proposition~\ref{prop:convergence_fw_mnl} and the last one because \eqref{eq:relaxation_mnl_suppliers} is a relaxation to our problem (Lemma~\ref{lemma:relaxation_mnl_suppliers}). 
\endproof

\subsubsection{Analysis under any Regime}\label{sec:approx_factor_FW_mnl_suppliers}
One natural question is how Algorithm~\ref{alg:frankwolfe_choice} performs in the worst-case scenario, i.e., not necessarily in the $\epsilon$-small probability regime. Formally, we show the following:
\begin{theorem}\label{thm:fw_factor_anyregime}
Assume that customers follow general choice models satisfying Assumption~\ref{assumption:oracle} and each suppliers follow different MNL choice models. Then, Algorithm~\ref{alg:frankwolfe_choice} guarantees a $1/4$-approximation factor.
\end{theorem}
The proof of this theorem relies on the  following key lemma.
\begin{lemma}\label{lemma:supplier_mnl_bound2}
Consider an instance $I$ where suppliers have different MNL choice models with values $w_{j,i}\geq 0$. Consider a {new instance} $\tilde{I}$ in which suppliers' weight preferences over customers are capped as follows $\tilde{w}_{j,i} = \min\left\{w_{j,i},\frac{\epsilon}{1-\epsilon}\right\}$ for all $j\in \S$, $i\in \C$ for a given $\epsilon\in(0,1)$. Then, for any assortment family $\mathbf{S}$, the following inequalities hold
\[
\EE[\tilde{\Ma}_{\mathbf{S}}] \leq \EE[\Ma_{\mathbf{S}}] \leq \frac{1}{\epsilon}\cdot \EE[\tilde{\Ma}_{\mathbf{S}}],
\]
where $\tilde{\Ma}_{\mathbf{S}}$ and $\Ma_{\mathbf{S}}$ are the random variables that indicate the total revenue of the final matching in $\tilde{I}$ and $I$, respectively, under the given assortment profile $\mathbf{S}$.
\end{lemma}
This lemma is a generalization of Lemma 4.1 in \cite{ashlagi_etal19}. Since \cite{ashlagi_etal19} prove it only for a special case of the MNL model and for $\epsilon=1/2$, we provide a proof of Lemma \ref{lemma:supplier_mnl_bound2} in Appendix~\ref{app:proofs_general_mnl} for completeness.

\proof[Proof of Theorem~\ref{thm:fw_factor_anyregime}.]
Fix $\epsilon\in(0,1)$. First, we construct a new instance of the problem where suppliers' values over customers are capped $\tilde{w}_{j,i}=\min\{w_{j,i},\epsilon/(1-\epsilon)\}$. Let $\OPT$ and $\tilde{\OPT}$ be the optimal values of the original and the new instance, respectively. From Lemma~\ref{lemma:supplier_mnl_bound2} we know that
\(
\tilde{\OPT}\geq \epsilon\cdot\OPT.
\)
Since $\tilde{w}_{j,i}\leq \epsilon/(1-\epsilon)$, we can use Theorem~\ref{thm:fw_nearoptimality_epsregime} to conclude that, in the new instance, Algorithm~\ref{alg:frankwolfe_choice} outputs a random assortment family $\mathbf{S}$ sampled from a distribution $\bm{\tau}^T$ such that
\(
\EE_{\mathbf{S}\sim\bm{\tau}^T}[\tilde{\Ma}_{\mathbf{S}}] \geq (1-\epsilon)\cdot\tilde{\OPT}.
\)
By putting everything together, we obtain 
\[
\EE_{\mathbf{S}\sim\bm{\tau}^T}[\Ma_{\mathbf{S}}]\geq \EE_{\mathbf{S}\sim\bm{\tau}^T}[\tilde{\Ma}_{\mathbf{S}}]  \geq  (1-\epsilon)\cdot\tilde{\OPT}\geq \epsilon\cdot(1-\epsilon)\cdot\OPT.
\]
where in the first inequality we use Lemma~\ref{lemma:supplier_mnl_bound2}.
We conclude the proof by noting that the factor $\epsilon\cdot(1-\epsilon)$ is maximized in $\epsilon=1/2$. For $\epsilon=1/2$, we need to cap the values at 1, i.e., $\tilde{w}_{j,i} = \min\{w_{j,i},1\}$ which essentially tells us that we need to distinguish between those suppliers that value more the customers than the outside option (recall that $w_{j,0}=1$) and those who value the outside option more.
\endproof

\section{Efficient Suboptimal Algorithms}\label{sec:efficient_algorithms}
In this section, we propose efficient variants of our main algorithms whose approximation guarantees are also constant but suboptimal. To derive these algorithms, we follow the ideas in Meta-algorithm~\ref{alg:meta_algorithm}. Specifically, in Section~\ref{subsubsec:choice_aware_greedy} we analyze a discrete variant of the choice-aware continuous greedy algorithm. In Section~\ref{sec:unconstrained_mnl_mnl}, we study a method specifically tailored for MNL-MNL instances without constraints on the assortments. Finally, in Section~\ref{sec:computational_study}, we provide a detailed empirical analysis of our methods' performance.
\subsection{The Choice-aware Discrete Greedy Algorithm}\label{subsubsec:choice_aware_greedy}
Observe that our choice-aware continuous greedy algorithm can be modified to avoid the estimation step of the expected marginal values $\EE_{C\sim\mathbf{z}_j(t)}\big[f_j(C+i)-f_j(C)\big]$. To do this, we will sequentially simulate the choices of each customer and, consequently, these marginal values will be determined by the updated backlogs after each customer choice. Once all the customers' choices have been simulated, the algorithm outputs exactly those assortments used during the simulation step.

Formally, our \emph{choice-aware discrete greedy} method, formalized in Algorithm~\ref{alg:choiceaware_greedy}, considers an arbitrary order over the customers and greedily selects an assortment for each customer $i\in\C$ in the given order. To select the assortment for customer $i$, our algorithm weights each supplier $j$ with the marginal value $\theta_{i,j}=f_j(C+i)-f_j(C)$ where $C$ is the backlog ``observed'' so far, i.e., the set of customers before $i$ whose simulated choice was $j$. Again, due to Assumption~\ref{assumption:oracle}, we can efficiently compute an assortment for customer $i$ by solving a single customer revenue maximization problem using these weights $\theta_{i,j}$. Note that this approach also fits in the framework given in Section~\ref{sec:meta_approach} since the resulting output are trivial distributions that place weight only on one assortment. The main difference with the choice-aware continuous greedy method is that the discrete variant builds one assortment at each step, while the continuous version fractionally builds an assortment distribution using $n$ assortments simultaneously.
Several online variants of the choice-aware discrete greedy have been previously studied in~\cite{golrezaei2014real,aouad2020online}. In particular, we emphasize that the analysis provided in \cite{aouad2020online} cannot be used in our setting since the authors compare against the clairvoyant optimum that offers a feasible solution to our problem, but not necessarily the optimal one.
\begin{algorithm}[htpb]
\caption{Choice-aware Discrete Greedy}\label{alg:choiceaware_greedy}
 \begin{algorithmic}[1]
\Require  Choice models $\phi$, feasibility assortment family $\mathcal{I}_i$ for all $i\in\C$, and revenues $r_j$.
\Ensure An assortment $S_i\in\I_i$ for each $i\in\C$.
\State Consider an arbitrary permutation $i_1,\ldots,i_n$ over the set of customers $\C$.
\State For each $j\in \S$, set $C^0_j=\emptyset$.
\For{$t=1,\ldots,n$} 
\State $S_{i_t} \in \argmax \left\{\sum_{j\in S}r_j\cdot\left[f_j(C^{t-1}_j+ i_t)-f_j(C^{t-1}_j)\right]\cdot\phi_{i_t}(j,S): \ S\in\mathcal{I}_i\right\}$
\State Simulate the choice of customer $i_t$ by sampling $\ell\in S_{i_t}\cup\{0\}$ w.p. $\phi_{i_t}(\ell,S_{i_t})$.
\State If $\ell\neq 0$, $C_{\ell}^t \leftarrow C_{\ell}^{t-1}+i_t$ and $C_j^{t} \leftarrow C_j^{t-1}$ for all $j\in\S\setminus\{\ell\}$. Otherwise, $C_j^{t} \leftarrow C_j^{t-1}$ for all $j\in\S$.
\EndFor
\end{algorithmic}
\end{algorithm}

\begin{theorem}\label{thm:choiceaware_greedy}
Algorithm~\ref{alg:choiceaware_greedy} guarantees (in expectation) a $1/2$-approximation factor for Problem~\eqref{def:opt_problem}.
\end{theorem}

\proof[Proof of Theorem~\ref{thm:choiceaware_greedy}.]
Consider an arbitrary permutation. With a slight abuse of notation, we write $t\in\{1,\ldots,n\}$ for the $t$-th customer in the given permutation, i.e., $t$ corresponds to $i_t$.
Let $S_1,\ldots, S_n$ and $S^\star_1,\ldots, S^\star_n$ be the assortments output by Algorithm~\ref{alg:choiceaware_greedy} and an optimal solution, respectively.\footnote{Recall that the optimal assortments do not depend on the permutation.} 
Let $\Phi^{1:t-1}\in[0,1]^{\C\times \S}$ be the vector of choice probabilities determined by the assortments presented to the first $t-1$ customers, i.e., $\Phi^{1:t-1}_{i,j} := \phi_{i}(j,S_{i})$ for  $i=1,\ldots, t-1$, $j\in S_i$ and zero in the rest of the components.\footnote{Note that  $\Phi^{1:t-1}$ does not consider the outside option in its components.} Let us condition on the choices made by the first $t-1$ customers, i.e., we condition on the random set of customers $C^{t-1}_j$ that chose supplier $j\in \C$ up to the end of iteration $t-1$ according to the choice probabilities in $\Phi^{1:t-1}$. Let $\mathbf{C}^{t-1}$ be the collection of the sets $C^{t-1}_j$ for all $j\in\S$ and $\uno_{\mathbf{C}^{t-1}}$ be the indicator vector whose $(i,j)$-th component is 1 if $C^{t-1}_j\ni i$ and 0 otherwise.
Let $\Phi^{\star}_{t}\in[0,1]^{\S}$ be the vector of the choice probabilities determined only by the optimal assortment of the $t$-th customer, i.e., $\Phi^{\star}_{t,j} = \phi_{t}(j,S^\star_{t})$ for all $j\in S^\star_t$ and 0 in the rest of the components. Finally, define the function $F(\mathbf{z})=\sum_{j\in\S}r_j\cdot F_j(\mathbf{z}_j)$ for $\mathbf{z}\in[0,1]^{\C\times\S}$ where $F_j$ is the multilinear extension of $f_j$. Given the above notation, we have the following sequence of inequalities:
\begin{align*}
F(\uno_{\mathbf{C}^{t-1}}\vee \Phi^{\star}_{t}) - F(\uno_{\mathbf{C}^{t-1}}) &=\sum_{j\in\S}r_j\cdot\left[F_j(\uno_{C_j^{t-1}}\vee \Phi^{\star}_{t,j}) - F_j(\uno_{C_j^{t-1}})\right]\\
& = \sum_{j\in\S}r_j\cdot\left[f_j(C_j^{t-1}+t)\cdot\phi_t(j,S^\star_t) + f_j(C_j^{t-1})\cdot(1-\phi_t(j,S^\star_t))  - f_j(C_j^{t-1})\right]\\
&= \sum_{j\in \S}r_j\cdot [f_j(C^{t-1}_j + t) - f_j(C^{t-1}_j)]\cdot \phi_{t}(j,S_t^\star)\\
&\leq \sum_{j\in \S}r_j\cdot[f_j(C^{t-1}_j + t) - f_j(C^{t-1}_j)]\cdot \phi_{t}(j,S_t)\\
& = F(\uno_{\mathbf{C}^{t-1}}\vee \Phi_{t}) - F(\uno_{\mathbf{C}^{t-1}}),
\end{align*}
where the first equality is due to the definition of the function $F(.) = \sum_{j\in\S}r_j\cdot F_j(.)$, the second equality is due to the random choice of customer $t$ given assortment $S^\star_t$, the first inequality is due to the choice of greedy, the last equality can be obtain by reversing the computations made in the second equality and $\Phi_t\in[0,1]^{\S}$ is the choice probability vector where the $j$-th component equals $\phi_t(j,S_t)$ for all $j\in S_t$ and zero in the remaining components. Taking expectation over $\mathbf{C}^{t-1}\sim\Phi^{1:t-1}$, we get
\begin{equation}\label{eq:aux_random_greedy}
F(\Phi^{1:t-1}\vee \Phi^{\star}_{t}) - F(\Phi^{1:t-1}) \leq F(\Phi^{1:t-1}\vee \Phi_{t}) - F(\Phi^{1:t-1}) = F(\Phi^{1:t}) - F(\Phi^{1:t-1}). 
\end{equation}
Summing over $t=1,\ldots,n$ on the left-hand side of~\eqref{eq:aux_random_greedy} results in
\[
\sum_{t=1}^nF(\Phi^{1:t-1}\vee \Phi^{\star}_{t}) - F(\Phi^{1:t-1})\geq F(\Phi^{1:n}\vee \Phi^{\star}) -F(\Phi^{1:n})\geq F(\Phi^{\star}) -F(\Phi^{1:n}),
\]
where in the first inequality we applied submodularity and the next inequality is due to the monotonicity of the function (Fact~\ref{fact:ML}).
Summing over $t=1,\ldots,n$ on the right-hand side of~\eqref{eq:aux_random_greedy}, we obtain
\[
\sum_{t=1}^n F(\Phi^{1:t}) - F(\Phi^{1:t-1})  = F(\Phi^{1:n}) -F(\Phi^{1:0}) = F(\Phi^{1:n}), 
\]
where we define $\Phi^{1:0} = \mathbf{0}$. Therefore, we conclude that
\(
F(\Phi^{\star})\leq 2\cdot F(\Phi^{1:n}). 
\)
\endproof
\begin{remark}
To the best of our knowledge, the analysis above is novel in the context of offline two-sided assortment optimization. Moreover, our $1/2$-approximation guarantee coincides with the optimal worst-case competitive ratio shown for the online variant of the problem~\cite{golrezaei2014real,aouad2020online}. 
\end{remark}

\subsection{The MNL-MNL Method for the Unconstrained Setting}\label{sec:unconstrained_mnl_mnl}
So far, all of our algorithms apply to any constrained setting as long as Assumption~\ref{assumption:oracle} is satisfied.
Our goal in this section is to study Problem~\eqref{def:opt_problem} in the unconstrained setting when both suppliers and customers follow MNL choice models. Specifically, we show that there is an algorithm which outputs a distribution for each customer supported on at most $m+1$ assortments and guarantees a constant-factor approximation (Theorem~\ref{thm:mnl_mnl}). The main advantage of this algorithm is two-fold:  First, it solves a (novel) polynomially-sized concave relaxation and, second, it samples an assortment for each customer from a family of at most $m+1$ nested assortments, where $m$ is the number of suppliers. \revised{Consequently, this algorithm is both efficient and effective, which we empirically confirm in Section~\ref{sec:computational_study}.}

As in Definition~\ref{def:MNL}, we denote the value of customer $i$ for each supplier $j$ as  $v_{i,j}\geq0$ and the value for the outside option is normalized $v_{i,0}=1$. Therefore, the MNL choice model of customer $i$ is such that for any assortment $S\subseteq\S$, $\phi_i(j,S) = \frac{v_{i,j}}{1+v_i(S)}$ for $j\in S\cup\{0\}$ and zero otherwise, where $v_i(S) = \sum_{\ell\in S}v_{i,\ell}$. Given this, for our relaxation, we consider the following feasible region
\begin{align}\label{eq:feasible_region_mnl_mnl}
\R = \Bigg\{\mathbf{y}\in\RR^{\C\times(\S\cup\{0\})}:& \quad  y_{i,0} + \sum_{j\in \S}v_{i,j}y_{i,j} = 1 \hspace{3.7em}\text{for all} \ i\in \C, \notag \\
&\quad y_{i,j} \leq y_{i,0} \hspace{8em}\text{for all} \ i\in \C, \ j\in S \notag \\
&\quad y_{i,j},y_{i,0}\geq 0 \hspace{7.1em} \text{for all} \ i\in \C, \ j\in \S \Bigg\}.
\end{align}
A single-customer variant of this feasible region has been previously studied in the literature of assortment optimization, see e.g.~\cite{sumida2021revenue}. We now show that given a point $\mathbf{y}\in\R$, we can construct, for each customer, a lossless distribution over at most $m+1$ nested assortments.
\begin{lemma}\label{lemma:rounding_algorithm}
Let $\mathbf{y}\in\R$ be any point. For each customer $i\in\C$, there exists a distribution $\bm{\tau}_{i}$ over at most $m+1$ different nested assortments such that $\PP_{\bm{\tau}_i}(i\text{ chooses supplier }j) = v_{i,j}y_{i,j}$ for every $i\in\C$, $j\in\S$.
\end{lemma}
\proof[Proof.]
Fix customer $i\in\C$. For the given point $\mathbf{y}$, order the components in the $i$-th row in decreasing order: $y_{i,0}\geq y_{i,j_1}\geq\cdots\geq y_{i,j_m}$. Then, construct the following sets: 
\[
S_{i,0} = \emptyset, \ S_{i,1} = \{j_1\}, \ S_{i,2} = \{j_1,j_2\},\ldots, S_{i,m} = \S
\]
Note that this construction is possible since there are no constraints on the assortments. Moreover, these assortments are nested which is a common feature in the literature of assortment optimization. We now construct a distribution $\bm{\tau}_i$. For each $\ell \in\{0,\ldots,m-1\}$ define 
\[
\tau_{i,S_{i,\ell}}  = (y_{i,\ell} - y_{i,\ell+1})\cdot (1+v_i(S_{i,\ell}))
\]
and $\tau_{i,S_{i,m}} = y_{i,m}\cdot (1+v_i(S_{i,m}))$. For any other subset $S\neq S_{i,\ell}$ define $\tau_{i,S} = 0$.
Given the ordering on the values, we have that $\tau_{i,S}\geq0$ for every subset $S\subseteq\S$. More importantly, we have
\begin{align*}
\sum_{S\subseteq\S }\tau_{i,S} &= \sum_{\ell=0}^{m-1}(y_{i,j_\ell} - y_{i,j_{\ell+1}})\cdot (1+v_i(S_{i,\ell})) +  y_{i,j_m}\cdot (1+v_i(S_{i,m})) =y_{i,0} + \sum_{\ell=1}^mv_{i,j_\ell}y_{i,j_\ell} = 1.
\end{align*}
where in the last equality we used that $\mathbf{y}\in\R$. Finally, let us compute the probability that a customer $i$ chooses supplier $j_\ell$ with $\ell\in\{1,\ldots,m\}$
\begin{align*}
\PP(i\text{ chooses }j_\ell) &= \sum_{S:S\ni j_\ell}\PP(i\text{ chooses }j_\ell\text{ in }S)\cdot\PP(i\text{ sees assortment }S) \\
&= \sum_{S:S\ni j_\ell}\tau_{i,S}\cdot\frac{v_{i,j_\ell}}{1+v_i(S)}=\sum_{\sigma=\ell}^{m-1}(y_{i,j_\sigma}-y_{i,j_{\sigma+1}})v_{i,j_\sigma} + y_{i,j_m}v_{i,j_m}=v_{i,j_\ell}y_{i,j_\ell}
\end{align*}
where the first equality is due to, once we conditioned on the assortment $S$, the choice of $i$ is independent from everything else. The third equality is because $j_{\ell}$ appears in the $\ell$-th assortment and onward (recall they are nested).
\endproof
Note that the construction given in Lemma~\ref{lemma:rounding_algorithm} has been previously considered in the literature of single-customer assortment optimization, see e.g.~\cite{topaloglu2013joint,gallego2019revenue}. However, to the best of our knowledge, it has not been used before as a randomized rounding tool. Following the framework in our Meta-algorithm~\ref{alg:meta_algorithm}, we consider the following relaxation: 
\begin{align}\label{eq:relaxation_mnl_mnl}
\max &\quad \sum_{j\in \S}r_j\cdot\frac{z_j}{1+z_j}\\
s.t. &\quad z_j = \sum_{i\in\C}w_{j,i}v_{i,j}y_{i,j}\notag\\
&\quad \mathbf{y}\in\R.\notag
\end{align}
We note that this relaxation is similar to the one used by the choice-aware Frank-Wolfe algorithm. In fact, as we show next, both problems are equivalent.
\begin{lemma}\label{lemma:relaxation_mnl_mnl}
Assume that customers and suppliers follow different MNL choice models and there are no constraints on the assortments. Then, Problem~\eqref{eq:relaxation_mnl_mnl} is equivalent to Problem~\eqref{eq:relaxation_mnl_suppliers}.
\end{lemma}
We defer the proof of this lemma to Appendix~\ref{sec:appendix_mnl_mnl}. Note that we can efficiently solve Problem~\eqref{eq:relaxation_mnl_mnl} using standard ascent algorithms from concave optimization. Our main method for this section is formalized in Algorithm~\ref{alg:mnl_mnl}. Note that Algorithm~\ref{alg:mnl_mnl} also roughly follows the structure of our Meta-algorithm~\ref{alg:meta_algorithm}, since it computes an assortment distribution for each customer and samples one from it. \revised{The main advantage of Algorithm~\ref{alg:mnl_mnl} is that it does not iteratively build the assortment distribution while iteratively solving the relaxation. Instead, the distributions are obtained ex-post by using the fractional solution of the relaxation. This makes it particularly appealing for solving large-scale instances.}
\begin{algorithm}[htpb]
\caption{The MNL-MNL Method}\label{alg:mnl_mnl}
\begin{algorithmic}[1]
\Require  Non-negative values $v_{i,j}$ and $w_{j,i}$ for every $i\in\C$, $j\in\S$.
\Ensure An assortment $S_i\subseteq\S$ for each $i\in\C$.
\State Obtain an optimal solution $\mathbf{y}^\star$ by solving Problem~\eqref{eq:relaxation_mnl_mnl}.
\State Construct the assortment distributions $\bm{\tau}$ as in Lemma~\ref{lemma:rounding_algorithm} using $\mathbf{y}^\star$.
\State For each $i\in\C$, independently sample an assortment according to distribution $\bm{\tau}_i$.
\end{algorithmic}
\end{algorithm}

\proof[Proof of Theorem~\ref{thm:mnl_mnl}.]
The expected objective value of the random assortment profile output by Algorithm~\ref{alg:mnl_mnl} is such that
\[
\EE_{\mathbf{S}\sim\bm{\tau}}[\Ma_{\mathbf{S}}] \geq \frac{1}{4}\cdot \OPT_{\eqref{eq:relaxation_mnl_mnl}} = \frac{1}{4}\cdot \OPT_{\eqref{eq:relaxation_mnl_suppliers}}\geq \frac{1}{4}\cdot \OPT,
\]
where the first inequality can be shown in the same way as shown in the proof of Theorem~\ref{thm:fw_factor_anyregime}, the next equality is due to Lemma~\ref{lemma:relaxation_mnl_mnl} and the last inequality is due to Lemma~\ref{lemma:relaxation_mnl_suppliers}.
\endproof
\begin{remark}
Note that we could have constructed a continuous submodular relaxation instead of the concave relaxation~\eqref{eq:relaxation_mnl_mnl} which would lead to a $(1-1/e)$-approximation factor. However, solving the continuous submodular relaxation is computationally more challenging than solving \eqref{eq:relaxation_mnl_mnl}.
\end{remark}

\subsection{Computational Study}\label{sec:computational_study}

In this section, we empirically evaluate our proposed approximation schemes. In particular, the main goal is to not only demonstrate that our methods perform well in practice relative to their worst-case guarantees, but also to show that the efficient methods with suboptimal approximation guarantees presented earlier can outperform (both in time and objective value) the algorithms with optimal or near-optimal guarantees.
\vspace{0.5em}

\noindent{\bf Experimental setup.} We  restrict our attention to the unconstrained MNL-MNL model, where customers and suppliers follow (possibly) different multinomial logit choice models.\footnote{Our preliminary analysis did not show significant  differences in the conclusions when cardinality constraints were considered.} Also, we focus on maximizing the expected total number of matches, i.e., we set $r_j=1$ for all $j\in\S$. Recall from the MNL-MNL model, the notation $v_{i,j}\geq 0$ for customer $i$'s values over suppliers and $w_{j,i}>0$ for supplier $j$'s values over customers. For the MNL-MNL model, we consider 3 classes of instances:
\begin{itemize}
    \item {\it Heterogeneous.} This is the most general class of MNL-MNL instances, where customers and suppliers may have different attraction values $v_{i,j}$ and $w_{j,i}$. We divide this class into three sub-classes: (i) {\it Low-value regime}, where we impose $v_{i,j}\leq 1$ for all $i\in \C, j\in \S$ (this means that suppliers are valued less than the outside option);  (ii) {\it High-value regime}, where we impose $v_{i,j}\geq 1$ for all $i\in \C, j\in \S$; (iii) {\it Mixed}, where values can be either greater or lower than 1. In all the sub-classes, we do not impose any restrictions on the values $w_{j,i}$.
    \item {\it Homogeneous-Uniform.} For this class of instances, we consider the model in \cite{ashlagi_etal19}, where the values satisfy $v_{i,j} = v_j$ (i.e., customers' valuations are homogeneous) and $w_{j,i} = 1/w_{j}$ for all $i\in \C, j\in \S$ (i.e., values do not depend on the customer).
    \item  {\it Small Probability Regime.} In this class of instances, all the parameters $v_{i,j}$ and $w_{j,i}$ may be different, but we assume that, for all $i\in \C, j\in \S$,  $w_{j,i}\leq \epsilon/(1-\epsilon)$ where $\epsilon\in[0,1/2]$ is a given parameter.
\end{itemize}
\vspace{0.5em}
In Appendix~\ref{sec:details_experiments}, we provide more details on the design of our instances. For the first  2 classes: we fix the number of suppliers to $m=100$, we vary the number of customers $n\in\{50,75,100,150,200\}$ and for each combination $(m,n)$ we generate 20 random instances. For the last class of instances: we fix $m=n=100$, we vary $\epsilon\in\{0.01,0.05,0.1,0.250.5\}$ and, for each $\epsilon$, we generate 20 random instances.  

\vspace{0.5em}
\noindent{\bf Algorithms and Benchmarks.} For the unconstrained setting we tested: (i) choice-aware continuous greedy (Algorithm~\ref{alg:continous_greedy_choice} with $\delta =0.1$)\footnote{We chose $\delta=0.1$ to improve efficiency. We did not observe any substantial change in our results for smaller $\delta$.}; (ii) choice-aware Frank-Wolfe (Algorithm~\ref{alg:frankwolfe_choice} with $T=10$)\footnote{The results were similar for larger values of $T$.}; (iii) choice-aware discrete greedy (Algorithm~\ref{alg:choiceaware_greedy}); and (iv) the MNL-MNL method (Algorithm~\ref{alg:mnl_mnl}). For the homogeneous instances, we also tested the main algorithm in \cite{ashlagi_etal19}. As a benchmark, we solved the concave relaxation~\eqref{eq:relaxation_mnl_mnl} which provides an upper bound of the true optimal value under the MNL-MNL setting. 
\vspace{0.5em}
\noindent{\bf Simulations.} Observe that most of our methods are randomized thus, for each distribution output, we sample 10 families of assortments and average their objective values. Also, since the objective value corresponds to the expected matching size then, for a given family of assortments, we average the resulting matching size over 100 simulations. Note that the choice-aware continuous greedy algorithm performs a sampling step to compute $\EE_{C\sim\mathbf{z}_j(t)}\big[f_j(C+i)-f_j(C)\big]$ for each $j\in\S,i\in\C$. To compute a rough estimation of this expected value, we run 10 simulations and average the resulting marginal values. 

\begin{figure}[htb!]
\begin{center}
\begin{subfigure}{0.49\textwidth}
\caption{\footnotesize Heterogeneous, high-value}
\label{fig:heterogeneous_high}
\begin{center}
\includegraphics[width=0.8\textwidth]{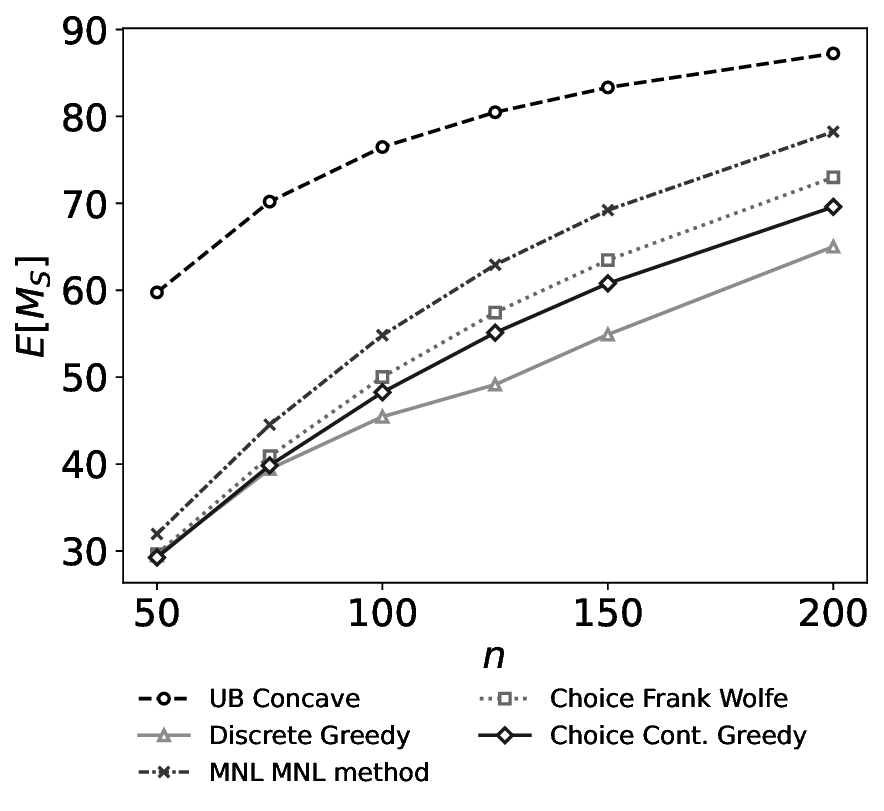}
\end{center}
\end{subfigure}
\begin{subfigure}{0.49\textwidth}
\caption{\footnotesize Heterogeneous, low-value}
\label{fig:heterogeneous_low}
\begin{center}
\includegraphics[width=0.8\textwidth]{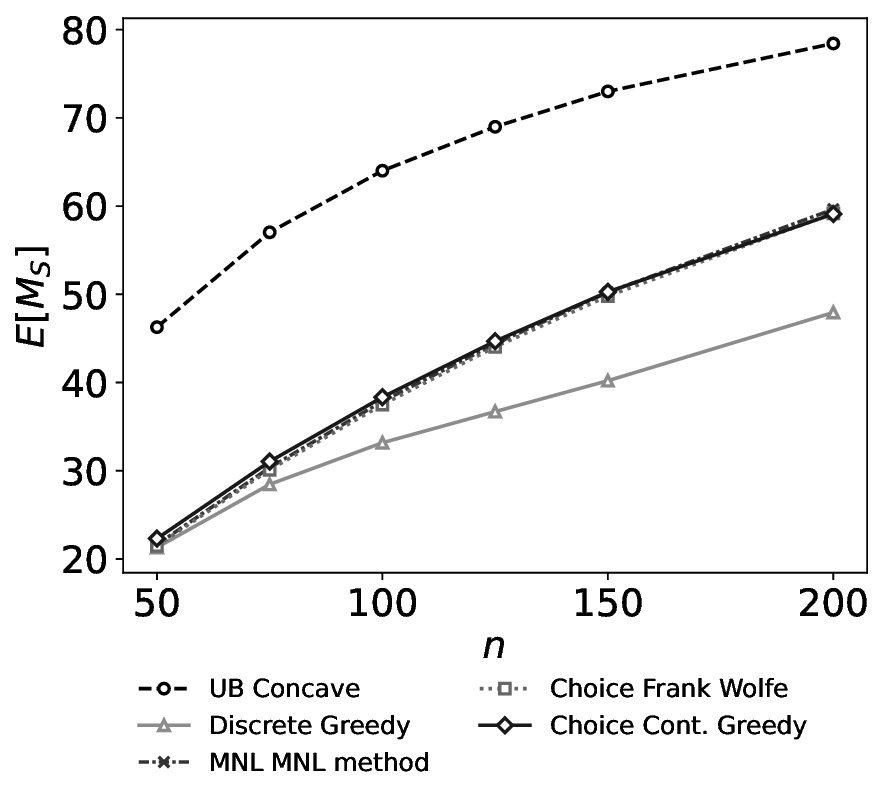}
\end{center}
\end{subfigure}
\begin{subfigure}{0.49\textwidth}
\caption{\footnotesize Heterogeneous, mix}
\label{fig:heterogeneous_mix}
\begin{center}
\includegraphics[width=0.8\textwidth]{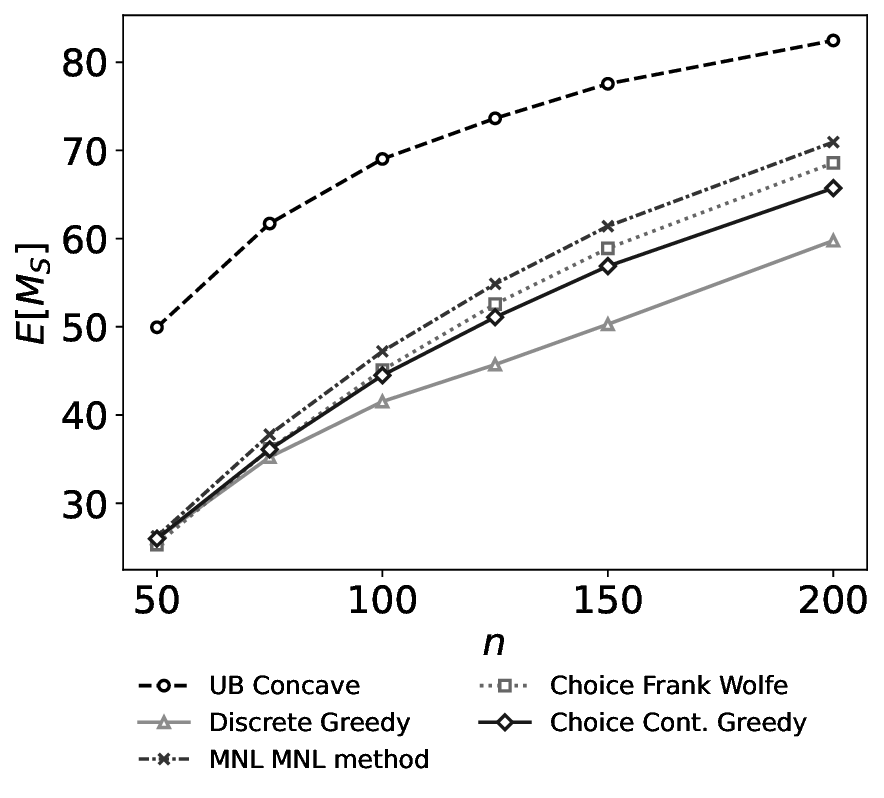}
\end{center}
\end{subfigure}
\begin{subfigure}{0.49\textwidth}
\caption{\footnotesize Homogeneous-Uniform}
\label{fig:homogeneous}
\begin{center}
\includegraphics[width=0.8\textwidth]{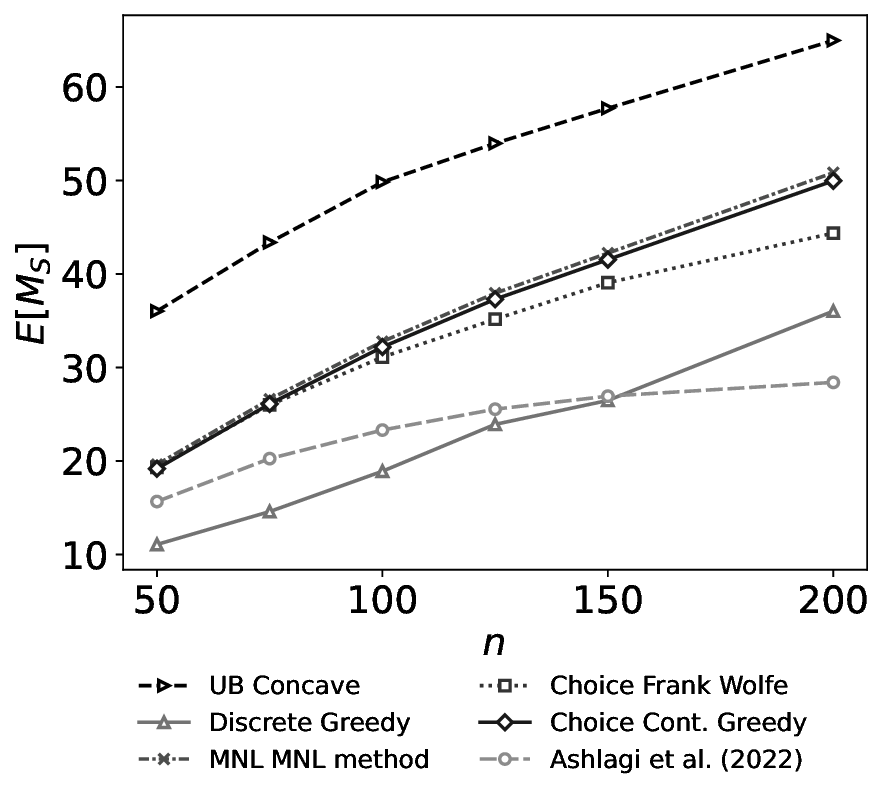}
\end{center}
\end{subfigure}
\caption{Results for heterogeneous and homogeneous instances}\label{fig:heterogeneous_homogeneous_results}
\end{center}
\end{figure}
\vspace{0.5em}
\noindent{\bf Results: Heterogeneous and Homogeneous Instances.}  The performances of each algorithm and benchmark in heterogeneous and homogeneous instances are shown in Figure~\ref{fig:heterogeneous_homogeneous_results}. One clear observation from each of these figures is the strong performance of the MNL-MNL method across all types of instances. The only type of instances where the choice-aware continuous greedy and Frank-Wolfe algorithms are competitive is the Heterogeneous low-value regime class. One possible explanation for this behavior is that when customers have a low valuation for each supplier, then denser assortments tend to perform better. In other words, when customers are pickier (lower values $v_{ij}$), then the optimal assortments tend to be larger in size since this would attract customers to stay in the market and not choose the outside option. In general, both the choice-aware continuous greedy and Frank-Wolfe have a fairly similar performance across all instances, perhaps with some differences under the Homogeneous-Uniform case when $n$ grows. On the other hand, the discrete greedy algorithm performs the worst among our methods, except when the number of customers $n$ decreases. Finally, for Homogeneous-Uniform instances, we note that the algorithm of \cite{ashlagi_etal19} is competitive for smaller values of $n$, however, its performance does not scale with $n$ and, instead, it is even worse than the Discrete Greedy method.

We also tested the performance of each algorithm in terms of running time, which we report in Figure~\ref{fig:time_performance} in Appendix~\ref{sec:details_experiments}. From these results, we note that the most efficient methods are the MNL-MNL algorithm, discrete greedy and the algorithm in \cite{ashlagi_etal19}. Optimal algorithms such as the choice-aware continuous greedy method suffer from scalability issues since each of the $n$ sub-problems (Step 3 in Algorithm~\ref{alg:continous_greedy_choice}) needs to estimate the expected marginal gains for each supplier $j\in\S$. Similar issues may be faced by the choice-aware Frank-Wolfe method since, for larger markets, more iterations (i.e., larger values of $T$) may be necessary for convergence. In conclusion, the MNL-MNL method is an efficient algorithm with a strong empirical performance (much better than its worst-case guarantee), therefore it is a promising tool for a market where customers and suppliers follow MNL choice models (arguably the most used choice model in practice). Also, the discrete greedy method is a versatile approach that performs reasonably well, it is efficient and it can be used for a wider class of choice models and constrained settings. For decision-makers whose computing resources are not constrained, then the choice-aware Frank-Wolfe algorithm can be an appealing alternative. Finally, the algorithm of \cite{ashlagi_etal19} has a strong time performance, however, it is constrained to a certain class of instances, which makes its usage quite limited.

\vspace{0.5em}
\noindent{\bf Results: Epsilon-Small Probability Regime.} The results for this class of instances are shown in Table~\ref{tab:small_probability_regime}. The results in this table show that, when $\epsilon$ decreases, the performance of the choice-aware Frank-Wolfe algorithm significantly improves, almost matching the value of the upper bound. These empirical results confirm our theoretical understanding under the $\epsilon$-small probability regime. This is not the case for the choice-aware continuous greedy algorithm and the discrete greedy method which still have a noticeable gap relative to the upper bound. Perhaps, surprisingly, the MNL-MNL method's performance is also nearly optimal. We leave as an interesting future direction to prove whether the MNL-MNL method has a near-optimal guarantee under this class of instances.

\begin{table}[htp!]
    \caption{Average objective values for different epsilons}\label{tab:small_probability_regime}
    \centering
    \scalebox{1.0}{\begin{tabular}{lrrrrr}
      \toprule
       &  \multicolumn{5}{c}{\large $\epsilon$} \\
       \cmidrule(lr){2-6}
         & 0.01 & 0.05 & 0.1 & 0.25 & 0.5\\
            \midrule
       Concave UB &  0.794 & 3.963 & 7.970 & 20.471 & 43.396\\
       \midrule
       {Choice Frank-Wolfe} & 0.792 & 3.843 & 7.492 & 17.675 & 32.730\\
      {Choice C. Greedy} & 0.765 & 3.720 & 7.251 & 17.084 & 31.776\\
      {MNL-MNL method} & 0.792 & 3.843 & 7.499 & 17.802 & 33.696\\
       {Discrete Greedy} &  0.789 & 3.817 & 7.369 & 16.541 & 29.300\\
      \bottomrule
    \end{tabular}}
\end{table}

\section{Extensions and Further Discussion}\label{sec:other_results}
In this section, we briefly discuss some extensions of our previous results. Moreover, in Appendix~\ref{sec:operational_insights}, we expand our discussion further by deriving insights from the matching model.

\subsection{Asymptotic Optimality in Other Settings}\label{sec:nested_logit}
One natural question is whether we can obtain improved guarantees under the $\epsilon$-small probability regime for suppliers' preferences that are beyond MNL choice models. For example, in the i.i.d. online arrival model, \cite{aouad2020online} showed (see their Theorem 2) that a competitive ratio better than $1-1/e$ can be obtained for the nested-logit model, but empirically show that is no better than $\approx 0.7$ (see their Figure 3). In the following, we show that in the offline two-sided assortment optimization model, there is an algorithm that is asymptotically optimal in the $\epsilon$-small probability regime. We finish this subsection by briefly discussing asymptotic optimality for general choice models. 
\paragraph{The Nested-Logit Model.} This model was introduced in~\cite{williams1977formation} which considers correlation between customers that belong to the same \emph{nest} and, subsequently, alleviates the IIA property of the MNL model. Formally, a supplier $j\in\S$ follows the $\nu$-Nested Logit model ($\nu$-NL) if there exists a partition $\C_1,\ldots,\C_K$ of the set of customers $\C$, where each $\C_k$ is called a nest and $\nu = (\nu_1,\ldots,\nu_K)$, with $\nu_k\in(0,1]$ for all $k\in[K]$, are called the dissimilarity coefficients. Supplier $j$ has a value $w_{j,i}$ for each customer $i\in\C$ and her choice behavior is as follows: For a given subset of customers $C\subseteq \C$, first, it chooses a nest $k\in[K]$ with probability
\[
\frac{w_j(C_k)^{\nu_k}}{1+\sum_{q\in[K]} w_j(C_q)^{\nu_q}}
\]
where the outside option is normalized to 1, $C_k = C\cap\C_k$ and $w_j(C_k) = \sum_{i\in C_k}w_{j,i}$. After choosing a nest, then supplier $j$ chooses a customer $i$ from that nest with probability proportional to the weight of the nest, i.e.,
\(
w_{j,i}/w_j(C_k).
\)
Therefore, for any $i\notin C$, the probability of $j$ choosing $i$ is zero. Given this, the demand function of supplier $j$ is given by
\begin{equation}\label{eq:demand_nl}
f_j(C) = \sum_{k\in[K]} \sum_{i\in C_k} \frac{w_{j,i}}{w_j(C_k)}\cdot \frac{(w_j(C_k))^{\nu_k}}{1+\sum_{q\in[K]}(w_j(C_q))^{\nu_q}} = \frac{\sum_{k\in[K]}(w_j(C_k))^{\nu_k}}{1+\sum_{k\in[K]}(w_j(C_k))^{\nu_k}}.
\end{equation}
Note that when $\nu_k\in(0,1)$ for all $k\in[K]$, the demand function is monotone and submodular as it is the composition of a concave function with a linear function. The value of the parameter $\nu_k$ models the competition among customers in nest $k$. In particular, when $\nu_k$ is small, the contribution of an extra customer in the nest is marginal as individuals of that nest are more similar. With this in mind, our result is the following:
\begin{proposition}\label{prop:approx_nl}
Assume that every suppliers has a $\nu$-NL model with $\nu_k\in(0,1)$ for all $k\in[K]$ and satisfies the $\epsilon$-small probability regime. Denote by $\bar{\nu} =\max_{k}\nu_k$. Then, there exists a $\Big(1-\frac{1}{n}-\frac{1}{n^2}\Big)$-approximation algorithm with $\epsilon\in(0,\frac{1}{n^{2\bar{\nu}}+1}]$, where $n$ is the number of customers.
\end{proposition}
We defer the proof of Proposition~\ref{prop:approx_nl} to Appendix~\ref{sec:proofs_nl}. As in Theorem~\ref{thm:fw_nearoptimality_epsregime}, Proposition~\ref{prop:approx_nl} shows that, in a large market with sufficiently picky suppliers, an appropriate concave relaxation leads to a solution whose objective value is asymptotically optimal.

\begin{remark}
We leave as an open question whether asymptotic optimality can be shown for a choice model other than MNL and the nested-logit. First, we remark that, in the general case, this is not possible due to Proposition~\ref{prop:optimal_factor} (Appendix~\ref{app:hardness}) which states that, unless NP $=$ P, there is no randomized algorithm that achieves a factor guarantee better than $1-1/e$ even under the $\epsilon$-small probability regime. Despite this, in Appendix~\ref{app:curvature}, we briefly discuss the relation between the notion of curvature of a set function, the $\epsilon$-small probability regime and a regime which we name $\xi$-large marginal probability regime: For any $j\in\S$ and $i\in\C$, we define as $f_j(\C)-f_j(\C\setminus\{i\})\geq \xi\cdot f_j(\C)$.
\end{remark}
\begin{remark}\label{remark:inventory}
We can also obtain near-optimality for the offline variant of the problem consider in \cite{golrezaei2014real}. In this case, each supplier has an aggregate function $f_j(C) = \min\{|C|,c_j\}$. The $\epsilon$-small probability regime would be equivalent to $1=f_j(\{i\})\leq \epsilon\cdot f_j(\C) =\epsilon\cdot \min\{n,c_j\}$ for all $i\in\C, j\in\S$. Therefore, if $\epsilon\leq 1/n$, this condition implies that $c_j\geq n$ for all $j\in\S$ and the objective~\eqref{eq:objective_alternative} can be equivalently written as 
\[
 \EE\left[\sum_{j\in\S} r_j\cdot f_j(C^{\mathbf{S}}_j)\right] = \EE\left[\sum_{j\in\S} r_j\cdot \min\{|C^{\mathbf{S}}_j|,c_j\}\right]= \sum_{j\in\S}r_j\cdot \EE\left[|C^{\mathbf{S}}_j|\right] = \sum_{j\in\S}r_j\cdot \sum_{i\in\C}\phi_i(j,S_i),
\]
where the second equality is because $|C^{\mathbf{S}}_j|\leq n$ and $c_j\geq n$. Thanks to Assumption~\ref{assumption:oracle}, this objective can be optimized efficiently as the function is separable.
\end{remark}

\subsection{Beyond Submodularity and Monotonicity}\label{sec:beyond_submod_mon}
So far we have assumed that the demand function of each supplier is a monotone submodular function. However, some of these properties may not hold in more general settings. One particular case is when the revenues in Problem~\eqref{def:opt_problem} are edge-dependent, i.e., if a customer $i\in\C$ and a supplier $j\in\C$ mutually select each other then $r_{i,j}\geq0$ is collected. In this case, functions $f_j(.)$ would correspond to the expected revenue defined as $f_j(C) = \sum_{i\in C}r_{i,j}\phi_i(j,C)$. It is well-known that, even for a simple choice model as the MNL, function $f_j$  is neither monotone nor submodular. Therefore, we expect Problem~\eqref{def:opt_problem} to be significantly harder under this revenue setting, even for the MNL-MNL case. For a simpler matching model, parametric guarantees are given in~\cite{ahmed2022parameterized} and we leave as an open question whether there exist constant approximation factors in the setting we study in the present manuscript.

Let us come back to our initial supplier-dependent revenue setting, i.e., $r_{i,j} = r_j$ for all $i\in\C,\; j\in\S$. As we pointed out in our discussion of Assumption~\ref{assumption:suppliers_monotone_submod}, the demand function of any random utility choice model satisfies submodularity. One natural question is whether we can obtain guarantees for a choice model whose demand function is not submodular. In this scenario, we can still study the case when the demand functions are monotone and $\eta$-weakly submodular, see e.g.~\cite{das2011submodular,bian2017guarantees,chen2018weakly}. Specifically, a set function $f:2^\U\to\RR_+$ is $\eta$-weakly submodular for $\eta\in(0,1]$ if for any sets $A,B\subseteq\U$ we have 
\[
\sum_{e\in B}f(A\cup\{e\}) - f(A) \geq \eta\cdot[f(A\cup B) - f(A)].
\]
Note that for $\eta=1$ the function is submodular. Given this, we can use Algorithm~\ref{alg:continous_greedy_choice} to obtain a $(1-e^{-\eta})$-factor for monotone $\eta$-weakly submodular demand functions, as long as Assumption~\ref{assumption:oracle} is met. 

Regarding monotonicity, recall from our discussion of Assumption~\ref{assumption:suppliers_monotone_submod} that any choice model that satisfy the substitutability axiom has a monotone demand function. This axiom is widely accepted in the literature of assortment optimization, however, we can think of a setting where functions $f_j$'s are  aggregate set functions which are not necessarily derived from a choice model. There are several technical challenges for the case when all functions $f_j$'s are submodular but not monotone. The analysis done for Algorithm~\ref{alg:continous_greedy_choice} cannot be extended to this setting for two reasons: First, the weights defined  as $\theta_{i,j}(t) := r_j\cdot\EE_{C\sim\mathbf{z}_j(t)}[f_j(C+i)-f_j(C)]$ are not necessarily non-negative, thus Assumption~\ref{assumption:oracle} might no be satisfied. Second, the functions $F_j$'s do not satisfy $F_j(\mathbf{z}_j(t)\vee\Phi(\mathbf{S}^\star))\geq F_j(\Phi(\mathbf{S}^\star))$. One alternative would be to adapt the measured continuous greedy algorithm introduced in~\cite{feldman2011unified} for (non-monotone) submodular maximization. However, the analysis of this algorithm is valid for \emph{downward closed}\footnote{A feasible region $\P$ is downward closed if for any $\mathbf{z}\in\P$ and $\mathbf{z}'\leq \mathbf{z}$ then $\mathbf{z}'\in\P$.} regions, which is not our case due to~\eqref{eq:relation_z_tau}. We leave as a future research direction to study adaptations of other algorithms in the literature of submodular maximization, e.g.~\cite{buchbinder2015tight,kuhnle2019interlaced}.

\section{Conclusion}

This work studies a general framework for the offline two-sided assortment optimization problem in sequential matching markets. While the related literature focuses on the MNL-Uniform model, we significantly expanded it to heterogeneous agents with general choice models. Moreover, our approach allows us to consider the maximization of the expected revenue over different type of constraints.
In this setting, we are the first to draw connections with the literature of submodular optimization. More importantly, we provide an optimal $(1-1/e)$-approximation algorithm which consists of a two-sided version of the continuous greedy algorithm. We also design a more efficient greedy algorithm, however, with a suboptimal $1/2$-approximation factor.
We are able to beat the $1-1/e$ barrier for the setting in which suppliers' preferences have MNL models and the instances fall into the $\epsilon$-small probability regime. This results marks a significant difference with the online version of the problem as we observed in Table~\ref{table:summary}.
 Our algorithm consists of a two-sided version of the Frank-Wolfe algorithm which is asymptotically optimal under this regime. Finally, we provide a simple and effective algorithm that guarantees $1/4$-approximation under the unconstrained MNL-MNL setting. Our empirical analysis show that this method, despite its worst-case guarantee, significantly outperforms the rest of our algorithms.
 
As a future research direction, it would be interesting, and promising from a practical perspective, to look at models that allow for a second round of matches, namely, remaining unmatched suppliers can be matched. From a theoretical standpoint there are a couple of challenging open questions:  (i) design deterministic algorithms with constant guarantees better than $1/2$; (ii) adapt the tools in this work to improve the gap between the upper and lower bounds for the online one-sided setting; (iii) show constant guarantees for the setting with edge-dependent revenue values.

\paragraph{Acknowledgements} The second author thanks the support of  the Institut de valorisation des donn\'ees (IVADO) and Fonds de recherche du Qu\'ebec (FRQ) through the FRQ- IVADO Research Chair and of the Natural Sciences and Engineering Research Council of Canada (NSERC) under the grant 2019-04557.

\addcontentsline{toc}{section}{Bibliography}
\bibliographystyle{plainnat}
\bibliography{bibliography_assortment}

\changelocaltocdepth{1} 
\appendix

\section{Appendix to Section~\ref{sec:warmup}}\label{app:warmup}

\begin{example}\label{example:not_monotone_nor_submodular}
Consider 1 customer $i$ and 2 suppliers $j,j'$ both with MNL choice models. The preference weights are as follows: $v_{i,j}=1/2$, $v_{i,j'}=1$, $w_{j,i} = 1$ and $w_{j',i} = 1/6$. Since there is only one customer, then an assortment profile $\mathbf{S}$ is composed only by $S_i$. If we consider $S_i = \{j\}$ then the expected number of matches is $f(S_i) = 1/6$. Now, consider $\hat{S}_i=\{j,j'\}$ then we obtain
\[
f(\hat{S}_i) = \frac{1/2}{1+1+1/2}\cdot\frac{1}{1+1}  + \frac{1}{1+1+1/2}\cdot \frac{1/6}{1+1/6} = \frac{11}{70} < \frac{1}{6}.
\]
To show that $f$ is not submodular, consider a third supplier $j''$ such that $v_{i,j''}$ and $w_{j'',i}$ are sufficiently large so, whenever $j''$ is in the assortment, then the probability of a match between $i$ and $j''$ is close to 1. Then, one can show that 
\[
0\approx f(\{j,j',j''\}) - f(\{j,j''\}) > f(\{j,j'\}) - f(\{j\}) .
\]
\end{example}

\subsection{Hardness.}\label{app:hardness}

Before showing Proposition~\ref{prop:optimal_factor}, we define the rank-based choice model.
\begin{definition}[Rank-based Choice Model]
A supplier $j\in\S$ is said to have a rank-based choice model if there exists a multinomial distribution $(\lambda^j_1,\ldots,\lambda^j_K)$ over $K$ ranked lists $L^j_1,\ldots,L^j_K$ where each list is a subset of customers with a strict order. For a given subset $C\subseteq \C$, supplier $j$ samples a ranked list $L^j_k$ according to the multinomial distribution and chooses the top rank customer among those in $C\cap L^j_k$; if empty, then the supplier remains unmatched. The demand function in this case corresponds to
\[
f_j(C) = \sum_{k=1}^K\lambda_k\cdot\min\{|C\cap L^j_k|,1\}.
\]
\end{definition}
In the next proof, we follow the ideas in~\cite{aouad2020online}.
\proof[Proof of Proposition~\ref{prop:optimal_factor}.]
Let be $(\N,\M,\{u_j\}_{j\in \M})$ be an instance of the offline welfare maximization problem, see e.g.~\cite{lehmann_etal06,vondrak2008optimal}, where $\N$ is the set of items, $\M$ the set of agents and $\{u_j\}_{j\in \M}$ the coverage valuations. For these valuations we assume there exists a universe of elements $\E$ and a subset system $\U=\{U_i\}_{i\in \N}$. Each item $i\in \N$ is associated with a set $U_i$ and each valuation $u_j$ is defined by a permutation $\sigma_j: \N\to \N$ over the sets in $\U$ such that $u_j(C) = \left|\bigcup_{i\in C}U_{\sigma_j(i)}\right|$. Now we construct an instance of our problem:
\begin{itemize}
\item For each item $i\in \N$ define a customer $i$, i.e. $\C =\N$. Each customer $i$ is: (i)  defined by a choice model $\phi_i$ with no outside option, i.e., $\phi_i(0,S) = 0$ whenever $S\neq\emptyset$; (ii) associated to the set $U_i$.
\item For each agent $j\in \M$ we define a supplier $j$, i.e., $\S =\M$. Each supplier $j$ has rank-based preferences defined as follows: Construct one preference list per element in $\U$, i.e., $L^j_1,\ldots, L^j_{|\U|}$, where $L^j_k = \{i: \ k\in U_{\sigma_j(i)}\}$ is ranked in arbitrary order and composed by customers who ``cover'' element $k\in \U$ with respect to permutation $\sigma_j$. Finally, the choice model of supplier $j$ is determined by a uniform distribution over the list $L^j_1,\ldots, L^j_{|\U|}$, i.e., for any set of customers $C$ we have 
\[
f_j(C) = \frac{1}{|\U|}\sum_{k\in \U}\min\{|C\cap L^j_k|,1\} = \frac{1}{|\U|}\sum_{k\in \U}\uno_{\{C\cap L^j_k\neq\emptyset\}}.
\]
\end{itemize}
Note then that for any $C\subseteq \N$ and $j\in \M$ we have
\[
u_j(C) = \left|\bigcup_{i\in C}U_{\sigma_j(i)}\right| = \sum_{k\in\U}\uno_{\{k\in \bigcup_{i\in C}U_{\sigma_j(i)}\}} =  \sum_{k\in \U}\uno_{\{C\cap L^j_k\neq\emptyset\}} = |\U|\cdot f_j(C).
\]
Consider an optimal solution of the offline welfare maximization problem $C^\star_j$ for each $j\in \M$. Recall that the sets $C^\star_j$ define a partition over $\N$. We now construct a feasible solution to our problem: Let $S_i = \{j\}$ be the assortment of customer $i$ where $j$ is such that item $i\in C^\star_j$. Therefore, each customer chooses with probability 1 the only  supplier in the assortment, so we obtain
\[
\OPT \geq \EE[ \Ma_{\mathbf{S}} ]=\sum_{j\in \S}f_j(C^\star_j) = \frac{1}{|\U|}\sum_{j\in \M}u_j(C^\star_j) = \frac{1}{|\U|}\OPT',
\]
where $\OPT'$ is the optimal value of the offline welfare maximization problem.
Assume by contradiction that there exists an $1-1/e+\delta$-approximation algorithm for our problem. Let $\mathbf{S}$ be the feasible solution output by that algorithm. We now construct the following randomized algorithm for the offline welfare maximization problem. Sample the choices made by the customers and let $C_j$ be the set of requests that supplier $j$ received. Assign the $C_j$ of items to agent $j$ in the original instance. Then note that the expected value of this algorithm is
\[
\EE_{C\sim\Phi}\left[\sum_{j\in \M}u_j(C)\right] = |\U|\EE_{C\sim\Phi}\left[\sum_{j\in\S}f_j(C)\right]\geq |\U|\cdot (1-1/e+\delta)\cdot \OPT \geq (1-1/e+\delta)\cdot \OPT',
\]
where $\Phi$ is the vector of choice probabilities determined by assortment family $\mathbf{S}$ and the expectation above is over customers' choices. The above contradicts the hardness of the offline welfare maximization problem, see e.g.~\cite{khot2008inapproximability,kapralov2013online}. 

The proof for the $\epsilon$-small probability regime follows by noting the details given in~\cite{kapralov2013online} for the hardness of the Max-$k$ cover problem. Specifically, the hardness result holds even if every subset is of size $s$ and $|\U|$ is greater than or equal to $N_0$. Using this in the reduction above we note that
\[
f_j(\{i\}) = \frac{1}{|\U|}\sum_{k\in \U}\uno_{\{\{i\}\cap L^j_k\neq\emptyset\}}=\frac{1}{|\U|}\sum_{k\in \U}\uno_{\{k\in U_{\sigma_j(i)}\}} = \frac{|U_{\sigma_j(i)}|}{|\U|}\leq \frac{s}{N_0}.
\]
Note also that $f_j(\C) = 1$. Therefore, the $\epsilon$-small probability regime is satisfied when $\frac{s}{N_0}\leq \epsilon$. We conclude that, for every $\delta,\epsilon>0$ no polynomial-time algorithm (even randomized) achieves an approximation guarantee better than $1-1/e+\delta$.
\endproof

\subsection{Missing Details in Section~\ref{subsubsec:choice_oblivious_greedy}}\label{app:poor_performance}

\proof[Proof of Proposition \ref{prop:guarantee_singlecustomer}.]
Assume $|\C| = |\S| = n$. Suppose every customer follow the same MNL choice model, i.e. $v_{ij}= v_j$, with the following weights:  there is a single supplier $j'$ such that $v_{j'} = n$ and $v_j= \epsilon = 1/(n-1)$ for all $j \neq j'$. For suppliers preferences, assume that every $j\in\S$ follows a MNL choice model with values $w_{j,i}=1$. Finally, assume $r_j=1$ for all  $j\in \S$.

In this instance, Algorithm \ref{alg:single_customer} will output $S_i = \S$ for every customer $i\in \C$, since weights are all the same and the function $h(S) = \sum_{j\in S}v_j/(1+\sum_{j\in S}v_j)$ is maximized when $S=\S$ due to monotonicity. Now, let us evaluate the objective value of the assortment family $\mathbf{S} = (\S,\ldots,\S)$. Due to Jensen's inequality we obtain
\[
\EE[\Ma_{\mathbf{S}}] \leq \sum_{j\in \S}\frac{\sum_{i\in \C}\phi_{i}(j,\S)}{1+\sum_{i\in \C}\phi_{i}(j,\S)} = \frac{n^2}{2+n+n^2} + \frac{n\epsilon}{2+n+n\epsilon} \leq 1+\frac{1}{n}
\]
where the first term corresponds to $j'$ and the second to the sum over $j\neq j'$. Also, note that for all $i\in \C$ we have $\phi_{i}(j,\S) = v_j/(1+\sum_{\ell\in \S}v_\ell) = n/(2+n)$ for $j=j'$ and $\phi_{i}(j,\S) = \epsilon/(2+n)$ for all $j\neq j'$.

On the other hand, consider the following assortment family $\mathbf{S}'$: for customer $i=1$ show assortment $S'_1 = \{j'\}$ and for the rest of the customers $i\neq 1$ show assortment $S'_i = \S\backslash\{j'\}$. Now, we will show that $\EE[\Ma_{\mathbf{S}'}] \geq n/6$. First, note that for every $i\neq 1$ and $j\neq j'$, 
\[
\phi_{i}(j,\S\backslash\{j'\}) = \frac{v_j}{1+\sum_{\ell \in \S\backslash\{j'\}}v_\ell} = \frac{\epsilon}{1+\epsilon\cdot (n-1)},
\] 
which implies that for every $j\neq j'$ we have
\[
\sum_{i\in \C}\phi_{i}(j,S_i) = \sum_{i\neq 1}\frac{\epsilon}{1+\epsilon\cdot (n-1)}= \frac{(n-1)\cdot \epsilon}{1+\epsilon\cdot (n-1)} = \frac{1}{2}.
\]
On the other hand, for customer $i=1$ we have $\phi_{1}(j',\{j'\}) = n/(n+1)$. Therefore, Lemma \ref{lemma:supplier_mnl_bound1} with $\epsilon=1/2$ implies
\begin{align*}
\EE[\Ma_{\mathbf{S}'}] &\geq \frac{1}{2}\left(\frac{\phi_{1}(j',\{j'\})}{1+\phi_{1}(j',\{j'\})}+\sum_{j\neq j'}\frac{\sum_{i\in \C}\phi_{i}(j,S'_i)}{1+\sum_{i\in \C}\phi_{i}(j,S'_i)}\right) = \frac{1}{2}\left(\frac{n}{1+2n}+\frac{n-1}{3}\right)\geq \frac{n}{6}
\end{align*}
where the last inequality follows from $n/(1+2n)\geq 1/3$.
Finally, the guarantee of Algorithm \ref{alg:single_customer} can be upper bounded as follows
\[
\frac{\EE[\Ma_{\mathbf{S}}]}{\OPT}\leq \frac{\EE[\Ma_{\mathbf{S}}]}{\EE[\Ma_{\mathbf{S}'}]} = \frac{1+1/n}{n/6}\xrightarrow[n \to \infty]{} 0.
\]
\endproof

We now discuss the other choice-oblivious approach of solving
\[
\max\left\{\sum_{j\in\S}r_j\cdot f_j(C_j):\; C_j\subseteq \C, \ C_j\cap C_{j'}\neq\emptyset\text{ for }j\neq j'\in\S\right\}
\]
First, note that this is the submodular welfare problem, which is NP-hard to solve. However, even the optimal solution of this would perform arbitrarily bad in the worst-case as the next proof shows.
\proof[Proof of Proposition~\ref{prop:guarantee_oblivious_sw}.]
Consider $n$ customers and two suppliers $j,j'$ with $r_j=r_{j'}=1$. Both sides follow MNL choice models with preference weights: for customers $v_{i,j}=1$, $v_{i,j'}=\epsilon$ and for suppliers $w_{j,i}=1/2$, $w_{j',i}=1$ for all $i\in\C$. Note that the demand functions are
\[
f_j(C) = \frac{|C|/2}{1+|C|/2} \qquad\text{and}\qquad f_{j'}(C) = \frac{|C|}{1+|C|} 
\]
Thanks to the concavity of these functions, one can easily compute the optimal solution of the submodular welfare problem which is $C_j^\star =\emptyset$ and $C_{j'}^\star = \C$. Therefore, the assortment profile $\mathbf{S}$ constructed from this solution is such that $S_i =\{j'\}$ for all $i\in\C$ resulting in $\EE[\Ma_{\mathbf{S}}] \approx 1-(1-\epsilon)^n$ which goes to 0 when $\epsilon\to 0$. On the other hand, there is a solution $\mathbf{S}'$ such that $S'_i = \{j\}$ for all $i\in\C$ which gives  $\EE[\Ma_{\mathbf{S}'}] \approx 1/2$. 
\endproof

\section{Appendix to Section~\ref{sec:optimal_algorithms}}\label{sec:missing_details_optimal}
\subsection{Missing Details in Section \ref{sec:general_case}}\label{sec:proofs_general_case}
\proof[Proof of Claim~\ref{claim:multilinear_expression}.]
Note also that once all customers  pick a supplier or the outside option, subsets $C^{\mathbf{S}}_1,\ldots, C^{\mathbf{S}}_m, C^{\mathbf{S}}_0$, where the last set denotes the customers who select the outside option, form a partition of the set of customers $\C$. Namely, for all $\bigcup_{j\in \S\cup\{0\}} C^{\mathbf{S}}_j= \C$ and $C^{\mathbf{S}}_j\cap C^{\mathbf{S}}_{j'}=\emptyset$ for all $j,j'\in \C\cup\{0\}$ such that $j\neq j'$. 
Now, we can express the objective function of our problem as follows
\begin{align}
\EE[\Ma_{\mathbf{S}}]&=\EE\left[\sum_{j\in \S}r_j\cdot f_j(C^{\mathbf{S}}_j)\right]  \notag \\
&=\sum_{\substack{(C_0,C_1,\ldots, C_m):\\ \forall \ell\neq \ell', \  C_\ell\cap C_{\ell'} =\emptyset\\ C_0\cup C_1\cup\cdots\cup C_m= \C}} \left(\sum_{j\in \S}r_j\cdot f_j(C_j)\right)\cdot\PP(C^{\mathbf{S}}_0 = C_0, C^{\mathbf{S}}_1=C_1, \ldots, C^{\mathbf{S}}_{m}=C_m) \notag\\
&=\sum_{\substack{(C_0,C_1,\ldots, C_m):\\ \forall \ell\neq \ell', \  C_\ell\cap C_{\ell'} =\emptyset\\ C_0\cup C_1\cup\cdots\cup C_m= \C}} \left(\sum_{j\in V}r_j\cdot f_j(C_j)\right)\cdot \prod_{i\in C_0}\phi_{i}(0,S_i)\prod_{i\in C_1}\phi_{i}(1,S_i)\cdots \prod_{i\in C_m}\phi_{i}(m,S_i) \notag\\
&=\sum_{j\in \S}r_j\cdot \sum_{C_j\subseteq \C}f_j(C_j)\prod_{i\in C_j}\phi_{i}(j,S_i) \sum_{\substack{(C_0,\ldots, C_{j-1},C_{j+1},\ldots,C_m):\\ \forall \ell\neq \ell', \  C_\ell\cap C_{\ell'} =\emptyset\\ C_0\cup C_1\cup\cdots\cup C_m= \C}}  \prod_{\ell\neq j}\prod_{i\in C_\ell}\phi_{i}(\ell,S_i) \notag \\
& = \sum_{j\in \S}r_j\cdot \sum_{C_j\subseteq \C}f_j(C_j)\prod_{i\in C_j}\phi_{i}(j,S_i)\prod_{i\in\C\setminus C_j}(1-\phi_{i}(j,S_i)), \notag \\
&= \sum_{j\in \S}r_j\cdot \sum_{C\subseteq \C}f_j(C)\prod_{i\in C}\phi_{i}(j,S_i)\prod_{i\in\C\setminus C}(1-\phi_{i}(j,S_i)), \notag
\end{align}
where the third equality is the explicit expectation over all possible partitions, the following equality follows from the independent choices of customers, the fifth equality is obtained by reordering the sum, the sixth expression results from fixing $C_j$ and summing over all partitions of the remaining customers and finally we note that $C_j$ does not depend on $j$.
\endproof

\proof[Proof of Lemma~\ref{lemma:relaxation_general_case}.]
Given an optimal solution $\mathbf{S}^\star$ of Problem~\eqref{def:opt_problem}, we construct the following feasible solution of Problem~\eqref{eq:relaxation_general_case}: For every $i\in \C$, consider $\tau_{i,S^\star_i} =1$ and $\tau_{i,S} =0$ for every $S\neq S^\star_i$. For every $i\in \C$ and $j\in S_i^\star$  consider $z_{i,j} = \phi_i(j,S_i^\star)$ and 0 otherwise. Note that $z_{i,j} = \sum_{S\in\mathcal{I}_i: S\ni j}\tau_{i,S}\cdot\phi_i(j;S)$, then clearly $\mathbf{z}$ and $\tau$ satisfy the constraints in~\eqref{eq:feasible_region_meta_relaxation}. Moreover, the objective value in Problem~\eqref{eq:relaxation_general_case} coincides with $\OPT$.
\endproof

In the following, we prove the approximation guarantee of Algorithm~\ref{alg:continous_greedy_choice}.
\proof[Proof of Theorem~\ref{thm:main_general}.]
To ease the exposition, in the following we consider $r_j=1$ for all $j\in\S$; the proof carries over for the case when the revenues $r_j$ are different. In this proof, we follow the outline of the proof for the approximation guarantee of the standard continuous greedy algorithm, specifically Lemma 3.3 in \cite{calinescu2011maximizing}. We assume that we have access to the exact value of the expected marginal values; otherwise, we can always estimate it to a desired close multiplicative factor as explained in~\cite{calinescu2011maximizing} with a polynomial number of samples.

First, it is straightforward to check that $\mathbf{z}^t$ and $\bm{\tau}^t$ are feasible for all $t$.
Consider $\mathbf{z}^{t+\delta}$ and $\mathbf{z}^t$ solutions from two consecutive iterations. Let $C^{t+\delta}_j$ the subset of customers who chose supplier $j$ according to choice probabilities $\mathbf{z}_j^{t+\delta}$. Similarly, define $C^{t}_j$. Also, let $D^{t}_j$ be the random subset of customers who chose supplier $j$ according to choice probabilities given by $\delta\Phi_j(\mathbf{S}^t)$ determined by the solution found in Step 3. Due to monotonicity of the demand functions,
\[
F(\mathbf{z}^{t+\delta}) = \sum_{j\in\S}\EE[f_j(C^{t+\delta})] \geq \sum_{j\in\S}\EE[f_j(C^t\cup D^t)],
\]
where the expectation is over the independent choices of customers. More formally, for each $j\in\S$ the set $C_j^{t+\delta}$ contains customers with independent probabilities $\mathbf{z}_j^{t} + \delta\Phi_j(\mathbf{S}^t)$, while the set $C^t\cup D^t$ contains customers with smaller independent probabilities $1-(1-z_{i,j}^{t})(1-\phi_i(j,S_i^t))$ for each $i\in\C$.

 Given the above, we have
\begin{align*}
F(\mathbf{z}^{t+\delta}) - F(\mathbf{z}^{t}) &\geq \EE\left[\sum_{j\in\S}f_j(C^t\cup D^t) - f_j(C_j^t)\right]\geq \sum_{j\in\S}\sum_{i\in\C}\PP(D^t=\{i\})\cdot \EE[f_j(C^t_j+i) - f_j(C^t_j)]\\
&=\sum_{i\in\C}\sum_{j\in S_i^t}\delta\cdot\phi_i(j,S_i^t)\prod_{k\neq i}(1-\delta\cdot\phi_k(j,S_k))\cdot \EE[f_j(C^t_j+i) - f_j(C^t_j)]\\
&\geq \delta\cdot(1-\delta)^{|\C|}\cdot\sum_{i\in\C}\sum_{j\in S_i^t}\phi_i(j,S_i^t)\cdot \EE[f_j(C^t_j\cup\{i\}) - f_j(C^t_j)]\\
&\geq \delta\cdot(1-\delta)^{|\C|}\cdot\sum_{i\in\C}\sum_{j\in S_i^\star}\phi_i(j,S_i^\star)\cdot \EE[f_j(C^t_j\cup\{i\}) - f_j(C^t_j)]\\
& = \delta\cdot(1-\delta)^{|\C|}\cdot\sum_{j\in\S}F_j(\mathbf{z}^t\vee\Phi_j(\mathbf{S}^\star))-F_j(\mathbf{z}^t)\\
& \geq \delta\cdot(1-n\delta)\cdot(\OPT - F(\mathbf{z}^t))\geq \delta\cdot\overline{\OPT} - F(\mathbf{z}^t),
\end{align*}
where $\overline{\OPT} = (1-n\delta)\OPT$ and $|\C|=n$. The third inequality follows by noting that $\phi_k(\cdot,\cdot)\leq 1$, the next inequality is due to optimality of Step 3, the fifth inequality is due to Fact~\ref{fact:ML} and the last one from monotonicity. Then, by solving the recurrence above we get
\[
\overline{\OPT} - F(\mathbf{z}^1) \leq (1-\delta)^{1/\delta}\cdot\overline{\OPT}\leq \frac{1}{e}\cdot\overline{\OPT},
\]
so by taking $\delta = O(1/n^2)$ we can show that 
\[
F(\mathbf{z}^1)\geq (1-1/e-o(1))\cdot\OPT.
\]
The remaining part of the proof (i.e., sampling $\mathbf{S}\sim\bm{\tau}^1$) follows from the sketch proof given before. 
Note that one can get a clean $1-1/e$ by doing a more refined analysis; we refer the interested reader to Appendix A in~\cite{calinescu2011maximizing}.
\endproof

\subsection{Missing Details in Section \ref{sec:general_mnl}}\label{app:proofs_general_mnl}
\proof[Proof of Lemma~\ref{lemma:relaxation_mnl_suppliers}.]
Consider an optimal solution $\mathbf{S}^\star$ of Problem~\eqref{def:opt_problem}. Then
\[
\EE[\Ma_{\mathbf{S}^\star}] =\EE\left[\sum_{j\in\S}r_j\cdot f_j(C^{\mathbf{S}^\star}_j)\right] = \sum_{j\in \S} \EE\left[r_j\cdot\frac{w_j(C^{\mathbf{S}^\star}_j)}{1+w_j(C^{\mathbf{S}^\star}_j)}\right]\leq \sum_{j\in V}r_j\cdot\frac{\EE[w_j(C^{\mathbf{S}^\star}_j)]}{1+\EE[w_j(C^{\mathbf{S}^\star}_j)]}
\]
where the last inequality follows by using Jensen's inequality on the concave function $x/(1+x)$. Note also that $\EE[w_j(C^{\mathbf{S}^\star}_j)]$ can be written as the multilinear extension of a linear function which is, in fact, linear. More formally, we have
\[
\EE[w_j(C^{\mathbf{S}^\star}_j)] = \sum_{C\subseteq\C}\left[\sum_{i\in C}w_{j,i}\right]\prod_{k\in C}\phi_k(j,S^\star_k)\prod_{k\in\C\setminus C}(1-\phi_k(j,S^\star_k)) = \sum_{i\in\C}w_{j,i}\cdot\phi_{i}(j,S^\star_i).
\]
We conclude the proof by constructing the following feasible solution to Problem~\eqref{eq:relaxation_mnl_suppliers}: Consider $z_j = \sum_{i\in\C}w_{j,i}\cdot\phi_{i}(j,S^\star_i)$, $\tau_{i,S^\star_i}=1$ and $\tau_{i,S} = 0$ for every $i\in\C$ and $S\neq S_i^\star$.
\endproof

\proof[Proof of Lemma~\ref{lemma:feasibility_fw}.]
Let us proceed by induction. For $t=0$, we know that $\mathbf{z}^0=\mathbf{0}$ and $\tau_{i,S}=0$ for all $i\in \C$ and $S\in\I_i$ such that $S\neq\emptyset$, thus the first family of constraints of the feasible region is satisfied. Also, we have initially $\tau_{i,\emptyset} = 1$ for all $i\in \C$, thus the second family of constraints is also satisfied. Now, let us assume that at the end of step $t-1$, points $\mathbf{z}^{t-1}$ and $\bm{\tau}^{t-1}$ are feasible. Consider $\mathbf{z}^t$ and $\bm{\tau}^t$ the updates at the end of the step $t$. We check the second family of constraints: For any $i\in \C$ we have
\begin{align*}
\sum_{S\in\I_i} \tau^t_{i,S} &= \sum_{S\in\I_i:S\neq S_i^t} \tau^t_{i,S} + \tau^t_{i,S_i^t} \\
&= (1-\gamma_t)\cdot \sum_{S\in\I_i:S\neq S_i^t} \tau^{t-1}_{i,S} + (1-\gamma_t) \tau^{t-1}_{i,S_i^t} + \gamma_t \\
&= (1-\gamma_t)\cdot\sum_{S\in \I_i}\tau^{t-1}_{i,S} + \gamma_t = 1
\end{align*}
where in the last equality we used that $\bm{\tau}^{t-1}$ is feasible. Now we check the first family of constraints: For any $j\in\S$
\begin{align*}
z_j^t &= (1-\gamma_t)\cdot z_j^{t-1} + \gamma_t\cdot\sum_{i\in \C}w_{j,i}\cdot\phi_i(j,S_i^t) \\
&= (1-\gamma_t)\cdot \sum_{i\in \C}w_{j,i}\cdot\sum_{S\in \I_i:S\ni j}\tau^{t-1}_{i,S}\phi_i(j,S)  + \gamma_t\cdot\sum_{i\in \C}w_{j,i}\cdot\phi_i(j,S_i^t)\\
&=(1-\gamma_t)\cdot \sum_{i\in \C}w_{j,i}\cdot\sum_{S\in \I_i:S\ni j,S\neq S_i^t}\tau^{t-1}_{i,S}\cdot\phi_i(j,S) +(1-\gamma_t)\cdot  \sum_{i\in \C}w_{j,i}\cdot\tau^{t-1}_{i,S_i^t}\cdot\phi_i(j,S_i^t) \\
&\quad+ \gamma_t\cdot\sum_{i\in \C}w_{j,i}\cdot\phi_i(j,S_i^t)\\
&=\sum_{i\in \C}w_{j,i}\cdot\sum_{S\in \I_i:S\ni j,S\neq S_i^t}\tau^{t}_{i,S}\cdot\phi_i(j,S) +\sum_{i\in \C}w_{j,i}\cdot\tau^{t}_{i,S_i^t}\cdot\phi_i(j,S_i^t) \\
&=\sum_{i\in \C}w_{j,i}\cdot\sum_{S\in \I_i:S\ni j}\tau^{t}_{i,S}\cdot\phi_i(j,S).
\end{align*}
\endproof

\proof[Proof of Proposition~\ref{prop:convergence_fw_mnl}.]
Let us define $g:\RR_+^\S\to\RR_+$ as the following function
\[
g(\mathbf{z}) = \sum_{j\in \S}r_j\cdot\frac{z_j}{1+z_j}
\]
which is continuous, concave and differentiable. Moreover, by the Mean Value Theorem, for any pair $\mathbf{z}, \mathbf{z}'$ there exists a point $\xi\in\RR_+^\S$ such that 
\[
g(\mathbf{z}') = g(\mathbf{z}) + \langle\nabla g(\mathbf{z}) ,\mathbf{z}'-\mathbf{z}\rangle + (\mathbf{z}'-\mathbf{z})^\top \nabla^2 g(\xi)(\mathbf{z}'-\mathbf{z}).
\]
Since the Hessian of $g$ is a diagonal matrix with values $-r_j/(1+\xi_j)^3$ for each $j\in \S$, we can obtain the following lower bound
\begin{equation}\label{eq:aux_lemma_fw}
g(\mathbf{z}') \geq g(\mathbf{z}) + \langle\nabla g(\mathbf{z}) ,\mathbf{z}'-\mathbf{z}\rangle - R\cdot\|\mathbf{z}'-\mathbf{z}\|^2
\end{equation}
where $R = \max_{j\in\S}r_j$.
By taking $\mathbf{z}' = \mathbf{z}^{t+1}$ and $\mathbf{z} = \mathbf{z}^t$ in \eqref{eq:aux_lemma_fw} and using the update in Step 4 of Algorithm~\ref{alg:frankwolfe_choice}, we get
\[
g(\mathbf{z}^{t+1}) - g(\mathbf{z}^{t}) \geq  \gamma_t\cdot \langle\nabla g(\mathbf{z}^t) ,\mathbf{z}^{t+1}-\mathbf{z}^t\rangle - R\cdot\|\mathbf{z}^{t+1}-\mathbf{z}^t\|^2\geq \gamma_t\cdot\langle\nabla g(\mathbf{z}^t) ,\tilde{\mathbf{w}}^t-\mathbf{z}^t\rangle - \gamma_t^2\cdot R\cdot\|\tilde{\mathbf{w}}^t-\mathbf{z}^t\|^2
\]
where $\tilde{\mathbf{w}}^t\in\RR_+^\S$ is the vector whose $j$-th component equals $\sum_{i\in\C}w_{j,i}\cdot\phi_{i}(j,S_i^t)$ and $S_i^t$ is the solution found in Step 3 for customer $i\in\C$. Note also due to the diameter of the feasible region, we know that $\|\tilde{\mathbf{w}}^t-\mathbf{z}^t\|^2\leq W$ where $W=\sum_{j\in\S}\left(\sum_{i\in\C}w_{j,i}\right)^2$. Therefore,
\[
g(\mathbf{z}^{t+1}) - g(\mathbf{z}^{t})\geq \gamma_t\cdot\langle\nabla g(\mathbf{z}^t) ,\tilde{\mathbf{w}}^t-\mathbf{z}^t\rangle - \gamma_t^2RW
\]
By taking $\gamma_t = \min\left\{\frac{\langle\nabla g(\mathbf{z}^t) ,\tilde{\mathbf{w}}^t-\mathbf{z}^t\rangle}{2\sqrt{mRW}},1\right\}$ where $m=|\S|$, we have
\begin{align*}
g(\mathbf{z}^{t+1}) - g(\mathbf{z}^{t})&\geq \min\left\{\frac{\langle\nabla g(\mathbf{z}^t) ,\tilde{\mathbf{w}}^t-\mathbf{z}^t\rangle^2}{2\sqrt{mRW}},\langle\nabla g(\mathbf{z}^t) ,\tilde{\mathbf{w}}^t-\mathbf{z}^t\rangle\right\} - \min\left\{\frac{\langle\nabla g(\mathbf{z}^t) ,\tilde{\mathbf{w}}^t-\mathbf{z}^t\rangle^2}{4m},RW\right\}\\
&\geq \frac{\langle\nabla g(\mathbf{z}^t) ,\tilde{\mathbf{w}}^t-\mathbf{z}^t\rangle^2}{2\sqrt{mRW}}- \frac{\langle\nabla g(\mathbf{z}^t) ,\tilde{\mathbf{w}}^t-\mathbf{z}^t\rangle^2}{4m}\\
&=\frac{\langle\nabla g(\mathbf{z}^t) ,\tilde{\mathbf{w}}^t-\mathbf{z}^t\rangle^2}{4\sqrt{mRW}},
\end{align*}
where in the second inequality we used that $\langle\nabla g(\mathbf{z}^t) ,\tilde{\mathbf{w}}^t-\mathbf{z}^t\rangle\leq\|\nabla g(\mathbf{z}^t)\|\cdot\|\tilde{\mathbf{w}}^t-\mathbf{z}^t\|\leq \sqrt{mRW}$.

On the other hand, by concavity of $g$ we know that for any $t\in[T]$,
\[
g(\mathbf{z}^\star) - g(\mathbf{z}^{t}) \leq \langle\nabla g(\mathbf{z}^t),\mathbf{z}^\star-\mathbf{z}^t\rangle\leq\langle\nabla g(\mathbf{z}^t) ,\Phi(\mathbf{S}^t)-\mathbf{z}^t\rangle
\]
where in the second inequality we use optimality of Step 3 in Algorithm~\ref{alg:frankwolfe_choice}. By putting the previous two inequalities together we get,
\[
g(\mathbf{z}^{t+1}) - g(\mathbf{z}^{t})\geq \frac{[g(\mathbf{z}^\star) - g(\mathbf{z}^{t})]^2}{4\sqrt{mRW}}
\]
which is equivalent to
\[
h(\mathbf{z}^t) - h(\mathbf{z}^{t+1}) \geq \frac{h(\mathbf{z}^t)^2}{4\sqrt{mRW}}
\]
where we defined $h(\mathbf{z}) = g(\mathbf{z}^\star) - g(\mathbf{z})$. Solving this classic  recurrence, we get 
\[
h(\mathbf{z}^t) \leq \frac{4\sqrt{mRW}}{t+2}
\]
which in $t=T$ implies that
\[
g(\mathbf{z}^T)\geq g(\mathbf{z}^\star) - \frac{4\sqrt{mRW}}{T+2}.
\]
Therefore, after sufficiently enough iterations $T\gg\sqrt{mRW}$, the error term becomes negligible. Finally, note that due to Lemma~\ref{lemma:feasibility_fw} the points $\mathbf{z}^T$ and $\bm{\tau}^T$ are feasible.
\endproof

\proof[Proof of Lemma~\ref{lemma:supplier_mnl_bound2}.]
Fix supplier $j\in \S$. Since suppliers have MNL choice models, then the demand function of supplier $j$ is $f_j(C)= w_{j}(C)/(1+w_{j}(C))$, where $w_{j}(C) := \sum_{i\in C}w_{j,i}$ and $C\subseteq \C$. Define $C^1_j = \{i\in \C: \ w_{j,i}>\epsilon/(1-\epsilon)\}$ and denote by $C^{\mathbf{S}}_j$ the random set of customers that selected supplier $j$ under assortment profile $\mathbf{S}$. First, observe that 
\[
\EE[f_j(C^{\mathbf{S}}_j)] = \EE\left[\frac{w_j(C^{\mathbf{S}}_j)}{1+w_j(C^{\mathbf{S}}_j)}\right] =\EE\left[\frac{w_j(C^{\mathbf{S}}_j\setminus C_j^1)+w_j(C_j^1)}{1+w_j(C^{\mathbf{S}}_j\setminus C_j^1)+w_j(C_j^1)}\right] > \EE\left[\frac{\tilde{w}_j(C^{\mathbf{S}}_j)}{1+\tilde{w}_j(C^{\mathbf{S}}_j)}\right],
\]
where the expectation is taken over the randomness of $C^{\mathbf{S}}_j$ and the inequality follows by noting that the function $f(z) = z/(1+z)$ is increasing in $z$ and $w_{j,i} > \frac{\epsilon}{1-\epsilon} = \tilde{w}_{j,i}$ for all $i\in C^1_j$. Therefore, we conclude the first inequality
\[
\EE[\Ma_{\mathbf{S}}] = \sum_{j\in \S}r_j\cdot \EE[f_j(C^{\mathbf{S}}_j)] \geq \sum_{j\in \S}r_j\cdot  \EE\left[\frac{\sum_{i\in C^{\mathbf{S}}_j}\tilde{w}_{j,i}}{1+\sum_{i\in C^{\mathbf{S}}_j}\tilde{w}_{j,i}}\right] = \EE[\tilde{\Ma}_{\mathbf{S}}].
\]
To prove the other inequality note that
\[
\EE[f_j(C^{\mathbf{S}}_j)] = \EE\left[\frac{w_j(C^{\mathbf{S}}_j)}{1+w_j(C^{\mathbf{S}}_j)}\right] =\EE\left[\frac{w_j(C^{\mathbf{S}}_j\backslash C^1_j)+w_j(C^1_j)}{1+w_j(C^{\mathbf{S}}_j\backslash C^1_j)+w_j(C^1_j)}\right] \leq \frac{1}{\epsilon}\cdot \EE\left[\frac{\tilde{w}_j(C^{\mathbf{S}}_j)}{1+\tilde{w}_j(C^{\mathbf{S}}_j)}\right],
\]
where the inequality results from observing that 
\[
\frac{w_j(C^{\mathbf{S}}_j\backslash C^1_j)+w_j(C^1_j)}{1+w_j(C^{\mathbf{S}}_j\backslash C^1_j)+w_j(C^1_j)}\leq \frac{1}{\epsilon}\cdot \frac{w_j(C^{\mathbf{S}}_j\backslash C^1_j)+\sum_{i\in C^1_j}\frac{\epsilon}{1-\epsilon}}{1+w_j(C^{\mathbf{S}}_j\backslash C^1_j)+\sum_{i\in C^1_j}\frac{\epsilon}{1-\epsilon}} = \frac{1}{\epsilon}\cdot \frac{\tilde{w}_j(C^{\mathbf{S}}_j)}{1+\tilde{w}_j(C^{\mathbf{S}}_j)}.
\]
By multiplying by $r_j$ and summing over $j\in \S$, we conclude that $\EE[\Ma_{\mathbf{S}}] \leq \frac{1}{\epsilon}\cdot \EE[\tilde{\Ma}_{\mathbf{S}}]$.
\endproof
\section{Appendix to Section~\ref{sec:efficient_algorithms}}\label{sec:appendix_efficient}

\subsection{Missing Details in Section~\ref{sec:unconstrained_mnl_mnl}}\label{sec:appendix_mnl_mnl}

\proof[Proof of Lemma~\ref{lemma:relaxation_mnl_mnl}.]
Consider an optimal solution $\mathbf{z}^\star$ and $\bm{\tau}^\star$ of Problem~\eqref{eq:relaxation_mnl_suppliers}. Define for each $i\in\C$, $j\in\S$
\[
y_{i,j} = \sum_{S:S\ni j}\tau_{i,S}^\star\cdot\frac{1}{1+v_i(S)}  \quad \text{and} \quad
y_{i,0} = \sum_{S\subseteq\S}\tau_{i,S}^\star\cdot\frac{1}{1+v_i(S)}.
\]
Clearly, $y_{i,0}\geq y_{i,j}$ for every $i\in\C,j\in\S$. Also,
\begin{align*}
y_{i,0} + \sum_{j\in \S}v_{i,j}y_{i,j} &= \sum_{S\subseteq \S}\tau_{i,S}^\star\cdot\frac{1}{1+v_i(S)} + \sum_{j\in \S}\sum_{S:S\ni j}\tau_{i,S}^\star\cdot\frac{v_{i,j}}{1+v_i(S)}\\
& = \sum_{S\subseteq \S}\tau_{i,S}^\star\cdot\frac{1}{1+v_i(S)} +\sum_{S\subseteq \S}\tau_{i,S}^\star\cdot\sum_{j\in S}\frac{v_{i,j}}{1+v_i(S)}\\
& = \sum_{S\subseteq \S}\tau_{i,S}^\star\cdot\frac{1}{1+v_i(S)} +\sum_{S\subseteq \S}\tau_{i,S}^\star\cdot\frac{v_{i}(S)}{1+v_i(S)}\\
& = \sum_{S\subseteq \S}\tau^\star_{i,S} = 1
\end{align*}
where in the last equality we used the feasibility of $\bm{\tau}^\star$. Note as well that
\[
\sum_{i\in\C}w_{j,i}v_{i,j}y_{i,j} = \sum_{i\in\C}w_{j,i}\sum_{S:S\ni j}\tau_{i,S}^\star\cdot\frac{v_{i,j}}{1+v_i(S)} = z_j
\]
due to the feasibility of $\mathbf{z}$. This concludes the first direction. For the opposite direction, consider an optimal solution $\mathbf{z}^\star$ and $\mathbf{y}^\star$ of Problem~\eqref{eq:relaxation_mnl_mnl}. We construct $\bm{\tau}$ as in Lemma~\ref{lemma:rounding_algorithm}. Given this, we already know that $\sum_{S\subseteq\S}\tau_{i,S}=1$ for every $i\in\C$. Moreover, as shown in the proof of Lemma~\ref{lemma:rounding_algorithm}
\[
v_{i,j}y^\star_{i,j} = \sum_{S:S\ni j}\tau_{i,S}\cdot\frac{v_{i,j}}{1+v_i(S)}.
\]
Therefore,
\[
z^\star_j=\sum_{i\in\C}w_{i,j}v_{i,j}y^\star_{i,j} = \sum_{i\in\C}w_{i,j}\cdot\sum_{S:S\ni j}\tau_{i,S}\cdot\frac{v_{i,j}}{1+v_i(S)},
\]
which implies that $\mathbf{z}^\star$ and $\bm{\tau}$ are feasible in Problem~\eqref{eq:relaxation_mnl_suppliers}. We conclude the proof by noting that both optimal objective values are equal.
\endproof

\subsection{Missing Details in Section \ref{sec:computational_study}}\label{sec:details_experiments}
Let us denote Unif$(a,b)$ the uniform random variable in the interval $[a,b]$ and Exp($\lambda$) the exponential random variable with parameter $\lambda$. In the following, we describe how each instance was generated:
\begin{itemize}
    \item {\it Heterogeneous.} First, the compatibility between each customer $i$ and supplier $j$ was sample by drawing two Bernoulli random variables with probability $0.707$ (so the product is 1 with probability $1/2$). Then, parameters $w_{j,i}\sim\text{Exp}(1)$ and $v_{i,j}$ were sampled differently depending on the subclass: (i) {\it Low-value regime}, $v_{i,j}\sim\text{Unif}[0,1]$; (ii) {\it High-value regime}, $v_{i,j}\sim\text{Unif}[2,5]$; (iii) {\it Mix}, with probability 0.8 we sample $v_{i,j}\sim\text{Unif}[0,1]$ and, otherwise,  $v_{i,j}\sim\text{Unif}[2,5]$.
    \item {\it Homogeneous-Uniform.} Each value $v_{i,j}=v_j$ was sampled as in the Heterogeneous-mix instances above. On the other hand, we consider $w_{j,i}=1/w_{j}$ where $w_j\sim\text{Exp}(1)$.
    \item {\it Small Probability Regime.} These instances were generated as in Heterogeneous-mix, however, $w_{j,i}\sim\text{Unif}[0,\epsilon/(1-\epsilon)]$.
\end{itemize}

\begin{figure}[htb!]
\begin{center}
\begin{subfigure}{0.49\textwidth}
\caption{\footnotesize Heterogeneous, mix}
\label{fig:heterogeneous_mix_time}
\begin{center}
\includegraphics[width=0.8\textwidth]{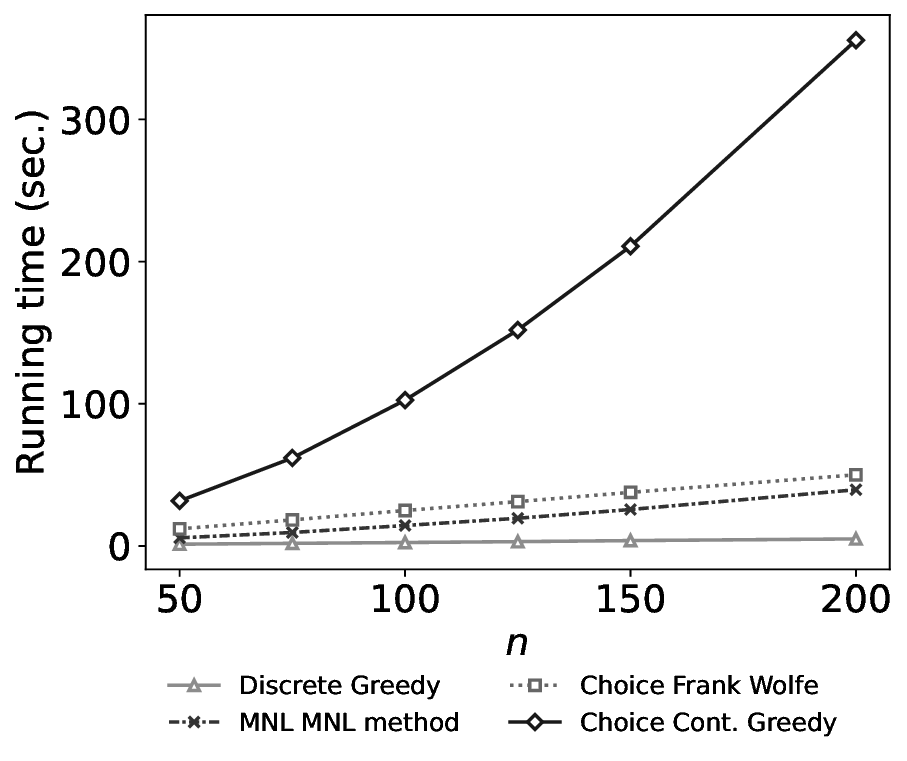}
\end{center}
\end{subfigure}
\begin{subfigure}{0.49\textwidth}
\caption{\footnotesize Homogeneous-Uniform}
\label{fig:homogeneous_time}
\begin{center}
\includegraphics[width=0.8\textwidth]{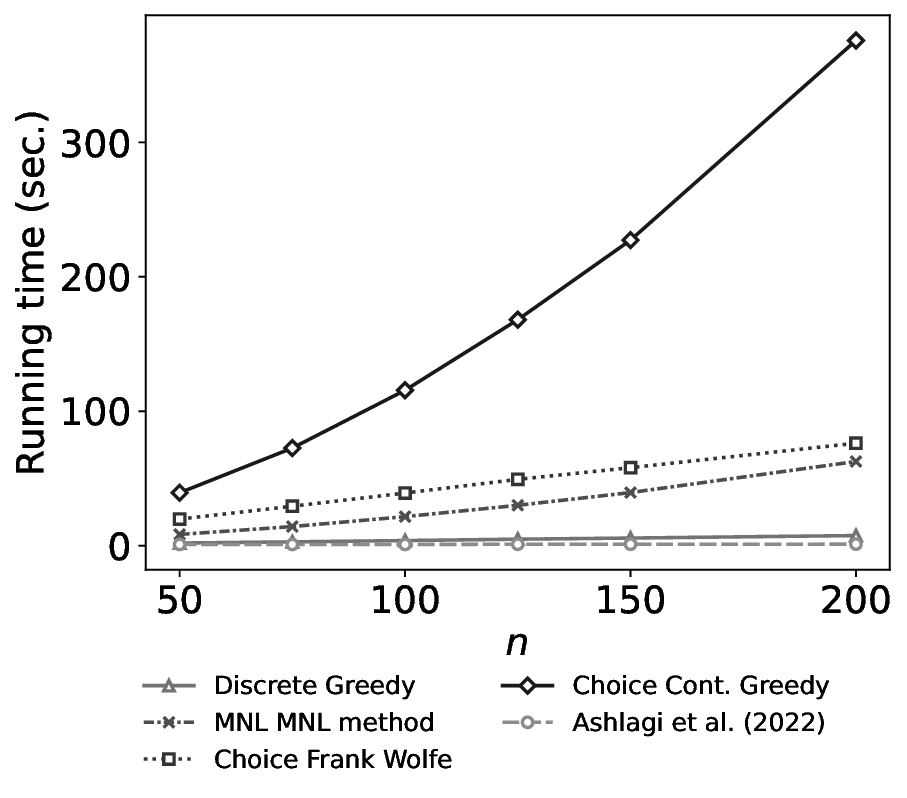}
\end{center}
\end{subfigure}
\caption{Time performance results}\label{fig:time_performance}
\end{center}
\end{figure}

\section{Appendix to Section \ref{sec:other_results}}\label{sec:proofs_managerial} 

\subsection{Missing Proofs in Section~\ref{sec:nested_logit}}\label{sec:proofs_nl}
First, note that for the nested-logit model under the $\epsilon$-small probability regime with $\epsilon\in(0,1)$, we have that for any $i\in \C$ and $j\in\S$
\begin{equation}\label{eq:aux_nested}
f_j(\{i\}) = \frac{w_{j,i}^{\nu_k}}{1+w_{j,i}^{\nu_k}}\leq \epsilon\cdot f_j(\C) \qquad \Rightarrow \qquad w_{j,i}\leq \left(\frac{\epsilon}{1-\epsilon}\right)^{1/\nu_k}.
\end{equation}
Similar to our approach in Section~\ref{sec:general_mnl}, we consider the following relaxation
\begin{align}\label{eq:relaxation_nl_suppliers}
\max &\quad \sum_{j\in \S}r_j\cdot \frac{\sum_{k\in[K]}z_{j,k}^{\nu_k}}{1+\sum_{k\in[K]}z_{j,k}^{\nu_k}}\\
s.t. &\quad \sum_{i\in \C_k}w_{j,i}\cdot\sum_{S\in\mathcal{I}_i: S\ni j}\tau_{i,S}\cdot\phi_i(j;S)=z_{j,k}, \qquad \text{for all}  \ j\in \S, k\in[K] \notag\\
&\quad \sum_{S\in\I_i} \tau_{i,S} =1, \hspace{12em} \text{for all} \ i\in \C, \notag \\
&\quad z_j, \tau_{i,S}\geq 0 \hspace{13.2em} \text{for all} \ i\in \C, \ S\in\mathcal{I}_i, \ j\in \S. \notag
\end{align}
Note that the objective is concave and separable as for each $j\in\S$, the values in $\mathbf{z}_j = (z_{j,1},\ldots,z_{j,K})$ do not appear in the objective of a different supplier $j'\neq j$. More importantly, by Jensen's inequality we can show:
\begin{lemma}\label{lemma:relaxation_nl_suppliers}
Problem~\eqref{eq:relaxation_nl_suppliers} is a relaxation of Problem~\eqref{def:opt_problem} when suppliers have $\nu$-NL choice models.
\end{lemma}
Given the above, we can easily adapt Algorithm~\ref{alg:frankwolfe_choice} to obtain the following result:
\begin{proposition}\label{prop:convergence_fw_nl}
The adaptation of Algorithm~\ref{alg:frankwolfe_choice} to the $\nu$-NL choice model outputs a feasible solution $(\mathbf{z}^T,\tau^T)$ such that
\[
\sum_{j\in \S}r_j\cdot\frac{\sum_{k\in[K]}(z^T_{j,k})^{\nu_k}}{1+\sum_{k\in[K]}(z^T_{j,k})^{\nu_k}} \geq \OPT_{\eqref{eq:relaxation_nl_suppliers}} - \; O\left(\frac{1}{T}\right)
\]
where $\OPT_{\eqref{eq:relaxation_nl_suppliers}}$ is the optimal value of Problem~\eqref{eq:relaxation_nl_suppliers}.
\end{proposition}
Therefore, our goal now is to prove an equivalent version of Lemma~\ref{lemma:supplier_mnl_bound1}.
\begin{lemma}\label{lemma:supplier_nl_bound1}
Assume suppliers have $\nu$-NL choice models with values satisfying~\eqref{eq:aux_nested} for all $j\in \S$, $i\in \C$ and $\epsilon\in(0,1/(n^{2\bar{\nu}}+1)]$ where $\bar{\nu}=\max_k\nu_k$. Then, for any vector $\mathbf{d}\in[0,1]^{\C\times \S}$, we have
\begin{equation}\label{eq:lowerbound_nl}
\sum_{j\in \S}r_j\cdot F_j(\mathbf{d}_j)\geq \left(1-\frac{1}{n}-\frac{1}{n^2}\right)\cdot\sum_{j\in \S}r_j\cdot\frac{\sum_{k\in[K]}\left(\sum_{i\in\C_k}w_{j,i}d_{i,j}\right)^{\nu_k}}{1+\sum_{k\in[K]}\left(\sum_{i\in\C_k}w_{j,i}d_{i,j}\right)^{\nu_k}}.
\end{equation}
\end{lemma}
The proof of this lemma is somehow similar to the proof of Lemma~\ref{lemma:supplier_mnl_bound1}, but we need to account for the presence of multiple nets.
\proof[Proof.]
Fix $j\in \S$ and consider $i\in \C$ such that $d_{i,j}>0$. Then, conditioned on the event that $i$ chose $j$ (which happens with probability $d_{ij}$), the expected probability (with respect to the product distribution) that $j$ selects $i\in C_k$ back (as given in the first equality of Expression~\eqref{eq:demand_nl}) can be lower bounded as follows
\begin{align*}
\EE_{C\sim\mathbf{d}_j^{-i}}&\left[\phi_{j}(i,C_k)\big| \text{$i$ chose $j$}\right] \\
&=\EE\left[\frac{w_{j,i}}{w_{j,i} + w_j(C_k\setminus\{i\})}\cdot \frac{\left(w_{j,i} + w_j(C_k\setminus\{i\})\right)^{\nu_k}}{1+\left(w_{j,i} + w_j(C_k\setminus\{i\})\right)^{\nu_k}+\sum_{q\neq k}(w_j(C_q))^{\nu_q}} \Big| \text{$i$ chose $j$}\right]\\
& = \EE\left[\frac{w_{j,i}}{(w_{j,i} + w_j(C_k\setminus\{i\}))^{1-\nu_k}}\cdot \frac{1}{1+\left(w_{j,i} + w_j(C_k\setminus\{i\})\right)^{\nu_k}+\sum_{q\neq k}(w_j(C_q))^{\nu_q}} \right]\\
& \geq \EE\left[\frac{w_{j,i}}{(w_{j,i} + w_j(C_k\setminus\{i\}))^{1-\nu_k}}\cdot \frac{1}{1+\frac{1}{n}+\sum_{q\neq k}(w_j(C_q))^{\nu_q}} \right]\\
& = \EE\left[\frac{w_{j,i}}{(w_{j,i} + w_j(C_k\setminus\{i\}))^{1-\nu_k}}\right]\cdot \EE\left[\frac{1}{1+\frac{1}{n}+\sum_{q\neq k}(w_j(C_q))^{\nu_q}} \right]\\
& \geq  \frac{\left(1-\frac{1}{n}\right)\cdot w_{j,i}}{(w_{j,i} + \EE\left[w_j(C_k\setminus\{i\})\right])^{1-\nu_k}}\cdot \frac{1}{1+\sum_{q\neq k}(\EE\left[w_j(C_q)\right])^{\nu_q}} \\
& =  \frac{\left(1-\frac{1}{n}\right)\cdot w_{j,i}}{(w_{j,i} + \sum_{\ell\in\C_k\setminus\{i\}}w_{j,\ell}d_{j,\ell})^{1-\nu_k}}\cdot \frac{1}{1+\sum_{q\neq k}(\sum_{\ell\in\C_q}w_{j,\ell}d_{j,\ell})^{\nu_q}} \\
& \geq  \frac{\left(1-\frac{1}{n}\right)\cdot w_{j,i}}{(w_{j,i} + \sum_{\ell\in\C_k\setminus\{i\}}w_{j,\ell}d_{j,\ell})^{1-\nu_k}}\cdot \frac{1}{1+\sum_{q\in[K]}(\sum_{\ell\in\C_q}w_{j,\ell}d_{j,\ell})^{\nu_q}} \\
& =  \frac{\left(1-\frac{1}{n}\right)\cdot w_{j,i}}{w_{j,i} + \sum_{\ell\in\C_k\setminus\{i\}}w_{j,\ell}d_{j,\ell}}\cdot \frac{(w_{j,i} + \sum_{\ell\in\C_k\setminus\{i\}}w_{j,\ell}d_{j,\ell})^{\nu_k}}{1+\sum_{q\in[K]}(\sum_{\ell\in\C_q}w_{j,\ell}d_{j,\ell})^{\nu_q}} \\
& \geq  \frac{\left(1-\frac{1}{n}\right)\cdot w_{j,i}}{w_{j,i} + \sum_{\ell\in\C_k\setminus\{i\}}w_{j,\ell}d_{j,\ell}}\cdot \frac{(\sum_{\ell\in\C_k}w_{j,\ell}d_{j,\ell})^{\nu_k}}{1+\sum_{q\in[K]}(\sum_{\ell\in\C_q}w_{j,\ell}d_{j,\ell})^{\nu_q}} \\
& \geq  \frac{\left(1-\frac{1}{n}\right)\cdot w_{j,i}}{\frac{1}{n^2} + \sum_{\ell\in\C_k\setminus\{i\}}w_{j,\ell}d_{j,\ell}}\cdot \frac{(\sum_{\ell\in\C_k}w_{j,\ell}d_{j,\ell})^{\nu_k}}{1+\sum_{q\in[K]}(\sum_{\ell\in\C_q}w_{j,\ell}d_{j,\ell})^{\nu_q}} \\
& \geq  \frac{\left(1-\frac{1}{n}-\frac{1}{n^2}\right)\cdot w_{j,i}}{\sum_{\ell\in\C_k\setminus\{i\}}w_{j,\ell}d_{j,\ell}}\cdot \frac{(\sum_{\ell\in\C_k}w_{j,\ell}d_{j,\ell})^{\nu_k}}{1+\sum_{q\in[K]}(\sum_{\ell\in\C_q}w_{j,\ell}d_{j,\ell})^{\nu_q}} \\
& \geq  \frac{\left(1-\frac{1}{n}-\frac{1}{n^2}\right)\cdot w_{j,i}}{\sum_{\ell\in\C_k}w_{j,\ell}d_{j,\ell}}\cdot \frac{(\sum_{\ell\in\C_k}w_{j,\ell}d_{j,\ell})^{\nu_k}}{1+\sum_{q\in[K]}(\sum_{\ell\in\C_q}w_{j,\ell}d_{j,\ell})^{\nu_q}} \\
\end{align*}
where the expectation is over $C\sim\mathbf{d}_j^{-i}$ and $\mathbf{d}_j^{-i}$ is the vector without  $\mathbf{d}_j$ without the $i$-th component. The first equality follows from the assumption that $i$ chose $j$. The first inequality is due to 
\[
w_j(C_k)\leq |C_k|\cdot \left(\frac{\epsilon}{1-\epsilon}\right)^{1/\nu_k} \leq n \cdot \left(\frac{1}{n^2}\right)^{\bar{\nu}/\nu_k}\leq \frac{1}{n}
\]
since $\epsilon/(1-\epsilon)$ is increasing in $\epsilon$ and $\epsilon\leq 1/(n^{2\bar\nu}+1)$, moreover, $(1/n^2)^{\bar{\nu}/\nu_k}\leq (1/n^2)^{\nu_k/\nu_k}$ because $\nu_k\leq \bar{\nu}$. The next equality is because of independence of the customers' choices and the partition structure of the nests. The second inequality is due to Jensen's inequality and $\EE[w_j(C_q)^\alpha]\leq\EE[w_j(C_q)]^\alpha$ for $\alpha\leq 1$. The next equality uses that 
\[
\EE[w_j(C_q)] = \sum_{\ell\in\C_q}w_{j,\ell}\cdot\PP(\ell\text{ chooses }j) =\sum_{\ell\in\C_q}w_{j,\ell}\cdot d_{\ell,j},
\]
The third inequality is because we added one more term in the denominator of the second fraction.
The fourth inequality is because $w_{j,i}\geq w_{j,i}d_{i,j}$. In the fifth inequality we use that $w_{j,i}\leq (\epsilon/(1-\epsilon))^{1/\nu_k}\leq 1/n^2$. In the final inequality we added one more term in the denominator of the first fraction.

Therefore, the demand function of supplier $j$ can be lower bounded as follows
\begin{align*}
\EE_{C\sim\mathbf{d}_j}[f_j(C)] &= \EE_{C\sim\mathbf{d}_j}\left[\sum_{i\in C} \phi_{j}(i,C) \right] = \EE_{C\sim\mathbf{d}_j}\left[\sum_{k\in[K]}\sum_{i\in \C_k} \phi_{j}(i,C) \cdot \uno_{\{i\in C\}} \right]\\
& = \sum_{k\in[K]}\sum_{i\in \C_k} \EE_{C\sim\mathbf{d}_j^{-i}}\left[\phi_{j}(i,C)\cdot  \uno_{\{i\in C\}}  \big| \text{$i$ chooses $j$}\right]\PP(\text{$i$ chooses $j$})\\
& =\sum_{k\in[K]}\sum_{i\in \C_k} \EE_{C\sim\mathbf{d}_j^{-i}}\left[\phi_{j}(i,C)  \big| \text{$i$ chooses $j$}\right]\cdot d_{i,j} \\
&\geq \left(1-\frac{1}{n}-\frac{1}{n^2}\right)\cdot \sum_{k\in[K]}\sum_{i\in \C_k}\frac{ w_{j,i}}{\sum_{\ell\in\C_k}w_{j,\ell}d_{j,\ell}}\cdot \frac{(\sum_{\ell\in\C_k}w_{j,\ell}d_{j,\ell})^{\nu_k}}{1+\sum_{q\in[K]}(\sum_{\ell\in\C_q}w_{j,\ell}d_{j,\ell})^{\nu_q}}\cdot d_{i,j}\\
& =\left(1-\frac{1}{n}-\frac{1}{n^2}\right)\cdot\frac{\sum_{k\in[K]}\left(\sum_{i\in\C_k}w_{j,i}d_{i,j}\right)^{\nu_k}}{1+\sum_{k\in[K]}\left(\sum_{i\in\C_k}w_{j,i}d_{i,j}\right)^{\nu_k}}.
\end{align*}
Finally, by multiplying by $r_j$ and summing over $j\in \S$, we obtain the desired inequality. 
\endproof
\proof[Proof of Proposition~\ref{prop:approx_nl}.]
The proof is analogous to the proof of Theorem~\ref{thm:fw_nearoptimality_epsregime}.
\endproof

\subsubsection{Relationship between Curvature and Epsilon-small Probability Regime}\label{app:curvature}
In the following, we use the shorthand $f(S-e)$ to denote $f(S\setminus\{e\})$ and $f(e)$ for $f(\{e\})$. The curvature $c\in[0,1]$ of a monotone submodular function $f:2^\U\to\RR_+$ measures how close the function is to be linear. Formally, assuming that $f(\emptyset)=0$, $c$ is such that such that $f(A+e)-f(A)\geq (1-c)\cdot f(e)$ for all $A\subseteq\U$ and $e\notin A$ and, since $f$ is also monotone submodular, we know that $f(A+e)-f(A)\leq f(e)$, thus $c=0$ means that $f$ is linear and $c=1$ is valid for any monotone function. Moreover, one can easily see that $c$ is attained when
\[
c = 1 - \min_{e\in\U}\left\{\frac{f(\U)-f(\U-e)}{f(e)}\right\}
\]
\cite{vondrak2010curvature} shows that for a monotone submodular function with curvature $c$, the continuous greedy algorithm for submodular optimization achieves a $(1-\exp(-c))/c$ approximation factor for the matroid constrained setting. Therefore, if $c$ is close to 0 then this factor goes to 1.

We now connect the above to the $\epsilon$-small probability regime. Assume that for every supplier $j\in\S$ and customer $i\in\C$, we have that $f_j(i)\leq \epsilon\cdot f_j(\C)$. Then, the curvature $c_j$'s of the demand functions $f_j$'s are such that 
\[
1-c_j = \min_{i\in\C}\left\{\frac{f_j(\C)-f_j(\C-i)}{f_j(i)}\right\} \geq \frac{1}{\epsilon}\cdot\min_{i\in\C}\left\{\frac{f_j(\C)-f_j(\C-i)}{f_j(\C)}\right\}.\]
Therefore, the curvature of $f_j$ depends on the marginal demand contribution of every customer $i$ relative to the entire set of customers. Suppose the suppliers' demand functions satisfy the $\xi$-large marginal probability regime then for every $j\in\S$ and $i\in\C$,  $f_j(\C)-f_j(\C-i) \geq \xi\cdot f_j(\C)$ which implies that $1-c_j\geq \frac{\xi}{\epsilon}$. From this, we could then adapt the proof of Theorem~\ref{thm:main_general} to show that, under a $(\xi,\epsilon)$-regime, Algorithm~\ref{alg:continous_greedy_choice} achieves a $\frac{1-\exp(-(1-\xi/\epsilon))}{1-\xi/\epsilon}$ approximation factor. More importantly, when $\xi=\epsilon$ then this factor goes to 1.

\subsection{Comparison between Matching Models}\label{sec:operational_insights}

In the following, we briefly discuss how the order of the matching process in Problem~\eqref{def:opt_problem} impact the quality of the solution when varying selectivity and the size of each side. Specifically, note that the model  assumes that customers initiate the matching process and suppliers ``respond'', however, an analogous model can be proposed in which suppliers choose first and customers respond back. Our goal is to analyze the ratio between the optimal values of both variants of the model depending on two aspects of the market: (1) the attractiveness of the agents (customers and suppliers); (2) the imbalance on the size of each side.  For the former, we show that, in the worst-case, the optimal matching size significantly differs when the more selective agents (i.e., whose opposite side is less attractive) initiate the process.
\begin{proposition}\label{prop:selectivity1}
In the worst-case, the ratio between the optimal value when customers initiate the matchmaking process and the optimal value when suppliers initiate the process can be arbitrarily close to zero.
\end{proposition}
This drastic difference appears because, from a design perspective, one can obtain more matches in expectation by controlling the assortments of those agents that are more selective. As we show next, even in a simple instance (not necessarily worst-case) we observe differences on the optimal values.
\begin{proposition}\label{prop:selectivity2}
Let $w, v\in(0,1]$ be two constants with respect to $n$. Assume that customers and suppliers follow MNL choice models with preference weights defined as follows: $w_{j,i} = w$ and $v_{i,j} = v$ for all $i\in \C$, $j\in \S$. If $w\leq v/3$, then for sufficiently large $n$ the optimal value when suppliers initiate the matchmaking process is at least the optimal value when customers initiate the process.
\end{proposition}
Finally, as we show next, selectivity does not play a significant role if there are enough agents on the less selective side.

\begin{proposition}\label{prop:size_analysis}
Let $w,v\in(0,1]$ two constants with respect to $m$. Assume that customers and suppliers follow MNL choice models with preference weights defined as follows: $v_{i,j} = v$ and $w_{j,i} = w$ for all $j\in \S$, $i\in \C$. If $n\geq m/w$ and $v\leq 1/3$, then for sufficiently large $m$ the optimal value when customers initiate the matching process is at least the optimal value when suppliers initiate the process. 
\end{proposition}

\subsubsection{Proofs}
For simplicity, in all of the results above, we consider the problem of maximizing the expected number of matches, i.e., $r_j = 1$ for all $j\in \S$. We also assume that there are no constraints on the assortments. 
Before we provide the proofs, we need to define the necessary notation for the setting in which suppliers initiate the matchmaking process. First, this process can be analogously defined as it was made for customers in the Introduction. We denote by $\mathbf{C}=(C_1,\ldots,C_m)$ an assortment profile of customers which is presented to suppliers. We analogously define $\Ma_{\mathbf{C}}$ the random variable indicating the revenue obtained in the final matching when suppliers start choosing. Therefore, the main problem that we focus for this setting is
\[
\max\{\EE[\Ma_{\mathbf{C}}]: \ C_j\in \I_j \text{ for all } j\in \S\},
\]
where $\I_j$ is the family of feasible assortments for supplier $j\in \S$. For the remainder of this section, we denote by $\OPT'$ the optimal value of this problem. For simplicity, we consider the unconstrained setting in both sequential models, i.e., $\I_i=2^\S$ and $\I_j=2^\C$. From Jensen's inequality, when customers follow an MNL choice model, we have the following bound for a given assortment family $\mathbf{C}$ and $\Ma_{\mathbf{C}}$:
\begin{align}\label{eq:inverted_bound1}
\EE[\Ma_{\mathbf{C}}]\leq \sum_{i\in \C}\frac{\sum_{j\in \S}v_{i,j}\cdot\phi_{j}(i,C_j)}{1+\sum_{j\in \S}v_{i,j}\cdot\phi_{j}(i,C_j)}.
\end{align}
Moreover, if customers' values over suppliers are such that $v_{i,j}\leq 1$ for all $i\in \C$, $j\in \S$, we also have (equivalent to Lemma~\ref{lemma:supplier_mnl_bound1} with $\epsilon=1/2$)
\begin{align}\label{eq:inverted_bound2}
\EE[\Ma_{\mathbf{C}}]\geq\frac{1}{2}\cdot\sum_{i\in \C}\frac{\sum_{j\in \S}v_{i,j}\cdot\phi_{j}(i,C_j)}{1+\sum_{j\in \S}v_{i,j}\cdot\phi_{j}(i,C_j)}.
\end{align}
Inequalities \eqref{eq:inverted_bound1} and \eqref{eq:inverted_bound2}, will be crucial for the results in this section.
Finally, we keep the notation for the process in which customers start choosing: for a given assortment family $\mathbf{S}$ of suppliers that will be presented to the customers, $\Ma_{\mathbf{S}}$ corresponds to the random variable indicating the revenue obtained in the final matching. Recall that $\OPT$ denotes the optimal value of the problem when customers initiate the process. 

\proof[Proof of Proposition \ref{prop:selectivity1}.]
Assume $|\S|=|\C|=n$. Consider customers and suppliers that follow MNL choice models with the following values: (1) there exists a constant $\alpha>0$ such that $\sum_{j\in \S}v_{i,j}\geq \alpha n$; (2) $w_{j,i}=1/n$ for every $j\in \S$, $i\in \C$.

First, let us lower bound $\OPT'$. For this, we will study the following assortment family $\mathbf{C} = (\C,\ldots,\C)$, i.e., we show all the customers to every supplier. Then,
\[
\OPT'\geq\EE[\Ma_{\mathbf{C}}]\geq \frac{1}{4}\sum_{i\in \C}\frac{\sum_{j\in \S}v_{i,j}\cdot\phi_{j}(i,\C)}{1+\sum_{j\in \S}v_{i,j}\cdot\phi_{j}(i,\C)} = \frac{1}{4}\cdot\sum_{i\in \C}\frac{\sum_{j\in \S}v_{i,j}}{2n+\sum_{j\in \S}v_{i,j}}\geq \frac{1}{4}\cdot\sum_{i\in \C}\frac{\alpha}{2+\alpha} = \frac{\alpha n}{4(2+\alpha)}
\]
where the first inequality follows from Inequality \eqref{eq:inverted_bound2} (applied to any scores $v_{i,j}$'s so the inequality has a factor 1/4 instead of 1/2 as in the proof of Theorem~\ref{thm:fw_factor_anyregime}) and the first equality follows from noting that 
\[
\phi_{j}(i,\C) = \frac{w_{j,i}}{1+\sum_{i\in \C}w_{j,i}} = \frac{1}{2n}.
\]

In the second part of this proof, we will give an upper bound to $\OPT$.
\[
\OPT = \EE[\Ma_{\mathbf{S}^*}]\leq \sum_{j\in \S}\frac{\sum_{i\in \C}w_{j,i}\phi_{i}(j,S_i^\star)}{1+\sum_{i\in \C}w_{j,i}\phi_{i}(j,S_i^\star)} =\sum_{j\in \S}\frac{\frac{1}{n}\sum_{i\in \C}\phi_{i}(j,S_i^\star)}{1+\frac{1}{n}\sum_{i\in \C}\phi_{i}(j,S_i^\star)}.
\]
Also, since $w_{j,i}\leq 1$, Lemma \ref{lemma:supplier_mnl_bound1} with $\epsilon=1/2$ implies 
\[
\OPT = \EE[\Ma_{\mathbf{S}^\star}]\geq \frac{1}{2}\cdot\sum_{j\in \S}\frac{\sum_{i\in \C}w_{j,i}\phi_{i}(j,S_i^\star)}{1+\sum_{i\in \C}w_{j,i}\phi_{i}(j,S_i^\star)} =\frac{1}{2}\cdot\sum_{j\in \S}\frac{\frac{1}{n}\sum_{i\in \C}\phi_{i}(j,S_i^\star)}{1+\frac{1}{n}\sum_{i\in \C}\phi_{i}(j,S_i^\star)}
\]
We want to maximize the expression on the right. Define $g(z_1,\ldots,z_j) = \sum_{j\in \S}\frac{z_j}{n+z_j}$. Note that $f(.)$ is a Schur-concave function. The worst case for our claim is when the expression on the right is maximized. This happens when ``few'' customers pick the outside option (in expectation), therefore we can assume that in expectation all customers do not pick the outside option, i.e., 
\[
\sum_{j\in \S}\sum_{i\in \C}\phi_{i}(j,S_i^\star)= \sum_{i\in \C}\sum_{j\in \S}\phi_{i}(j,S_i^\star)= \sum_{i\in \C} 1 = n
\]
This implies that $\sum_{j\in \S}z_j = n$. Since $f$ is Schur-concave, then it is maximized when $z_1=\cdots = z_{|\S|}=1$; recall $|\S| = |\C| = n$. This condition means that every supplier gets chosen (in expectation) by one customer. For instance, one instance of these characteristics is the following: $v_{i,0}= 1$ and $v_{i,j} \gg 1$ for all $i\in \C$, $j\in \S$. Therefore, we obtain
\[
\OPT \leq \sum_{j\in \S}\frac{\frac{1}{n}\sum_{i\in \C}\phi_{i}(j,S_i^\star)}{1+\frac{1}{n}\sum_{i\in \C}\phi_{i}(j,S_i^\star)} = |\S|\cdot\frac{1}{n+1} = \frac{n}{n+1}
\]
Finally, the desired result follows by noting that
\[
\frac{\OPT}{\OPT'} \leq \frac{n/(n+1)}{\alpha n/(8+4\alpha)}\xrightarrow[n \to \infty]{} 0. 
\]
This result for $\OPT$ is intuitive because of the selectivity of suppliers, $w_{j,i} = 1/n$. We would need that the expected number of customers that choose each supplier is $O(n)$. In this way, each supplier will pick one customer back with constant probability and we would get a matching of expected size $O(n)$.
\endproof

\proof[Proof of Proposition \ref{prop:selectivity2}.]
First, let us upper bound $\OPT$. As in the proof of Proposition \ref{prop:selectivity1}
\[
\OPT = \EE[\Ma_{\mathbf{S}^\star}]\leq \sum_{j\in \S}\frac{\sum_{i\in \C}w_{j,i}\phi_{i}(j,S_i^\star)}{1+\sum_{i\in \C}w_{j,i}\phi_{i}(j,S_i^\star)} =\sum_{j\in \S}\frac{w\sum_{i\in \C}\phi_{i}(j,S_i^\star)}{1+w\sum_{i\in \C}\phi_{i}(j,S_i^\star)}.
\]
We want to maximize the expression on the right. Define $f(z_1,\ldots,z_{|\S|}) = \sum_{j\in \S}\frac{wz_j}{1+w z_j}$. Note that $f(.)$ is a Schur-concave function which is maximized when $z_1 = \cdots = z_{|\S|}$. Let us analyze if there exists such as solution 
\[
\phi_{i}(j,S_i^\star) = \frac{v_{i,j}}{1+\sum_{\ell\in S_i^\star}v_{i,\ell}} = \frac{v}{1+|S_i^\star|v}
\]
Since we want to make $\sum_{i\in \C}\phi_{i}(j,S_i^\star)$ as close to 1 as possible, then let us consider $S_i^\star = \S$ for all $i\in \C$. Therefore, $\phi_{i}(j,S_i^\star) = v/(1+nv)$ and 
\[
\sum_{i\in \C}\phi_{i}(j,S_i^\star) = \frac{v}{1+nv} |\{i\in \C: \ S_i^\star\ni j\}| = \frac{nv}{1+nv},
\]
where the second equality comes from observing that each supplier $j$ is present in every assortment. The last expression is close to 1 for sufficiently large $n$, since $v$ is a constant. We conclude that
\[
\OPT \leq \sum_{j\in \S}\frac{w\sum_{i\in \C}\phi_{i}(j,S_i^\star)}{1+w\sum_{i\in \C}\phi_{i}(j,S_i^\star)}  = n\cdot\frac{wv n}{1+nv+wv n} \approx n\cdot\frac{w}{1+w}.
\]
Now, let us study $\OPT'$. For this note that since $v_{i,j}\in[0,1]$, then for any assortment family $\mathbf{C}$ because of Inequality \eqref{eq:inverted_bound2} we have
\[
\OPT' \geq \EE[\Ma_{\mathbf{C}}]\geq \frac{1}{2}\cdot\sum_{i\in \C}\frac{\sum_{j\in \S}v_{i,j}\cdot\phi_{j}(i,C_j)}{1+\sum_{j\in \S}v_{i,j}\cdot\phi_{j}(i,C_j)}  =\frac{1}{2} \sum_{i\in \C}\frac{v\sum_{j\in \S}\phi_{j}(i,C_j)}{1+v\sum_{j\in \S}\phi_{j}(i,C_j)}.
\] 
Let us consider $C_j = \C$ for all $j\in \S$. Therefore, 
\[
\phi_{j}(i,C_j) = \frac{w_{j,i}}{1+\sum_{k\in \C}w_{j,k}}  = \frac{w}{1+nw}
\]
 and
\[
\sum_{j\in \S}\phi_{j}(i,C_j) = \frac{w}{1+nw} |\{j\in \S: \ C_j\ni i\}| = \frac{nw}{1+nw}.
\]
We conclude that
\[
\OPT' \geq \frac{1}{2}\cdot \sum_{i\in \C}\frac{v\sum_{j\in \S}\phi_{j}(i,C_j)}{1+v\sum_{j\in \S}\phi_{j}(i,C_j)}  = \frac{n}{2}\cdot\frac{nwv}{1+nw+nwv}\approx n\cdot \frac{v}{2(1+v)}.
\]
We conclude the proof by noting that when $n\to\infty$
\[
\frac{\OPT}{\OPT'} \leq\frac{\OPT}{\EE[\Ma_{\mathbf{C}}]}  \leq \frac{w/(1+w)}{v/2(1+v)} = 2\cdot \frac{w/(1+w)}{v/(1+v)}\leq 2\cdot\frac{1+v}{3+v}\leq 1
\]
where the second inequality comes from $u\leq v/3$ and the last inequality follows from $v\leq 1$. 
\endproof

\proof[Proof of Proposition \ref{prop:size_analysis}.]
First, let us lower bound $\OPT$. As in the proof of Proposition \ref{prop:selectivity2}, we can obtain a lower bound with a feasible solution. Consider $S_i = \S$ for all $i\in \C$, then $\phi_{i}(j,S_i)= v/(1+mv)$ for all $i\in \C$ and $j\in V$.
\[
\OPT \geq \EE[\Ma_{\mathbf{S}}]\geq \frac{1}{2}\sum_{j\in \S}\frac{\sum_{i\in \C}w_{j,i}\phi_{i}(j,S_i)}{1+\sum_{i\in \C}w_{j,i}\phi_{i}(j,S_i)} =\frac{1}{2}\sum_{j\in \S}\frac{wnv/(1+mv)}{1+wnv/(1+mv)} \geq \frac{m}{2}\cdot \frac{mv}{1+2mv},
\]
where in the last inequality we use that $n\geq m/w$.

Now, let us upper bound $\OPT'$.
\[
\OPT' = \EE[\Ma_{\mathbf{C}^\star}]\leq\sum_{i\in \C}\frac{\sum_{j\in \S}v_{i,j}\phi_{j}(i,C_j^\star)}{1+\sum_{j\in \S}v_{i,j}\phi_{j}(i,C_j^\star)}  = \sum_{i\in \C}\frac{v \sum_{j\in S}\phi_{j}(i,C_j^\star)}{1+v \sum_{j\in \S}\phi_{j}(i,C_j^\star)}.
\]
As in the previous proofs, we want to maximize the expression on the right. Define $f(z_1,\ldots,z_{n}) = \sum_{i\in \C}\frac{v z_i}{1+v z_i}$. Note that $f(.)$ is a Schur-concave function which is maximized when $z_1 = \cdots = z_{n}$. Let us analyze if there exists such as solution. Observe that
\[
\phi_{j}(i,C_j^\star)= \frac{w_{j,i}}{1+\sum_{k\in C_j^\star}w_{j,k}} = \frac{w}{1+|C_j^\star|w}.
\]
Since we want to make $\sum_{j\in \S}\phi_{j}(i,C_j^\star)$ as close to 1 as possible, then let $\C'\subseteq \C$ a subset of customers such that $|\C'| = m$ and consider $C_j^\star = \C'$ for all $j\in \S$. Recall that the maximum expected number of matches is upper bounded by $\min\{n,m\} = m$ since agents select only one option. Therefore, $\phi_{j}(i,C_j^\star)= w/(1+mw)$ and 
\[
\sum_{j\in \S}\phi_{j}(i,C_j^\star) = \frac{w}{1+mw} |\{j\in \S: \ C_j^\star\ni i\}| = \frac{mw}{1+mw}.
\]
Since only customers in $\C'$ will get chosen, then 
\[
\OPT' \leq \sum_{i\in \C'}\frac{v \sum_{j\in \S}\phi_{j}(i,C_j^\star)}{1+v \sum_{j\in \S}\phi_{j}(i,C_j^\star)} = m\cdot\frac{vmw}{1+mw+vmw}.
\]
We conclude the proof by noting that
\[
\frac{\OPT}{\OPT'} \geq \frac{m^2v/2(1+2mv)}{m^2vw/(1+mw+vmw)} = \frac{1+mw+vmw}{2w(1+2mv)} \xrightarrow[m \to \infty]{} \frac{w+vw}{4wv} =\frac{1+v}{4v}\geq 1
\]
where the last inequality follows from $v\leq 1/3$.  
\endproof

\end{document}